\documentstyle[12pt]{article}

\begin{document}
% --------------------------------------------------------------------------
\hyphenation{bi-quad-ra-tic}
\hyphenation{co-or-di-na-te}
\newcommand{\beq}{\begin{equation}}
\newcommand{\eeq}{\end{equation}}
\newcommand{\calE}{{\cal E}}
\newcommand{\calK}{{\cal K}}
\newcommand{\calL}{{\cal L}}
\newcommand{\calQ}{{\cal Q}}
\newcommand{\calP}{{\cal P}}
\newcommand{\sn}{{\rm sn}}
\newcommand{\cn}{{\rm cn}}
\newcommand{\dn}{{\rm dn}}
\newcommand{\const}{{\it const}}
\newcommand{\etal}{{\it et al. }}
\newcommand{\dg}{^{\rm o}}
%\lta and \gta produce < and > signs with twiddle underneath:
\def\spose#1{\hbox to 0pt{#1\hss}}
\def\lta{\mathrel{\spose{\lower 3pt\hbox{$\mathchar"218$}}
     \raise 2.0pt\hbox{$\mathchar"13C$}}}
\def\gta{\mathrel{\spose{\lower 3pt\hbox{$\mathchar"218$}}
     \raise 2.0pt\hbox{$\mathchar"13E$}}}
\parskip=0pt
\parindent=-24pt
% --------------------------------------------------------------------------
\normalsize
\pagestyle{empty}

{\Large\bf I.~~A star orbiting around a supermassive rotating black~hole:
 Free motion and corrections due to star-disc collisions}
\bigskip\bigskip

\noindent
{\it revised manuscript accepted by Mon. Not. R. astr. Soc.}

\bigskip\bigskip\bigskip

{\Large\bf II.~Relativistic precession of the orbit of a star near a
 supermassive black hole}

\bigskip\bigskip

\noindent
{\it accepted by Astrophys. J.}

\bigskip\bigskip\bigskip
\parindent=-8pt

\noindent
{\Large V. Karas$^{1,2}$ and D. Vokrouhlick\'y$^{2,3}$}
\bigskip\bigskip\bigskip\bigskip

$^1$ {\it Scuola Internazionale Superiore di Studi Avanzati,
 Via Beirut 2-4,\par\noindent
 I-34013 Trieste, Italy}\par
$^2$ {\it Astronomical Institute, Charles University,
 \v Sv\'edsk\'a 8, CS-150 00 Prague,\par\noindent
 Czech Republic}\par
$^3$ {\it Observatoire de la ${\it C\!\hat{o}te}$ d'Azur, dept. CERGA,
 Av. Nicolas Copernic,\par\noindent
 F-06130 Grasse, France}\par

\bigskip\bigskip\bigskip\bigskip\bigskip\bigskip
\noindent
Ref. SISSA 69/93/A

\parindent=15pt
\parskip=5pt
\newpage
\pagestyle{plain}
\pagenumbering{arabic}
\noindent
{\Large\bf A star orbiting around a supermassive rotating black hole:
 Free motion and corrections due to star-disc collisions}
\bigskip\bigskip

\noindent
{\large\bf ABSTRACT}

\noindent
Our aim is to study the evolution of the orbit of a
star under the influence of interactions
with an accretion disc in an AGN. The model considered consists of a
low-mass compact object orbiting a supermassive black hole and colliding
periodically with the accretion disc.
Approximate calculations based mostly on Newtonian theory of gravity
have been carried out by several authors to estimate the effects
of circularization of initially eccentric
orbits and their dragging to the disc plane. Here, we present the first step to
a more adequate general relativistic approach in which the gravitational
field of the nucleus is described by the Kerr metric.
The star is assumed to move along a geodesic arc between successive
interactions with an equatorial accretion disc.
We solved relevant formulae for the geodesic motion in terms
of elliptic integrals
and constructed a fast numerical code which, after specifying details
of the star-disc interaction, enables us to follow the trajectory of the star
for many revolutions and study the evolution of its eccentricity and
inclination with respect to the disc. Lense-Thirring
precession of the orbit is potentially a very
important effect for observational
confirmation of the presence of a rotating black hole in the nucleus.
Our approach takes effects of the
Lense Thirring precession into account with no approximation.

The model of star-disc interactions has been suggested to
explain the X-ray variability observed in the Seyfert galaxy NGC~6814.
We briefly discuss on this subject.

\bigskip

\bigskip\bigskip

\noindent
{\large\bf 1\quad INTRODUCTION AND MOTIVATION}
 %%%%%%%%%%%%%%%%%%%%%%%%%%%%%%%%%%%
%%%%%%%%%%%%%%%%%%%%%%%%%%%%%%%%%%%%%%%%%%%%%%%%%%%%%%%%%%%%%%%%%%%%%%%%%%%%%
\medskip

\noindent
General interest in studying star-disc interactions in the nuclei of
galaxies has greatly increased in the last years.
This is partly due to the fact that they appear important in explaining the
X-ray variability of active galactic nuclei (AGN).
Although it is generally believed that many galaxies, and active galaxies
in particular, harbour massive black holes in their cores, there is no
direct observational confirmation for this paradigm. The origin of this
difficulty is apparent: complicated plasma physics of
the matter swirling around the
black hole makes it difficult to distinguish the effects of general
relativity---although they may be essential for the mechanism of energy
generation itself.
Potentially very important observable is the X-ray data
on variability of active galactic nuclei (for the recent review
see Wallinder, Kato \& Abramowicz 1992). Although our understanding of the
 origin of X-rays
is not satisfactory, it is often accepted that they are generated in the
inner regions of the
accretion disc where general relativistic effects are significant.
When trying to describe these effects it is useful
to separate details of the mechanism generating
radiation (described in the local frame co-moving with the matter),
i.e. the local physics of the interaction on the one
hand, and observable effects
as seen by a distant observer on the other.
In our previous work (Karas, Vokrouhlick\'y \& Polnarev 1992, Paper~I)
we developed a code
which can be used in many astrophysically relevant situations to calculate
images of various effects occurring in the close vicinity of the rotating
(Kerr)
black hole. Our code deals efficiently with problems treated
originally by Cunningham~\& Bardeen (1973).
All relativistic effects on photons
(like gravitational and Doppler shift of frequency and bending of light
rays) were taken into account.
As an example, we applied the code
to the ``hot spot'' model of the Seyfert galaxy
NGC~6814 which tries to explain its X-ray variability on the scale of
approximately 3 hours
in terms of a bright orbiting spot (or spots) located on the accretion disc
(Abramowicz \etal 1992), and we also discussed the case of a large number
of spots with different intrinsic characteristics which may
be relevant in explaining the X-ray variability of AGN on still
shorter time scales (Abramowicz \etal 1991).

In this paper we present a method for calculating the evolution of the orbital
parameters of a compact star orbiting a massive black hole. Such a
star may come from a binary or a cluster tidally disrupted by the
central black hole (Hills 1988, Novikov, Pethick \& Polnarev 1992).
It can be deposited in a tightly bound orbit
with a close pericentre where relativistic effects are important.
The star interacts with the accretion disc only at the moment
when it crosses the equatorial plane and
this interaction weakly affects its motion. Cumulative effects of
successive tiny interactions circularize the trajectory and
change the orbital plane
into the plane of the disc. Mutual star-disc interactions have been discussed
by Syer, Clarke \& Rees (1991) as a possible origin of fueling and
variability of AGN. Particularly, they discussed relative
time-scales for circularization of the orbit and its ``grinding''
to the disc plane, and the final
radius of the embedded orbit. Naturally, quantitative estimates
depend on poorly known
details of the interaction (cf. Zurek, Siemiginowska \& Colgate 1992).
A compact
star coliding with an accretion disc is one of viable models for
NGC~6814 (Abramowicz 1992, Sikora~\& Begelman 1992, Rees 1993). All present
models possess
their advantages and difficulties when compared to observational data
and they need to be investigated in greater detail.
General relativistic precession
of orbital nodes (Lense-Thirring effect)---if detected---would
strongly support models
of AGN with rotating supermassive black holes. Lense-Thirring precession
affects the inclined trajectory of a star, but it has been discussed only
in special cases of free orbits with a large radius compared to the
gravitational radius of the central black hole (Lense~\& Thirring 1918)
and spherical orbits around an extreme Kerr black hole (Wilkins 1972).
We consider the more general case of eccentric orbits which interact
with the equatorial disc and we do not assume any particular value
of the angular momentum of the central black hole.

Star-disc interactions are a complex problem. Because we assume that the
disc produces only a weak perturbation of free motion of the star, we
can attack the problem in several steps.
The present paper concentrates on an effective method for calculating the free
motion of the star between its successive interactions with the disc.
Simple examples of how star-disc interactions may change the picture
are given in the last Section and they will be discussed in a
forthcoming paper in greater detail. Detailed calculation of the Lense-Thirring
frequency relevant for this model
will be given elsewhere (Karas \& Vokrouhlick\'y 1993).
Once the motion of the star and its interaction with the disc are
specified, we can apply the method of Paper~I to compute
the shape of the light curve or resultant spectrum. [See also
Cunningham~\& Bardeen (1973), de~Felice, Nobili~\& Calvani (1974),
Luminet (1978) and Laor~\& Netzer (1989)].
Recently, Fabian \etal (1989), Kojima (1991) and Laor (1991)
applied analogous approach to study line profiles from accretion discs.
For a more complete list of references cf. Paper~I.

Let us briefly describe the configuration of the model.
We consider a low-mass compact object (white dwarf, neutron star
or black hole) orbiting the central massive
black hole of the AGN. In our approach, we restrict ourselves
to the assumption that the orbiting object moves along a time-like
geodesic in the unperturbed background Kerr metric outside the equatorial
plane.
Thus we implicitly assume that (i) the orbiting object is sufficiently
compact and/or far from the central black hole (we neglect any coupling of
the star's higher multipoles to the background curvature), and (ii) its
mass is very small compared to the mass of the central black hole
(we do not consider perturbations of the background
metric). We also neglect the influence of radiating gravitational waves.
In several astrophysical situations such assumptions may not
be appropriate---see, e.g., Carter~\& Luminet (1983), Luminet~\& Marck (1985),
and Carter (1992) who studied tidal squeezing of the stars by the nearby
black hole. Hartle~\& Thorne (1985) and Suen (1986) developed a scheme
for multipole-tidal interactions of relativistic objects. As for
the motion of close and comparable in mass black holes, see, e.g.,
D'Eath (1975\/a,\/b). We do not consider such extreme situations in this paper.
Kates (1980) has shown that the star will
move closely along a geodesic in the unperturbed background
metric for sufficiently long time provided the ratio of its
mass and the characteristic reference length of the background metric
is a small parameter. We have in mind
situations where this parameter is of the order $10^{-5}$ or even
less. This is important to note because we will attempt to follow the
trajectory of the orbiting object for long periods of time.
Next, we assume that the accretion disc is in the equatorial plane of the
central black hole. We exclude thick
disc models from our present considerations.
Each time the object crosses the equatorial
plane it interacts with the disc (provided the intersection
is between the outer and the inner edges of the disc).
In other words, we assume that the trajectory consists of arcs of free
geodesic motion above (or below) the disc plane
and impulsive changes of the orbital parameters
at the moment of passages through the disc. The whole ``physics''
of the problem is compressed into the prescription governing the changes
of the orbital parameters when crossing the equatorial plane.
The interaction is assumed very weak which implies that relative changes
of energy, angular momentum and other quantities characterizing the orbit
of the star are much less than unity in each single event.
We should note that the geodesic motion in the Kerr space-time is
integrable; thus, we do not expect strong dependence of the shape of the orbit
on initial conditions which is typical for a chaotic motion.
An alternative approach which employs a statistical
description with appropriately averaged quantities is under preparation.
The separation of dynamical and
physical aspects, as we introduce it in the present paper, appears
very advantageous and it allows us to employ
a fast method for computing the evolution of orbital parameters.
\bigskip\bigskip

\noindent
{\large\bf 2\quad THE MAPPING -- DETAILS OF THE CALCULATION} %%%%%%%%%%%%%%%%%%
%%%%%%%%%%%%%%%%%%%%%%%%%%%%%%%%%%%%%%%%%%%%%%%%%%%%%%%%%%%%%%%%%%%%%%%%%%%%%%
\smallskip

\noindent
We consider the geodesic motion of a test particle (representing
a star or a low-mass black hole) in the fixed Kerr
background metric. We are
interested only in short arcs of geodesic motion with the following
boundary conditions: the initial point (indexed ``$i$\/'') lies in the
equatorial plane $(\theta=\pi/2$ in Boyer-Lindquist
coordinates), the final point (indexed ``$f$\/'') is the nearest successive
intersection of the orbit with
the equatorial plane.
We employ a `mapping' by which we understand an analytical algorithm to
evaluate the final position $(r_f,$$\phi_f)$ from
the initial position $(r_i,$$\phi_i)$ with constants of
motion assumed to be given.
In applications, we also need to
know the transformation from initial to final
velocities $dr/dt,$ $d\theta/dt,$ $d\phi/dt$
to obtain the full starting information for the
physical model of the interaction of the orbiting object with the
accretion disc. The whole
procedure is trivial in principle because the geodesic motion
is separable in the Kerr background metric and the equations
of motion can be reduced to a set of ordinary first order
differential equations (Carter 1968). Our main task is to
handle the problem efficiently.

Carter's equations involve squares of the
velocities. As the star crosses the equatorial plane,
the latitudinal velocity changes its sign
periodically. In order to treat the case of the radial velocity, we introduce
the sign function $\eta\equiv{\rm sgn}(dr/dt).$
Thus, our mapping is the analytical transformation
$$(r,\phi;\eta)_i\stackrel{\bf M}{\longrightarrow}(r,\phi;\eta)_f\;.$$
We will also be interested in analytical evaluation of the
delay in coordinate time which is necessary to pass from the initial to the
final configuration in the disc plane: $(t_f - t_i)$. In particular, this
is important for reconstruction of the AGN photometric curve,
provided similar periodic process as described here (star-disc
interactions) arise its variability (Karas \& Vokrouhlick\'y 1993).

It is worth noting that the code based on this mapping technique is
optimized as concerns {\it both} the speed and accuracy.
The effective step of the method is the whole
orbital arc and it cannot be made greater in principle. Moreover,
{\it the exact analytical solution} of the problem is chosen as
a sample function covering one integration step (instead of, for example,
polynomials in Runge-Kutta methods).
We remark that the name `mapping' comes from
analogous methods developed in celestial mechanics
(e.g. Wisdom 1982, Murray 1986).

We use the standard notation for the Kerr metric (Bardeen 1973).
Quantities with the dimension of length in geometrized units are divided
by the mass of the central black hole $M,$ and they are thus made
dimensionless.
Time-like geodesics in the Kerr space-time can be integrated in the form
(Carter 1968)
\beq %vvvvvvvvvvvvvvvvvvvvvvvvvvvvvvvvvvvvvvvvvvvvvvvvvvvvvvvvvvvvvvvvvvvvvvv
 t_f - t_i=\int_{r_i}^{r_f}\frac{r^2(r^2+a^2)\calE+2ar(a\calE-\Phi)}
 {\Delta R(r)^{1/2}}\,dr+
 \int_{\theta_i}^{\theta_f}\frac{a^2\calE\cos^2\theta}
 {\Theta(\theta)^{1/2}}\,d\theta,
\eeq %^^^^^^^^^^^^^^^^^^^^^^^^^^^^^^^^^^^^^^^^^^^^^^^^^^^^^^^^^^^^^^^^^^^^^^^
\beq %vvvvvvvvvvvvvvvvvvvvvvvvvvvvvvvvvvvvvvvvvvvvvvvvvvvvvvvvvvvvvvvvvvvvvvv
 \phi_f - \phi_i=\int_{r_i}^{r_f}\frac{r^2\Phi+2r(a\calE-\Phi)}
 {\Delta R(r)^{1/2}}\,dr+
 \int_{\theta_i}^{\theta_f}\frac{\Phi\cot^2\theta}{\Theta(\theta)
 ^{1/2}}\,d\theta,
\eeq %^^^^^^^^^^^^^^^^^^^^^^^^^^^^^^^^^^^^^^^^^^^^^^^^^^^^^^^^^^^^^^^^^^^^^^^
\beq %vvvvvvvvvvvvvvvvvvvvvvvvvvvvvvvvvvvvvvvvvvvvvvvvvvvvvvvvvvvvvvvvvvvvvvv
 \int_{r_i}^{r_f}\frac{dr}{R(r)^{1/2}}=
 \int_{\theta_i}^{\theta_f}\frac{d\theta}{\Theta(\theta)^{1/2}}.
\eeq %^^^^^^^^^^^^^^^^^^^^^^^^^^^^^^^^^^^^^^^^^^^^^^^^^^^^^^^^^^^^^^^^^^^^^^^
Here,
\beq %vvvvvvvvvvvvvvvvvvvvvvvvvvvvvvvvvvvvvvvvvvvvvvvvvvvvvvvvvvvvvvvvvvvvvvv
 R(r)=(\calE^2-1)r^4+2r^3+\left[ (\calE^2-1)a^2-\Phi^2-Q\right]r^2
 +2\calK r-a^2Q,
\eeq %^^^^^^^^^^^^^^^^^^^^^^^^^^^^^^^^^^^^^^^^^^^^^^^^^^^^^^^^^^^^^^^^^^^^^^^
\beq %vvvvvvvvvvvvvvvvvvvvvvvvvvvvvvvvvvvvvvvvvvvvvvvvvvvvvvvvvvvvvvvvvvvvvvv
 \Theta(\theta)=Q-\left[a^2(1-\calE^2)+\Phi^2\sin^{-2}\theta\right]
 \cos^2\theta, \qquad Q=\calK-(\Phi-a\calE)^2 ,
\eeq %^^^^^^^^^^^^^^^^^^^^^^^^^^^^^^^^^^^^^^^^^^^^^^^^^^^^^^^^^^^^^^^^^^^^^^^
and
$$ \Sigma = r^2 + a^2 \cos^2\theta, \qquad \Delta = r^2 + a^2
 -2r, $$
$$ A = (r^2 + a^2)^2 - \Delta a^2 \sin^2\theta. $$
The constants of motion, $\calE=-p_t,$ $\Phi=p_\phi,$ and $\calK,$
can be expressed in terms
of components of the four-momentum in the locally non-rotating frame
(LNRF):
\beq %vvvvvvvvvvvvvvvvvvvvvvvvvvvvvvvvvvvvvvvvvvvvvvvvvvvvvvvvvvvvvvvvvvvvvvv
 \calE=\left[{\left(\frac{\Sigma\Delta}{A}\right)}^{1/2}p^{\hat{t}}+
 \frac{2ar}{{\left(\Sigma A\right)}^{1/2}}\,p^{\hat{\phi}}\right]
 _{r_i,\theta_i=\frac{\pi}{2}}, \qquad
 \Phi=\left[\left(\frac{A}{\Sigma}\right)^{1/2}\,p^{\hat{\phi}}
 \right]_{r_i,\theta_i=\frac{\pi}{2}},
\eeq %^^^^^^^^^^^^^^^^^^^^^^^^^^^^^^^^^^^^^^^^^^^^^^^^^^^^^^^^^^^^^^^^^^^^^^^
\beq %vvvvvvvvvvvvvvvvvvvvvvvvvvvvvvvvvvvvvvvvvvvvvvvvvvvvvvvvvvvvvvvvvvvvvvv
 \calK=\left[(\Phi-a\calE)^2+\Sigma(p^{\hat{\theta}})^2\right]
 _{r_i,\theta_i=\frac{\pi}{2}}.
\eeq %^^^^^^^^^^^^^^^^^^^^^^^^^^^^^^^^^^^^^^^^^^^^^^^^^^^^^^^^^^^^^^^^^^^^^^^
Components of the four-momentum in terms of direction cosines in
the local sky of an observer at rest with respect to LNRF are
$$p^{\hat{t}}=\gamma,\qquad p^{\hat{r}}=\gamma v\cos\alpha,$$
\beq %vvvvvvvvvvvvvvvvvvvvvvvvvvvvvvvvvvvvvvvvvvvvvvvvvvvvvvvvvvvvvvvvvvvvvvv
p^{\hat{\theta}}=\gamma v\sin\alpha\cos\beta,\qquad
p^{\hat{\phi}}=\gamma v\sin\alpha\sin\beta,
\eeq %^^^^^^^^^^^^^^^^^^^^^^^^^^^^^^^^^^^^^^^^^^^^^^^^^^^^^^^^^^^^^^^^^^^^^^^
where $v$ is the tetrad velocity of the particle in
LNRF and the Lorentz factor
$$\gamma=\frac{1}{\sqrt{1-v^2}}.$$
Specification of
the initial conditions as a result of the star-disc interaction in the
local frame co-moving with the matter of the disc requires another boost to
the disc co-rotating frame (DCF). In the case of a Keplerian thin disc in the
equatorial plane the linear velocity of the DCF with respect to LNRF is
$$v_{\rm DCF}=\frac{r^2-2ar^{1/2}+a^2}{(r^{3/2}+a)\Delta}.$$
It is known that integrals (1)--(3) can be reduced to standard
elliptic integrals, but only the simplest cases have been discussed in the
literature. The explicit form of relevant formulae
depends on the value of the constants of motion and initial conditions.
We shall restrict ourselves to the most interesting astrophysical case:
stable, energetically bound trajectories which cross the
equatorial plane many times repeatedly. Thus, we assume $0<\calE<1.$
We exclude singular cases of orbits lying exactly in the equatorial plane
$(Q=0;$ Bardeen, Press \& Teukolsky 1972) or
those intersecting the rotation axis
$(\Phi=0;$ Stoghianidis~\& Tsoubelis 1987); we also exclude
the case of the extremely
rotating black hole, $a=1.$ In the Schwarzschild case
$(a=0)$ the geodesic is always planar, while $a\neq 0$ leads to the
Lense-Thirring precession of the orbit and we have to take
into consideration the dragging
of the nodes [Wilkins (1972) discussed this effect for
spherical orbits in the extreme Kerr case].

First, we simplify the integrals (1)--(3) to a form which can be
directly found in standard tables of elliptic integrals (Byrd \& Friedman 1971,
Gradshteyn \& Ryzhik 1980, Gr\" obner \& Hofreiter 1965).
For this purpose we need to find the roots of $R(r)$ and
$\Theta(\theta).$
Real roots of are the turning points of the radial
and latitudinal motion, respectively.
We specify the initial point of the geodesic
as $(t_i,r_i,\pi/2,\phi_i)$ and we look for the final position
$(t_f,r_f,\pi/2,\phi_f)$ which is in the equatorial plane again.
The polynomial $R(r)$ governing the radial motion
is of the fourth order with
$R(r_i)\geq 0$. In our case, $R(r)<0$ for $r\rightarrow\infty$
and $r=0.$ Thus we can find two real roots, $r_A\in (0,r_i),$
$r_B\in (r_i,\infty),$ and for the remaining two roots we obtain
the quadratic equation
\beq %vvvvvvvvvvvvvvvvvvvvvvvvvvvvvvvvvvvvvvvvvvvvvvvvvvvvvvvvvvvvvvvvvvvvvvv
(\calE^2-1)r^2+\left[(\calE^2-1)(r_A+r_B)+2\right]r-\frac{a^2Q}{r_Ar_B}=0.
\eeq %^^^^^^^^^^^^^^^^^^^^^^^^^^^^^^^^^^^^^^^^^^^^^^^^^^^^^^^^^^^^^^^^^^^^^^^
Supposing all roots are real we can denote them, in a descending sequence,
as $r_1>r_2>r_3>r_4.$
We exclude the possibilities of multiple roots because such
situations are singular in the sense that they
occur for precisely arranged values of $\calE$ and $\Phi$.
The probability that interaction with the
accretion disc will lead to such values is zero (in the measure
sense). Thus we have
\beq %vvvvvvvvvvvvvvvvvvvvvvvvvvvvvvvvvvvvvvvvvvvvvvvvvvvvvvvvvvvvvvvvvvvvvvv
 I_r\equiv\int\frac{dr}{R(r)^{1/2}}=\frac{1}{\sqrt{1-\calE^2}}\int
 \frac{dr}{\sqrt{\mid(r-r_1)(r-r_2)(r-r_3)(r-r_4)\mid}}\,.
\eeq %^^^^^^^^^^^^^^^^^^^^^^^^^^^^^^^^^^^^^^^^^^^^^^^^^^^^^^^^^^^^^^^^^^^^^^^
In the case of two real and two complex roots $(r_1>r_2$ and $r_3,$
$\overline{r_3},$ respectively), we obtain
\beq %vvvvvvvvvvvvvvvvvvvvvvvvvvvvvvvvvvvvvvvvvvvvvvvvvvvvvvvvvvvvvvvvvvvvvvv
 I_r=\frac{1}{\sqrt{1-\calE^2}}\int
 \frac{dr}{(r_1-r)(r-r_2)\sqrt{(r-\chi_1)^2+\chi_2^2}}\,,
\eeq %^^^^^^^^^^^^^^^^^^^^^^^^^^^^^^^^^^^^^^^^^^^^^^^^^^^^^^^^^^^^^^^^^^^^^^^
where
$$\chi_1=\Re e(r_3),\qquad \chi_2=\Im m(r_3).$$
The latitudinal motion is governed by the polynomial
$\Theta_\mu(\mu)\equiv\sin^2\theta\;\Theta(\theta)$
[Eq.~(5)].
Solving the bi-quadratic equation in $\mu\equiv\cos\theta,$
\beq %vvvvvvvvvvvvvvvvvvvvvvvvvvvvvvvvvvvvvvvvvvvvvvvvvvvvvvvvvvvvvvvvvvvvvvv
a^2(1-\calE^2)\mu^4-\left[Q+a^2(1-\calE^2)+\Phi^2\right]\mu^2+Q=0,
\eeq %^^^^^^^^^^^^^^^^^^^^^^^^^^^^^^^^^^^^^^^^^^^^^^^^^^^^^^^^^^^^^^^^^^^^^^^
we obtain the roots $\mu_+>\mu_->0;$
The latitudinal motion is only possible in the region
$\mu\in\;\langle-\mu_-,\mu_-\rangle$ and
\beq %vvvvvvvvvvvvvvvvvvvvvvvvvvvvvvvvvvvvvvvvvvvvvvvvvvvvvvvvvvvvvvvvvvvvvvv
 I_\mu\equiv\int\frac{d\mu}{\Theta_\mu(\mu)^{1/2}}=
 \frac{1}{a\sqrt{1-\calE^2}}\int\frac{d\mu}
 {\sqrt{(\mu_+^2-\mu^2)(\mu_-^2-\mu^2)}}\,.
\eeq %^^^^^^^^^^^^^^^^^^^^^^^^^^^^^^^^^^^^^^^^^^^^^^^^^^^^^^^^^^^^^^^^^^^^^^^
Analogously, the azimuthal motion can be solved in the form
\beq %vvvvvvvvvvvvvvvvvvvvvvvvvvvvvvvvvvvvvvvvvvvvvvvvvvvvvvvvvvvvvvvvvvvvvvv
 \phi_f-\phi_i=\left[2(a\calE-\Phi)A_++\Phi B_+\right]I_++
 \left[2(a\calE-\Phi)A_-+\Phi B_-\right]I_-+\Phi J_\mu\,,
\eeq %^^^^^^^^^^^^^^^^^^^^^^^^^^^^^^^^^^^^^^^^^^^^^^^^^^^^^^^^^^^^^^^^^^^^^^^
where
\beq %vvvvvvvvvvvvvvvvvvvvvvvvvvvvvvvvvvvvvvvvvvvvvvvvvvvvvvvvvvvvvvvvvvvvvvv
 I_\pm=\int\frac{dr}{(r-r_\pm)R(r)^{1/2}}\,,\qquad
 J_\mu=\int\frac{d\mu}{(1-\mu^2)\Theta_\mu(\mu)^{1/2}}\,,
\eeq %^^^^^^^^^^^^^^^^^^^^^^^^^^^^^^^^^^^^^^^^^^^^^^^^^^^^^^^^^^^^^^^^^^^^^^^
$$A_\pm=\pm\frac{r_\pm}{r_+-r_-}\,,
 \qquad B_\pm=\pm\frac{2r_\pm-a^2}{r_+-r_-}\,,$$
$$r_\pm=1\pm\sqrt{1-a^2}\,.$$
Finally, for the time coordinate we obtain
\begin{eqnarray} %vvvvvvvvvvvvvvvvvvvvvvvvvvvvvvvvvvvvvvvvvvvvvvvvvvvvvvvvvvv
 \lefteqn{\nonumber
 t_f-t_i=\calE\left(J_r+2K_r\right)+
 2\left[B_+r_+\calE+a(a\calE-\Phi)A_+\right]I_+ }\\
 & & +2\left[B_-r_-\calE+a(a\calE-\Phi)A_-\right]I_-+
 4\calE I_r+a^2\calE K_\mu
\end{eqnarray} %^^^^^^^^^^^^^^^^^^^^^^^^^^^^^^^^^^^^^^^^^^^^^^^^^^^^^^^^^^^^^^
with
\beq %vvvvvvvvvvvvvvvvvvvvvvvvvvvvvvvvvvvvvvvvvvvvvvvvvvvvvvvvvvvvvvvvvvvvvvv
 J_r\equiv\int\frac{r^2dr}{R^{1/2}}\,,\qquad
 K_r\equiv\int\frac{r\,dr}{R^{1/2}}\,,\qquad
 K_\mu\equiv\int\frac{\mu^2d\mu}{\Theta_\mu(\mu)^{1/2}}\,.
\eeq %^^^^^^^^^^^^^^^^^^^^^^^^^^^^^^^^^^^^^^^^^^^^^^^^^^^^^^^^^^^^^^^^^^^^^^^
It is straightforward but moderately tedious to derive the explicit form of
the mapping. We give relevant formulae in Appendix.

\bigskip\bigskip

\noindent
{\large\bf 3\quad THE EVOLUTION OF ORBITS -- SIMPLE EXAMPLES}
%%%%%%%%%%%%%%%%%%
%%%%%%%%%%%%%%%%%%%%%%%%%%%%%%%%%%%%%%%%%%%%%%%%%%%%%%%%%%%%%%%%%%%%%%%%%%%%%%%
\smallskip

\noindent
There are several interesting issues closely related to our problem
which have not been fully understood as yet.
In this Section we present simple examples of secular changes of
orbital parameters of a star orbiting a supermassive $(10^6$--$10^9\,M_\odot)$
black hole and interacting with the accretion disc. In particular, we
study the eccentricity and inclination of the orbit,
as they are introduced in Appendix. Differences from previous estimates
which have been made within the framework of Newtonian gravity are found
to be significant for nearly unstable orbits.
These differences can be
very subtle and interfere with the details of star-disc interaction and thus
at first we adopted an extremely simplified (and perhaps unrealistic)
description of the interaction. The Lense-Thirring
precession of the orbit was included fully relativistically
with no approximation. This is important for our particular example we
discuss below, NGC~6814. Mittaz~\& Branduardi-Raymont (1989) and Done \etal
(1993) determined the periodicity in modulation of the X-ray
emission from NGC~6814 to be $\approx 12,000$ seconds. Somewhat uncertain
value for the mass of the central
black hole was estimated to $\approx 10^7\,M_\odot$ and thus
the radius of the corresponding orbit is a few tens of the gravitational
radius of the black hole---rather close to
the horizon where approximation methods for the
Lense-Thirring precession are no longer satisfactory.
We should note, for completeness, that if the star-disc interaction is switched
off, the eccentricity of the orbit remains constant in time
and the value of the precession is exactly that of the Lense-Thirring
precession. In the following examples we tune the strength of the star-disc
interaction in such a way that the relative change of orbital parameters is of
the order $\approx 10^{-5}$ in each interaction and we follow $\approx
10^5$ revolutions. Each of the following Figures
shows the sequence of radial coordinates $r$ for successive intersections
of the trajectory with the disc---one point corresponds to one intersection,
two intersections correspond to one revolution of the orbiter.
(Alternatively, instead of the number of intersections $N$ we could use
coordinate time $t$
for labelling the $x$-axis; Figures remain very similar in shape
but the second possibility appears more adequate for
plotting computed light curves which should be related to the
observer's time at infinity.) Upper and lower boundaries of the distribution
of intersections are current values of the apocentre and the pericentre,
respectively. Intersections are (seemingly randomly) scattered
in the whole range between these boundaries due to the shift of the pericentre
and Lense-Thirring precession (if $a\neq 0).$ Frequencies corresponding to
both of these effects are quantitatively studied in Karas \& Vokrouhlick\'y
(1993).

In the first example we assume that, as a result of the interaction,
the {\it difference} in the azimuthal components of the
star and the disc material
(evaluated in DCF) is decreased:
\beq%vvvvvvvvvvvvvvvvvvvvvvvvvvvvvvvvvvvvvvvvvvvvvvvvvvvvvvvvvvvvvvvvvvvv
\Delta v_{\phi,\,{\rm DCF}}\longrightarrow\alpha_\star\,
\Delta v_{\phi,\,{\rm DCF}},
\eeq%^^^^^^^^^^^^^^^^^^^^^^^^^^^^^^^^^^^^^^^^^^^^^^^^^^^^^^^^^^^^^^^^^^^^
where $\alpha_\star$ is a phenomenological
parameter $\lta 1.$ Initially retrograde orbits (those with
$I>90\dg)$ decrease the absolute magnitude of their $\Phi$ component of
angular momentum
due to interactions with the disc and they either get captured by the black
hole
or become prograde. Once they are prograde, the star acquires
angular momentum from the disc and moves away from the black hole.
Simultaneously, both the eccentricity
and the inclination decrease, and the orbital period increases.
The effect of energy dissipation due to crashing through the disc
is not considered in Eq. (18).
The model is of course inadequate when the inclination reaches zero and
the star becomes a part of the disc. Figure~1 is an example of such an orbit.
As mentioned above, the unperturbed geodesic
motion in the Kerr metric is integrable and thus we do not expect any
dependence of the characteristic time for circularization or changing of the
trajectory to the disc plane on particular values of the initial position
or directions of velocity of the star.

Our second example is a modification of the model studied by Syer \etal
(1991). It is complementary to the previous one because energy dissipation is
now
considered while the difference in azimuthal velocities of the star relative
to the disc material is ignored.
We supposed that the star hits the disc supersonically and pulls
some amount of the material with a mass
$\Delta m\approx\rho_{\rm disc}\,h_{\rm disc}\,A_{\rm eff}\sin^{-1}\!I$
out of the disc;
here, $\rho_{\rm disc}$ and $h_{\rm disc}$ are the local density and
thickness of the disc, respectively,
and $A_{\rm eff}$ is the effective cross-section
for the star-disc interaction. The energy dissipated during the interaction
is proportional to the kinetic energy acquired by the disc material,
$\Delta\calE_{\rm diss}\propto\Delta m(\gamma_{\rm DCF}-1).$
We assumed that the
acceleration of the star which results from this interaction is
anti-parallel to the velocity of the star and the corresponding change of the
velocity is
\beq%vvvvvvvvvvvvvvvvvvvvvvvvvvvvvvvvvvvvvvvvvvvvvvvvvvvvvvvvvvvvvvvvvvvv
 \Delta{\vec v}=-\frac{\Delta\calE\,{\vec v}}
 {m_\star\,\gamma_{\rm DCF}^3v^2}
 \propto-\frac{\rho_{\rm disc}\,h_{\rm disc}\,A_{\rm eff}(\gamma_{\rm DCF}-1)
 \,\vec v}{m_\star\,\gamma_{\rm DCF}^3v^2\sin\! I}\,.
\eeq%^^^^^^^^^^^^^^^^^^^^^^^^^^^^^^^^^^^^^^^^^^^^^^^^^^^^^^^^^^^^^^^^^^^^
(We should note that the last equation for the star-disc drag becomes
inappropriate and must be modified
when the motion of the star is subsonic and the disc
material is directly accreted onto the star. The motion is highly supersonic
with the Mach number of the order $10^2-10^3$ under conditions we consider.)
Naturally, $\rho_{\rm disc}$ and $h_{\rm disc}$ depend on the disc model.
Because we do not want to enter into these additional details we assumed
that they, as well as $A_{\rm eff},$ are constant. [We have also carried out
computations using the density profiles corresponding to the Novikov~\& Thorne
(1973) relativistic thin disc model which yielded only
moderate modifications to the results.] Figure~2 illustrates
two typical cases---both orbits are initially prograde
with (a) $I=35\dg$ and (b) $I=80\dg$. In general,
the final radius of the orbit can be
either larger (for small values of the initial inclination)
or smaller (for large values) than the initial pericentre. We found that
initially retrograde orbits became captured in this model. This
feature can be naturally explained as follows. Dissipation of the orbital
energy in each intersection with the disc tends to increase the
binding energy of the
orbiting object. In the case of originally prograde orbits, however,
the orbiter obtains sufficient amount of angular momentum, which
saves it from being captured by the black hole. Finally, the object
settles into a circular orbit in the disc plane. However, an
object which started with retrograde orbit does not acquire enough
angular momentum during the period of nearly perpendicular intersections
with the disc. Due to continuous losses of energy it
typically becomes captured by the hole.

To clarify previous results based on the relativistic treatment
we compared them with the corresponding Newtonian ``elliptic"
mapping (see Appendix). To be consistent,
we also reduced formula for the star-disc interaction by eliminating the
Lorentz factors  $\gamma_{\rm DCF}$ in (19) and
instead of the Lorentz boost from LNRF to DCF
we used the Galilean transformation. Fig.~3a shows the
fully relativistic model with the Schwarzschild background metric, while
Fig.~3b is the Newtonian analogue. We have chosen formally
the same initial eccentricity and inclination in both Figures:
$e \simeq 0.83$ and $I=130\dg.$ The orbit is initially
retrograde and in the relativistic case it becomes captured by the central
black
hole. On the contrary, this does not occur in the Newtonian case
and the orbit is circularized to some definite radius. (In a realistic case,
however, the orbiter can be tidally disrupted before it is captured
but this depends on its internal structure and we do not consider such
possibility in the present paper.)
Because the interaction of the orbiter with the disc was always
chosen to be weak, the time scale for the precession of the pericentre is
much shorter than the time scale for the evolution of
the other orbital parameters.
As a consequence, the points of intersection with the disc fill the
interval between the current pericentre
and apocentre in the relativistic case (Fig.~3a). There is no precession
of the pericentre in the Newtonian case and thus only an `adiabatic'
evolution of the orbital parameters is seen (Fig.~3b).

We also show the results of integration with the  nearly extreme
Kerr black hole. In this example we have chosen $a=0.9981$ for definiteness
(Thorne 1974), the same initial eccentricity and inclination, and
the same initial apocentre and pericentre
(expressed in the gravitational radii) as in Fig~2.
We observed (Fig.~4) a slightly shorter circularization time than in the
corresponding Schwarzschild case, but the qualitative features
of the disc-orbit interaction remained unchanged. They may be changed,
however, when details of the structure of the accretion disc are
taken into account because the structure of the disc and the location of
its inner edge depend significantly on $a.$ Again,
we found that initially retrograde orbits get captured by the black hole.
\bigskip\bigskip

\noindent
{\large\bf 4\quad CONCLUSIONS} %%%%%%%%%%%%%%%%%%%%%%%%%%%%%%%%%%%%%%%%%%%%%%%%
%%%%%%%%%%%%%%%%%%%%%%%%%%%%%%%%%%%%%%%%%%%%%%%%%%%%%%%%%%%%%%%%%%%%%%%%%%%%%%%
\smallskip

\noindent
We assumed that the low-mass compact object interacts with the thin
accretion disc twice per each revolution---exactly when
it crosses the equatorial plane of the black hole (impulsive
approximation).
We described the relevant equations and we employed them in the fast
numerical code computing the evolution of the trajectory.
We found that the effective time of circularization is shorter
than the time to change the orbital plane into the plane of the disc.
This conclusion is in accordance with previous results of Syer \etal (1991)
based on Newtonian gravity.
However, we have also observed short periods during the evolution
when eccentricity increases.
In particular, this increase occurred in model described by Eq.~(18)
during the transition period where the
initially retrograde character of the orbit is
changed to a prograde one.

The star-disc interaction was described
by a phenomenological parameter characterizing the magnitude of the change
of orbital parameters in each collision. This phenomenological description is
satisfactory provided the disc remains thin and
the orbital parameters are changed only at the moment of transition of the
star through the equatorial plane of the
central black hole. A description of the
interaction which would result from a detailed
physical model is not crucial in this case; we want to improve our
understanding of the interaction in future work.
Effects of the dynamical friction and direct
accretion acting on an object moving through the gaseous medium were studied
by a number of authors under various conditions (recently by
Petrich \etal 1989). In our highly supersonic and turbulent case,
the approach outlined
by Zurek \etal (1992) appears the most appropriate one.

One can specify parameters of the model for the case of NGC~6814.
Each single long-duration
observation of {\it EXOSAT} or {\it Ginga} covers less than
30 revolutions of the orbiter.
The characteristic time-scale for the precession of nodes
is much longer than the orbital one. The estimate which
adopts the maximum value of $a=1$ and radius of the orbit
$\approx 50$ gravitational radii
of the central black hole yields the ratio of the Lense-Thirring to the orbital
frequency $\approx 0.005\,.$ This means that the orientation of the orbit
is not significantly changed during each observation. However, the interval
between {\it EXOSAT} observations and {\it Ginga} observations
was certainly long enough
and a resulting change in the orientation suggests a possibility to
understand the perfect stability of the orbital period detected by both
satellites and at the same time the puzzling change
in the light curve profile. We do not want to speculate further on this
important subject until the
star-disc collisions and dynamical friction acting on the star
are better understood (work in preparation).

To conclude, adopting the model of star-disc interactions as an explanation
of the origin of the NGC~6814 light curve,
we see one important contribution from general
relativistic effects which is due to the pericentre shift and Lense-Thirring
precession. These effects
drag the point on the orbit where the star crashes through the disc.
They also
modify the velocity at which the star hits the disc as well as the orientation
of the orbit with respect to the observer. This fact has two consequences:
(i) additional periodicities corresponding to the precession frequencies
are present and can potentially be revealed in the power
spectrum of the signal from
the source (Karas~\& Vokrouhlick\'y 1993); (ii) long-term evolution
of marginally stable and marginally bound orbits is very different
compared to the orbits with identical initial parameters treated in
Newtonian theory of gravity. The first consequence above
gives us a possibility to
detect Lense-Thirring precession induced near the core of an AGN.
If the corresponding frequency is not present we will be able to conclude
that the central black hole (if any) is non-rotating. This would be
extremely important information especially from the point of view
of electromagnetic scenarios of AGN which require a rotating black hole.
The second consequence is particularly important in describing the capture
of the star into a bound orbit around the central black hole.
Although many models assume a star located on such an orbit, the very
process of the capture is not well understood.

\bigskip
\newpage

\noindent
{\large\bf ACKNOWLEDGMENTS}
\medskip

\noindent
Our work was largely motivated by the discussions with M.~A. Abramowicz
and by the work by Syer \etal (1991). We thank the unknown referee
for very stimulating remarks which helped us to improve presentation
of our work. We also acknowledge discussions
with J.~Bi\v c\'ak, P.~Hadrava, A.~Lanza, O.~Semer\'ak and R.~Stark.
V.~K. acknowledges support from
SISSA, Trieste. D.~V. is grateful for kind hospitality of CERGA, Grasse.
\bigskip\bigskip

\parindent=-15pt
{\large\bf REFERENCES}%%%%%%%%%%%%%%%%%%%%%%%%%%%%%%%%%%%%%%%%%%%%%%%%%%%%%%%%%
%%%%%%%%%%%%%%%%%%%%%%%%%%%%%%%%%%%%%%%%%%%%%%%%%%%%%%%%%%%%%%%%%%%%%%%%%%%%%%%
\medskip

\parskip=0pt

Abramowicz M. A. 1992, in Holt S. S.,
 Urry C. M., eds, Testing the AGN Paradigm,
 Proceedings of the 2nd Maryland Astrophysical Conference.
 Am. Inst. Phys., New York, p.~69

Abramowicz M. A., Bao G., Lanza A., Zhang X.-H. 1991,
 A\&A, 245, 454

Abramowicz M. A., Lanza A., Spiegel E. A., Szuszkiewicz E. 1992,
 Nat, 356, 41

Bardeen J. M. 1973, in DeWitt C., DeWitt B. S., eds, Black Holes.
 Gordon and Breach, New York, p.~215

Bardeen J. M., Press W. H., Teukolsky S. A. 1972, ApJ,
 178, 347

Byrd P. F., Friedman M. D. 1971, Handbook of Elliptic Integrals
 for Engineers and Scientists. Springer-Verlag, Berlin

Carter B. 1968, Phys. Rev., 174, 1559

Carter B. 1992, ApJ, 391, L67

Carter B. \& Luminet J.-P. 1983, A\&A, 121, 97

Chandrasekhar S. 1983, The Mathematical Theory of Black Holes.
 Clarendon, Oxford

Cunningham C. T., Bardeen J. M. 1973, ApJ, 183, 237

D'Eath P. D. 1975a, Phys. Rev. D, 11, 1387

D'Eath P. D. 1975b, Phys. Rev. D, 12, 2183

de Felice F., Nobili L., Calvani M. 1974, A\&A 30, 111

Done C., Madejski G. M., Mushotzky R. F., Turner T. J. 1993,
 ApJ, 400, 138

Fabian A. C., Rees M. J., Stella L., White N. E. 1989,
 MNRAS, 238, 729

Gradshteyn I. S., Ryzhik I. W. 1980, Table of Integrals, Series, and
 Products. Academic, New York

Gr\" obner W., Hofreiter N. 1965, Integraltafel, Erster Teil,
Unbestimmte Integrale. Springer-Verlag, Vienna

Hills J. G. 1988, Nat, 331, 687

Karas V., Vokrouhlick\'y D. 1993, ApJ, in press (this preprint)

Karas V., Vokrouhlick\'y D., Polnarev A. 1992, MNRAS,
 259, 569

Kates R. E. 1980, Phys. Rev. D, 22, 1853

Kojima Y. 1991, MNRAS, 250, 629

Laor A. 1991, MNRAS, 376, 90

Laor A., Netzer H. 1989, MNRAS, 238, 897

Lense J., Thirring H. 1918, Physik Z., 19, 156

Luminet J.-P. 1979, A\&A, 75, 228

Luminet J.-P., Marck J.-A. 1985, MNRAS, 212, 57

Mittaz J. P. D., Branduardi-Raymont G. 1989, MNRAS,
 238, 1029

Murray C. D. 1986, Icarus, 65, 70

Novikov I. D., Pethick C. J., Polnarev A. G. 1992, MNRAS, 255, 276

Novikov I. D., Thorne K. S. 1973, in DeWitt C.,
 DeWitt B. S., eds, Black Holes. Gordon and Breach, New York, p.~343

%Page D. N. \& Thorne K. S. 1974, ApJ, 191, 499

Petrich L. I., Shapiro S. L., Stark R. F., Teukolsky S. A. 1989,
 ApJ, 336, 313

Press W. H., Teukolsky S. A. 1990, Computers in Physics, Jan/Feb, 92

Press W. H., Flannery B. P., Teukolsky S. A., Vetterling W. T. 1986,
 Numerical Recipes: The Art of Scientific Computing.
 Cambridge University Press, New York

Rees M. J. 1993, in Ellis G. F. R., Lanza A., Miller J. C., eds,
 The Renaissance of General Relativity and Cosmology: A survey meeting
 to celebrate 65th birthday of Dennis Sciama. Cambridge University Press,
 Cambridge, in press

Sikora M., Begelman M. C. 1992, Nat, 356, 224

Stoghianidis E., Tsoubelis D. 1987, Gen. Rel. Grav., 19, 1235

Suen W.-M. 1986, Phys. Rev. D, 34, 3633

Syer D., Clarke C. J., Rees M. J. 1992, MNRAS,
 250, 505

Thorne K. S. 1974, ApJ, 191, 507

Thorne K. S., Hartle J. B. 1985, Phys. Rev. D, 31, 1815

Wallinder F. H., Kato S., Abramowicz M. A. 1992, A\&AR, 4, 79.

Wilkins D. C. 1972, Phys. Rev. D, 5, 814

Wisdom J. 1982, AJ, 87, 577

Zurek W. H., Siemiginowska A., Colgate S. A. 1992, in Holt S. S.,
 Urry C. M., eds,
 Testing the AGN Paradigm, Proceedings of the 2nd Maryland Astrophysical
 Conference. Am. Inst. Phys., New York, p.~564

\bigskip\bigskip
\parindent=15pt
\parskip=5pt
\newpage

\noindent
{\large\bf APPENDIX A: THE MAPPING ALGORITHM} %%%%%%%%%%%%%%%%%%%%%%%%%%%%%%%%%
%%%%%%%%%%%%%%%%%%%%%%%%%%%%%%%%%%%%%%%%%%%%%%%%%%%%%%%%%%%%%%%%%%%%%%%%%%%%%%%
\medskip
\setcounter{equation}{0}
\renewcommand{\theequation}{A\arabic{equation}}

\noindent
This Appendix outlines the explicit form of the mapping
$(r,\phi;\eta)_i$ $\rightarrow$ $(r,\phi;\eta)_f$
from Section 2. Although our approach is straightforward, we believe that it
has not yet been employed by other authors. As we found it very
advantageous for practical purposes, we describe derivations
relevant for this work in some detail.
We constructed a numerical code which employs efficient routines
for evaluation of elliptic integrals and Jacobian elliptic functions
(Press \etal 1986, Press~\& Teukolsky 1990).
The code achieves better precision and
about two orders of magnitude higher
speed compared to direct numerical integration of the geodesic equation
in its equivalent form of first order differential equations. (We used
direct integration to check the code.)
The two cases,
$0<a<1$ and $a=0,$ are technically somewhat different and we discuss
them separately.

\medskip

\noindent
{\bf A1\qquad The case 0\mbox{\boldmath $\;<a<\;$}1}
 %%%%%%%%%%%%%%%%%%%%%%%%%%%%%%
%%%%%%%%%%%%%%%%%%%%%%%%%%%%%%%%%%%%%%%%%%%%%%%%%%%%%%%%%%%%%%%%%%%%%%%%%%%%%%
\smallskip

\noindent
(i) Evaluate the latitudinal integral between two successive intersections
with the equatorial plane:
\beq %vvvvvvvvvvvvvvvvvvvvvvvvvvvvvvvvvvvvvvvvvvvvvvvvvvvvvvvvvvvvvvvvvvvvvvv
I_\mu=\frac{2}{a\mu_+\sqrt{1-\calE^2}}K(\mu_-/\mu_+),
\eeq %^^^^^^^^^^^^^^^^^^^^^^^^^^^^^^^^^^^^^^^^^^^^^^^^^^^^^^^^^^^^^^^^^^^^^^^
where $K(k)$ denotes the complete elliptic integral of the first kind.

\noindent
(ii) Distinguish the three cases which may occur:
\begin{itemize}
\item{Case I---four real roots of $R(r)=0,$ $r_4<r<r_3.$}
\item{Case II---four real roots of $R(r)=0,$ $r_2<r<r_1.$}
\item{Case III---two real and two complex roots of $R(r)=0,$ $r_2<r<r_1.$}
\end{itemize}
Denote $\eta_i=1$ $(\eta_i=-1)$ if $r$ is increasing (decreasing) at $r_i;$
analogously $\eta_f$ for $r_f.$

\noindent
(iii) Evaluate the increase of $I_r$ between each two radial turning points:
\beq %vvvvvvvvvvvvvvvvvvvvvvvvvvvvvvvvvvvvvvvvvvvvvvvvvvvvvvvvvvvvvvvvvvvvvvv
 \delta I_r=\left\{\begin{array}{lll}
 \kappa K(k_1) & \mbox{(Case I and II),}\\ \\
 \frac{2}{\sqrt{pq(1-\calE^2)}}K(k_2) & \mbox{(Case III),}
 \end{array}\right.
\eeq %^^^^^^^^^^^^^^^^^^^^^^^^^^^^^^^^^^^^^^^^^^^^^^^^^^^^^^^^^^^^^^^^^^^^^^^
where
$$k_1=\frac{(r_1-r_2)(r_3-r_4)}{(r_1-r_3)(r_2-r_4)}\,,$$
$$k_2=\frac{(r_1-r_2)^2-(p-q)^2}{4pq}\,,$$
$$p^2=(\chi_1-r_1)^2+\chi_2^2\,,\qquad q^2=(\chi_1-r_2)^2+\chi_2^2\,,$$
$$\kappa=\frac{2}{\sqrt{(r_1-r_3)(r_2-r_4)(1-\calE^2)}}\,.$$

\noindent
(iv) Denote
$$
 f=\left\{\begin{array}{lll}
 \kappa F(\varphi,k_1) & \mbox{(Case I and II)},\\ \\
 \frac{1}{\sqrt{pq(1-\calE^2)}}F(\varphi,k_2) & \mbox{(Case III)},
 \end{array}\right.
$$
\beq %vvvvvvvvvvvvvvvvvvvvvvvvvvvvvvvvvvvvvvvvvvvvvvvvvvvvvvvvvvvvvvvvvvvvvvv
 \tilde{I}_r=\left\{\begin{array}{lll}
 m\;\delta I_r-\eta_ff & \mbox{if\quad $\eta_f\eta_i>0$},\\ \\
 (m-\eta_f)\;\delta I_r+\eta_ff & \mbox{if\quad $\eta_f\eta_i<0$}.
 \end{array}\right.
\eeq %^^^^^^^^^^^^^^^^^^^^^^^^^^^^^^^^^^^^^^^^^^^^^^^^^^^^^^^^^^^^^^^^^^^^^^^
Here,
$$\sin^2\varphi=\left\{\begin{array}{lll}
 {\frac{(r_1-r_3)(r_i-r_4)}{(r_3-r_4)(r_1-r_i)}} & \mbox{(Case I)},\\ \\
 {\frac{(r_1-r_3)(r_i-r_2)}{(r_1-r_2)(r_i-r_3)}}
 & \mbox{(Case II)},
 \end{array}\right.$$
$$\tan^2\frac{\varphi}{2}={\frac{p(r_i-r_2)}{q(r_1-r_i)}}
\qquad\mbox{(Case III)};$$
$F(\varphi,k)$ is the incomplete elliptic integral of the first kind
and $m$ is the number of turning points in $r$ between the two successive
intersections. In addition, one has to check whether the trajectory
still remains above the horizon if the lower turning point is
located below the horizon.

\noindent
(v) The radial coordinate of the intersection is
\beq %vvvvvvvvvvvvvvvvvvvvvvvvvvvvvvvvvvvvvvvvvvvvvvvvvvvvvvvvvvvvvvvvvvvvvvv
 r_f=\left\{\begin{array}{lll}
 \frac{(r_1-r_3)r_4+(r_3-r_4)r_1\sigma}{r_1-r_3+(r_3-r_4)\sigma}
 & \mbox{(Case I)},\\ \\
 \frac{(r_1-r_3)r_2+(r_1-r_2)r_3\sigma}{r_1-r_3-(r_1-r_2)\sigma}
 & \mbox{(Case II)},
 \end{array}\right.
\eeq %^^^^^^^^^^^^^^^^^^^^^^^^^^^^^^^^^^^^^^^^^^^^^^^^^^^^^^^^^^^^^^^^^^^^^^^
with
$$\sigma=\sn^2(u,k_1)\,,$$
$$u=\frac{1}{2}\sqrt{(r_1-r_3)(r_2-r_4)(1-\calE^2)}\left(I_\mu-\tilde
{I}_r\right),$$
or
\beq %vvvvvvvvvvvvvvvvvvvvvvvvvvvvvvvvvvvvvvvvvvvvvvvvvvvvvvvvvvvvvvvvvvvvvvv
r_f=\frac{qr_1\sigma+pr_2}{p+q\sigma}\qquad\mbox{(Case III)},
\eeq %^^^^^^^^^^^^^^^^^^^^^^^^^^^^^^^^^^^^^^^^^^^^^^^^^^^^^^^^^^^^^^^^^^^^^^^
with
$$\sigma=\frac{\sn^2(u,k_2)}{\left[1+\cn(u,k_2)\right]^2}\,,$$
$$u=\sqrt{pq(1-\calE^2)}\left(I_\mu-\tilde{I}_r\right);$$
$\sn(u,k)$ and $\cn(u,k)$ are Jacobian elliptic functions.
At this point we are able to compute the $r$-coordinates of the intersections,
which is sufficient to determine the evolution of eccentricity and
inclination of the orbit and the number of revolutions before the trajectory
becomes captured by the black hole or escapes to $\calE\geq 1$
(and then presumably to infinity); $\phi$
coordinates are also needed if we wish to study the precession. Finally,
we need coordinate time to relate the revolutions to time as measured by
a distant observer. In the Case III the orbit is in practice captured by
the black hole after a few revolutions. Thus we exclude this case from
further considerations.

\noindent
(vi) Evaluate the following quantities.

\noindent
Case I:
\beq %vvvvvvvvvvvvvvvvvvvvvvvvvvvvvvvvvvvvvvvvvvvvvvvvvvvvvvvvvvvvvvvvvvvvvvv
I_\pm=\kappa_\pm\left[(r_4-r_1)\Pi\left(\varphi,n_\pm,k_1\right)+
 (r_\pm-r_4)F(\varphi,k_1)\right]+\tilde{I}_\pm,
\eeq %^^^^^^^^^^^^^^^^^^^^^^^^^^^^^^^^^^^^^^^^^^^^^^^^^^^^^^^^^^^^^^^^^^^^^^^
$$n_\pm=\frac{(r_3-r_4)(r_\pm-r_1)}{(r_1-r_3)(r_\pm-r_4)}\,,$$
\begin{eqnarray} %vvvvvvvvvvvvvvvvvvvvvvvvvvvvvvvvvvvvvvvvvvvvvvvvvvvvvvvvv
\lefteqn{\nonumber
J_r+2K_r=\kappa r_4\left[\left(\frac{r_4\alpha_1^4}{\alpha^4}+
 2\frac{\alpha_1^2}{\alpha^2}\right)U+
 2\frac{\alpha^2-\alpha_1^2}{\alpha^2}\,V_1+       \right.}\\
& & \left. 2r_4\alpha_1^2\frac{\alpha^2-\alpha_1^2}{\alpha^4}\,V_1+
 r_4\frac{(\alpha^2-\alpha_1^2)^2}{\alpha^4}\,V_2\right]+\tilde{J}_r
 +2\tilde{K}_r,
\end{eqnarray} %^^^^^^^^^^^^^^^^^^^^^^^^^^^^^^^^^^^^^^^^^^^^^^^^^^^^^^^^^^^
with
$$U=F(\varphi,k_1)\,,$$
$$V_1=\Pi\left(\varphi,-\alpha^2,k_1\right),$$
\begin{eqnarray*} %vvvvvvvvvvvvvvvvvvvvvvvvvvvvvvvvvvvvvvvvvvvvvvvvvvvvvvvvv
\lefteqn{
V_2=\frac{1}{2(\alpha^2-1)(k_1^2-\alpha^2)}\Biggl[\alpha^2E(\varphi,k_1)+
 \left(2\alpha^2k_1^2+2\alpha^2-\alpha^4-
 3k_1^2\right)V_1    }\\
 & &  +\left(k_1^2-\alpha^2\right)U
 -\sn(U,k_1)\,\cn(U,k_1)\,\dn(U,k_1)\,\frac{\alpha^4}
 {1-\alpha^2\sn^2(U,k_1)}\Biggr],
\end{eqnarray*} %^^^^^^^^^^^^^^^^^^^^^^^^^^^^^^^^^^^^^^^^^^^^^^^^^^^^^^^^^^^
$$\alpha^2=\frac{r_4-r_3}{r_1-r_3}\,,\qquad
 \alpha_1^2=\frac{r_1(r_4-r_3)}{r_4(r_1-r_3)}\,,$$
$$\sin^2\varphi=\frac{(r_1-r_3)(r_i-r_4)}{(r_3-r_4)(r_1-r_i)}\,,$$
$$\kappa_\pm=\frac{2}{(r_\pm-r_1)(r_\pm-r_4)
 \sqrt{(r_1-r_3)(r_2-r_4)(1-\calE^2)}}\,.$$

\noindent
Case II:
\beq %vvvvvvvvvvvvvvvvvvvvvvvvvvvvvvvvvvvvvvvvvvvvvvvvvvvvvvvvvvvvvvvvvvvvvvv
I_\pm=\kappa_\pm\left[(r_2-r_3)\Pi\left(\varphi,n_\pm,k_1\right)+
 (r_\pm-r_2)F(\varphi,k_1)\right]+\tilde{I}_\pm,
\eeq %^^^^^^^^^^^^^^^^^^^^^^^^^^^^^^^^^^^^^^^^^^^^^^^^^^^^^^^^^^^^^^^^^^^^^^^
$$n_\pm=\frac{(r_2-r_1)(r_\pm-r_3)}{(r_1-r_3)(r_\pm-r_2)}\,,$$
\begin{eqnarray} %vvvvvvvvvvvvvvvvvvvvvvvvvvvvvvvvvvvvvvvvvvvvvvvvvvvvvvvvv
\lefteqn{\nonumber
J_r+2K_r=\kappa r_2\left[\left(\frac{r_2\alpha_1^4}{\alpha^4}+
 2\frac{\alpha_1^2}{\alpha^2}\right)U+
 2\frac{\alpha^2-\alpha_1^2}{\alpha^2}\,V_1+   \right.}\\
& & \left. 2r_2\alpha_1^2\frac{\alpha^2-\alpha_1^2}{\alpha^4}\,V_1+
 r_2\frac{(\alpha^2-\alpha_1^2)^2}{\alpha^4}\,V_2\right]+\tilde{J}_r
+2\tilde{K}_r,
\end{eqnarray}
with $U,$ $V_1$ and $V_2$ defined as above, and
$$\alpha^2=\frac{r_1-r_2}{r_1-r_3}\,,\qquad
 \alpha_1^2=\frac{r_3(r_1-r_2)}{r_2(r_1-r_3)}\,,$$
$$\sin^2\varphi=\frac{(r_1-r_3)(r_i-r_2)}{(r_1-r_2)(r_i-r_3)}\,,$$
$$\kappa_\pm=\frac{2}{(r_\pm-r_3)(r_2-r_\pm)
 \sqrt{(r_1-r_3)(r_2-r_4)(1-\calE^2)}}\,.$$
$E(\varphi,k)$ and $\Pi\left(\varphi,n,k\right)$ are incomplete
elliptic integrals of the
second and the third kind, respectively. Integration constants
$\tilde{I}_\pm,$ $\tilde{J}_r$ and $\tilde{K}_r$
are the values of integrals in (15) and (17) evaluated between $r_i$
and the last turning point. Thus, they depend on the
number of turning points in $r$ $(m)$ and the sign of initial radial
velocity $(\eta_i)$ and they can be given in terms analogous to Eq.~(A3).
We skip explicit expressions because they are rather lengthy.

\noindent
(vii) Finally,
\beq %vvvvvvvvvvvvvvvvvvvvvvvvvvvvvvvvvvvvvvvvvvvvvvvvvvvvvvvvvvvvvvvvvvvvvvv
J_\mu=\frac{2}{a\mu_+\sqrt{1-\calE^2}}\Pi\left(-\mu_-^2,\mu_-/\mu_+\right),
\eeq %^^^^^^^^^^^^^^^^^^^^^^^^^^^^^^^^^^^^^^^^^^^^^^^^^^^^^^^^^^^^^^^^^^^^^^^
\beq %vvvvvvvvvvvvvvvvvvvvvvvvvvvvvvvvvvvvvvvvvvvvvvvvvvvvvvvvvvvvvvvvvvvvvvv
K_\mu=\frac{2r_1}{a\sqrt{1-\calE^2}}
 \left[K(\mu_-/\mu_+)-E(\mu_-/\mu_+)\right]
\eeq %^^^^^^^^^^^^^^^^^^^^^^^^^^^^^^^^^^^^^^^^^^^^^^^^^^^^^^^^^^^^^^^^^^^^^^^
with $E(k)$ and $\Pi\left(n,k\right)$ being complete
elliptic integrals of the second and the third kind.
Now we have all necessary quantities required for complete mapping
of relevant trajectories in the Kerr metric.
\medskip

\noindent
{\bf A2\qquad The Schwarzschild case (\mbox{\boldmath $a=\;$}0)} %%%%%%%%%%%%%%
%%%%%%%%%%%%%%%%%%%%%%%%%%%%%%%%%%%%%%%%%%%%%%%%%%%%%%%%%%%%%%%%%%%%%%%%%%%%%%%
\smallskip

\noindent
We present the case of the Schwarzschild background metric
separately, even though the general formalism
developed for the Kerr metric can also be applied. The reason is twofold:
(i) the formulae valid for the general Kerr metric often include
the angular momentum parameter $a$ in denominators [e.g. Eqs.~(A10), (A11)].
These apparent singularities cancel out in the limit $a\longrightarrow 0,$
but they are the source of
difficulties in numerical evaluation; (ii) as
the symmetry of the space-time is now higher and geodesics in the Schwarzschild
metric remain always planar, we can avoid integration
of the latitude $\theta,$ restricting ourselves
to the current orbital
plane of the test particle spanning one loop of the trajectory
above/below the disc (the true orbital plane of the object
is changed due to
the interactions with the disc). Thus we reduce the
order of the mapping by evaluating the integrals in polar
coordinates {\it in the orbital plane}. The orbital plane differs from
{\it the disc plane} by the current value of the
inclination.

We consider the following coordinate systems: (i)
Schwarzschild spherical coordinates $(r,\theta,\phi)$;
latitude $\theta$ is measured
from the axis of the disc plane and polar
angle $\phi$ in the disc plane ($\phi=0$ direction can be chosen arbitrarily),
(ii) $(r,\vartheta)$ polar
coordinates in {\it the current orbital plane} of the orbiting
object where the angle $\vartheta$ is measured from the actual
nearest preceding apocentre of the unperturbed trajectory with
the current orbital parameters. Let us clarify better the concept
of the $\vartheta$ origin, as it is intimately connected with our
technique. The analytic integration of the geodesic motion
in the Schwarzschild space-time is advantageously
carried out if the polar angle in the orbital plane is measured from the
nearest preceding apocentre. In
each step of the mapping procedure we are interested only in
one orbital loop above/below the accretion disc; then
the interaction with the disc changes the orbital parameters for the
successive loop. It is this orbital loop, where the free motion
of the test particle in the Schwarzschild background is applied.
However, the orbital loop which is under consideration
may not necessarily contain the apocentre
of the orbit. Thus, apart from {\it the true trajectory} of the
object we introduce a {\it reference
trajectory} of the object with the same orbital parameters as for
the true one and coinciding with the true trajectory
just on the current segment. This fictitious reference orbit defines
the $\vartheta$ origin---it is
measured from the nearest preceding apocentre of the reference
trajectory.

The equations of motion covering a single mapping step are as
follows (e.g. Chandrasekhar 1983):
\beq%vvvvvvvvvvvvvvvvvvvvvvvvvvvvvvvvvvvvvvvvvvvvvvvvvvvvvvvvvvvvvvvvvvvv
 \vartheta_f - \vartheta_i\;(=\pi) = \int_{r_i}^{r_f} \frac{du}
 {U(u)^{1/2}},
\eeq%^^^^^^^^^^^^^^^^^^^^^^^^^^^^^^^^^^^^^^^^^^^^^^^^^^^^^^^^^^^^^^^^^^^^
\beq%vvvvvvvvvvvvvvvvvvvvvvvvvvvvvvvvvvvvvvvvvvvvvvvvvvvvvvvvvvvvvvvvvvvv
t_f - t_i = \frac{\calE}{\calL} \int_{\vartheta_i}^{\vartheta_f}
 \frac{d\vartheta}{u^2 (1-2u)},
\eeq%^^^^^^^^^^^^^^^^^^^^^^^^^^^^^^^^^^^^^^^^^^^^^^^^^^^^^^^^^^^^^^^^^^^^
where
$$ U(u) = 2u^3 - u^2 + 2\calL^{-2}u - (1-\calE^2)\calL^{-2}, $$
and $u=1/r.$ Constants of motion are defined as $\calE = -p_t$
and $\calL = p_{\vartheta}$ [note, that angular momentum $\calL$ is
defined with respect to the fictitious orbital plane, not with respect
to the disc plane like $\Phi$ in the Kerr case].
They are related to the tetrad components
of the four-momentum in the locally static frames by means of Eqs.~(6)--(7).
We {\it define} $I\equiv\frac{\pi}{2} - \beta$ as the inclination
of the fictitious orbit with respect to
the fixed reference disc plane [it can be
equivalently expressed using the LNRF tetrad components in the equatorial
plane:
$\tan I=p^{\hat{\theta}}/p^{\hat{\phi}};$
cf. Eq.~(8)].

Again, we will concentrate on orbits characterized by $\calE < 1$ and
$\calL \neq 0$. The signs of the first and the last term of the
polynomial $U(u)$ guarantee at least one positive root of the equation
$U(u)=0$ and, as $U(u=0)<0,$ we conclude that this root corresponds
to the apocentre of the orbit. The type of the orbit is
determined by the properties of the other two roots of the
equation $U(u)=0$. The roots cannot be real and negative at the same time
(Chandrasekhar 1983). We exclude the
possibility of multiple roots of as before.
Hence, we are left with the two kinds of orbits characterized by:
\begin{itemize}
\item the three positive real roots of $U(u)=0$ which we arrange according
 to magnitude: $u_1 < u_2 < u_3$,
\item one positive real root ($u_1$) and two complex conjugated roots
 ($u_c,\overline{u_c}$) of $U(u)=0$.
\end{itemize}
The first item still encompasses two types of orbits:
(i) those captured by the black hole in the sense that they have no
pericentre above the horizon
(the apocentre of such orbits is always less than $6$ --- where the last
stable orbit and presumably the inner edge of the accretion disc
are located in the Schwarzschild case; hence we
do not consider these orbits);
(ii) quasi-elliptic orbits bound between the turning points,
$u_1$ and $u_2;$ we will call them Case~I orbits.
In terminology used by Chandrasekhar (1983) our Case~I orbits correspond
to the orbits of the first kind. We call the Case~II orbits those
of the second item above. They have no pericentre and fall unavoidably
to the black hole (they correspond to the orbits with purely
imaginary eccentricity in Chandrasekhar's terminology).

One can easily find a simple rule dividing
both cases of the orbits: $\calL^2<12$ implies the Case~II orbit.
The Case~I orbits are characterized by $\calL^2>12$ and simultaneously
$$ (1-2u_{\star})(1+\calL^2 u_{\star}^2) > \calE^2 , $$
where $u_{\star} = 1+|\calL|^{-1}\sqrt{\calL^2-12}$.
In what follows, we will describe an algorithm for the
mapping of these two cases of orbits in detail. We will pay special
care to the Case~I orbits, as they will be shown to be the most
important in astrophysical applications.
\smallskip

\noindent
{\it A2.1 Case I orbits} %%%%%%%%%%%%%%%%%%%%%%%%%%%%%%%%%%%%%%%%%%%%%%%%%%%%%%
\smallskip

\noindent
In close analogy to the Newtonian case, Eq.~(A12) is advantageously
integrated in terms of the relativistic ``true anomaly''
$$ u(\chi) = \mu(1 + e \cos\chi) , $$
where
$$ \mu = \frac{u_1+u_2}{2}, \qquad e = \frac{u_2-u_1}{u_2+u_1}.$$
The quantity $e$ can be interpreted as the eccentricity of the orbit.
We do not write the explicit form of primitive functions
obtained by integration (Chandrasekhar
1983) but we give formulae for the mapping which we need in our present
work. After some manipulation we
arrive at the following form of the mapping:
\beq%vvvvvvvvvvvvvvvvvvvvvvvvvvvvvvvvvvvvvvvvvvvvvvvvvvvvvvvvvvvvvvvvvvvv
 u_f =u_2 -(u_2-u_1)\left[\frac{\sqrt{\Phi \Phi_{\alpha}}
 \sqrt{(1-k^2\Phi)(1-k^2\Phi_{\alpha})} + \eta_i \sqrt{\Psi \Psi_{\alpha}}}
 {1-k^2 \Phi \Phi_{\alpha}} \right]^2,
\eeq%^^^^^^^^^^^^^^^^^^^^^^^^^^^^^^^^^^^^^^^^^^^^^^^^^^^^^^^^^^^^^^^^^^^^
where now
$$ \Phi = \frac{u_i-u_1}{u_2-u_1}\, , \qquad \Psi = \frac{u_2-u_i}{u_2-
 u_1}\,, $$
$$ \Phi_{\alpha} = {\rm sn}^2 \left( \frac{\pi}{2} \omega , k\right) ,
 \qquad \Psi_{\alpha} = {\rm cn}^2 \left( \frac{\pi}{2} \omega , k\right),$$
$$ \omega = \sqrt{1 - 2u_2 - 4u_1}, \qquad k^2 = 2 (u_2 - u_1)\omega^{-2}, $$

\noindent
Mapping of the sign function $\eta$ is given as follows:
\beq%vvvvvvvvvvvvvvvvvvvvvvvvvvvvvvvvvvvvvvvvvvvvvvvvvvvvvvvvvvvvvvvvvvvv
 \eta_f = \left\{\begin{array}{ll}{\rm sgn}[\sigma(\chi_i)
 +\pi-\sigma(0)] & \mbox{if\quad $\eta_i=-1$,} \\ {\rm sgn}[\sigma(\chi_i) -
 \pi]   & \mbox{if\quad $\eta_i=1$,}\end{array} \right.
\eeq%^^^^^^^^^^^^^^^^^^^^^^^^^^^^^^^^^^^^^^^^^^^^^^^^^^^^^^^^^^^^^^^^^^^^
where
$$ \chi_i = \arccos(\Phi - \Psi)\,, \qquad \chi \in\;\langle 0,\pi\rangle\,, $$
$$ \sigma(\chi_i) =2\omega^{-1}\,F\left(\frac{\pi-\chi_i}{2},k\right).$$

It is instructive to discuss the Newtonian limit of the mapping formula (A14)
in which
the terms proportional to some power of $1/c$ are neglected. This limit
is now
obscured by the fact that we imposed widely used ``relativistic
convention $c=1$'' while now we want to suppress the terms containing
$c$ in the denominator. Careful bookkeeping of $c$ in the preceding
equations suggests that the Newtonian limit corresponds
to the fictitious procedure: $\omega \rightarrow 1$, $k \rightarrow 0$.
As a result we obtain
\beq%vvvvvvvvvvvvvvvvvvvvvvvvvvvvvvvvvvvvvvvvvvvvvvvvvvvvvvvvvvvvvvvvvvvv
u_f = - u_i + u_1^N + u_2^N \, ,
\eeq%^^^^^^^^^^^^^^^^^^^^^^^^^^^^^^^^^^^^^^^^^^^^^^^^^^^^^^^^^^^^^^^^^^^^
where the Newtonian boundaries are found to be
$$ u_{1,2}^N = \calL^{-2} \left(1 \pm \sqrt{1+2\calE^N\calL^2}\right)
\;, \qquad \calE^N = \frac{1}{2} \left(\calE^2 - 1\right) \; . $$
(we retain the negative value for the Newtonian energy of bound
orbits, as seen from the above definition of $\calE^N).$
Eq.~(A16) is the correct expression for the ``elliptic" mapping. Formula
(A16) is surprisingly simple, showing the {\it linearity} of the
elliptic mapping. Moreover, we also have the simple rule
$\eta_f = - \eta_i$. By contrast, the full relativistic mapping (A14)
for the orbits of the Case I is highly nonlinear and the mapping in $u$
coordinate is coupled with the mapping of $\eta$ function due to
$\eta_i$ in Eq.~(A14) and $u$ in Eq.~(A15).

Let us turn to Eq. (A13) describing the mapping
in the $t$-coordinate. We start by expressing the indefinite
integral on the right-hand-side. Changing the
$\vartheta$ variable to $\chi$ according to the relation
$$ \frac{d \chi}{d \vartheta} = - \omega \sqrt{1-k^2\cos^2(\chi/2)} \equiv
 -\omega \Delta(\chi,k) \; , $$
one arrives at the primitive function
\begin{eqnarray}%vvvvvvvvvvvvvvvvvvvvvvvvvvvvvvvvvvvvvvvvvvvvvvvvvvvvvvv
T(\chi) & = & \frac{2\calE}{\omega \calL}\Biggl\{\frac{2}{\mu(1-e)}
 \Pi\left(\frac{\pi-\chi}{2},\epsilon^{-1},k\right)  \nonumber \\
 & & +\frac{1}{[1-2\mu(1-e)]}
 \Pi\left(\frac{\pi-\chi}{2},\frac{4\mu e}{2\mu (1-e) - 1},k\right) \nonumber
\\
 & & +\frac{1}{2\mu^2(1-e^2)(1+\epsilon k^2)}\Biggl[
 \frac{\Delta(\chi,k)\sin\chi}{2[\epsilon + \cos^2(\chi/2)]}\nonumber \\
 & & +[1+2\epsilon (1+k^2)+3\epsilon^2k^2]\epsilon^{-1}
 \Pi\left(\frac{\pi-\chi}{2},\epsilon^{-1},k\right)  \nonumber \\
 & & + E\left(\frac{\pi-\chi}{2},k\right) -
 (1+\epsilon k^2) F\left(\frac{\pi-\chi}{2},k\right)
 \Biggr]\Biggl\} \, ,
\end{eqnarray}%^^^^^^^^^^^^^^^^^^^^^^^^^^^^^^^^^^^^^^^^^^^^^^^^^^^^^^^^^^
where
$$ \epsilon = \frac{u_1}{u_2 - u_1}\,.$$
Now, the algorithm for evaluating the
time step associated with the mapping is as follows:

\begin{description}
\item[\mbox{\boldmath $\eta_i = -$1:}]
\qquad introducing $\zeta_i = {\rm sgn}(\sigma(0)-
 \sigma(\chi_i) - \pi)$ we obtain
\beq%vvvvvvvvvvvvvvvvvvvvvvvvvvvvvvvvvvvvvvvvvvvvvvvvvvvvvvvvvvvvvvvvvvv
  t_f - t_i = \left\{ \begin{array}{ll} T(\chi_f) - T(\chi_i) &
  \mbox{if\quad $\zeta_i \geq 0$,}\\
  2T(0) - T(\chi_f) - T(\chi_i) & \mbox{if\quad $\zeta_i < 0$,}
  \end{array} \right.
\eeq%^^^^^^^^^^^^^^^^^^^^^^^^^^^^^^^^^^^^^^^^^^^^^^^^^^^^^^^^^^^^^^^^^^^
where $\chi_f$ is determined by relations
$$ \left\{\begin{array}{c} \sin \\ \cos \end{array}\right\} \chi_f =
   \left\{\begin{array}{c} \sn \\ \zeta_i\,\cn \end{array}\right\}
   \left[\frac{\omega}
  {2}(\sigma(\chi_i)+\pi),k\right]\,;$$
\item[\mbox{\boldmath $\eta_i =\;$1:}]
\qquad introducing
 $\zeta_i = {\rm sgn}(\sigma(\chi_i) - \pi)$
 we obtain
\beq%vvvvvvvvvvvvvvvvvvvvvvvvvvvvvvvvvvvvvvvvvvvvvvvvvvvvvvvvvvvvvvvvvvv
  t_f - t_i = \left\{ \begin{array}{ll} T(\chi_i) - T(\chi_f) &
  \mbox{if\quad $\zeta_i \geq 0$,} \\
  T(\chi_i) + T(\chi_f) & \mbox{if\quad $\zeta_i < 0$,}
  \end{array}\right.
\eeq%^^^^^^^^^^^^^^^^^^^^^^^^^^^^^^^^^^^^^^^^^^^^^^^^^^^^^^^^^^^^^^^^^^^
and now $\chi_f$ is determined by relations
$$ \left\{\begin{array}{c} \sin \\ \cos \end{array}\right\} \chi_f =
  \left\{\begin{array}{c} \zeta_i\,\sn \\ \cn \end{array}\right\} \left[
  \frac{\omega}{2}(\sigma(\chi_i)-\pi),k\right]\,.$$
\end{description}

The Newtonian limit of Eq.~(A18) is
\beq%vvvvvvvvvvvvvvvvvvvvvvvvvvvvvvvvvvvvvvvvvvvvvvvvvvvvvvvvvvvvvvvvvvvv
 t_f - t_i = \frac{2}{u_1^N u_2^N\calL} \left[ \frac{1}
 {\sqrt{1-e^2}}\arctan\left( \frac{\sqrt{1-e^2}}{e|\sin\vartheta_i|}
 \right) + \frac{e \sin\vartheta_i}{1-e^2\cos^2\vartheta_i} \right],
\eeq%^^^^^^^^^^^^^^^^^^^^^^^^^^^^^^^^^^^^^^^^^^^^^^^^^^^^^^^^^^^^^^^^^^^^
where
$$ \cos \vartheta_i = \Phi - \Psi \, , \qquad
 \sin\vartheta_i = \eta_i\, \sqrt{1-(\Phi-\Psi)^2}\; . $$
and functions $\Phi$ and $\Psi$ are defined as in (A14) and
evaluated for $u_1\equiv u_1^N,$ $u_2\equiv u_2^N$.
\smallskip

\noindent
{\it A2.2 Case II orbits} %%%%%%%%%%%%%%%%%%%%%%%%%%%%%%%%%%%%%%%%%%%%%%%%%%%%
%%%%%%%%%%%%%%%%%%%%%%%%%%%%%%%%%%%%%%%%%%%%%%%%%%%%%%%%%%%%%%%%%%%%%%%%%%%%%%%

\noindent
These orbits have no pericentre.
They are captured by the black hole in most cases, even
though the interaction with the accretion disc can in principle modify the
orbital parameters and change the type of the orbit. Consequently,
we do not perform the
mapping in full detail---we skip the derivation of the
time interval ($t_f - t_i$) which will not be needed
for this type of orbits. We still need to describe the
complete algorithm for the mapping
in the $u$-coordinate and associated $\eta$-function.
We were unsuccessful in finding a compact expression for the mapping
$u_i\longrightarrow u_f(u_i;u_1,u_c)$ similar to that
presented in formula (A14), and thus in the following we give the algorithm
of the mapping in several steps (well suited for programming).

\noindent
(i) Evaluate the following quantities:
$$ \mu=\frac{1}{4}(1-2u_1)\,, \qquad
 e = \sqrt{3-l(1-l{\cal L}^{-2})} \, ,$$
$$ \delta = \sqrt{(6\mu -1)^2 + 4 \mu^2 e^2}\,, \qquad
 \gamma_{\pm} = \delta \pm6\mu -1\, , \qquad
 k^2_{\pm} =\frac{ \gamma_{\pm}}{2\delta}\,  , $$
$$ \epsilon = \frac{4u_i+2u_1-1}{e(1-2u_1)}\,, \qquad
 \epsilon_c = \frac{1}{1+\epsilon^2}\, , $$
\[ \epsilon_s = \left\{ \begin{array}{ll} \sqrt{1-\epsilon_c^2}
 & \mbox{if\quad $u_i \geq \mu$,} \\
 - \sqrt{1-\epsilon_c^2} & \mbox{if\quad $u_i < \mu$,}
 \end{array}\right. \]
and the angle $\omega_i \in\;\langle-\frac{\pi}{2},\frac{\pi}{2}\rangle$
which is defined by
$$ \sin^2 \omega_i = 1- 2 \gamma_{+}^{-1} \epsilon_c \left[
 2e\mu \epsilon_s + (6\mu - 1)\epsilon_c \right]  , $$
where ${\rm sgn}(\omega_i) = {\rm sgn}(\epsilon_s - k_{-})$.

\noindent
(ii) Exclude the captured trajectories by evaluating
$$ \vartheta_i = [K(k_{+}) - F(\omega_i,k_+)]\,\delta^{-1/2}  ;$$
the particle will be
captured by the black hole before reaching the equatorial plane
if $\vartheta_i \leq \pi$ and $\eta_i = -1$.
Otherwise we continue to the following step.

\noindent
(iii) Define
$$ \lambda_{\star} = F(\omega,k_+) - \eta_i \pi \delta^{1/2}  ,
\qquad \zeta_i = {\rm sgn}(K(k_+) + \lambda_{\star})\, , $$
and
$$ \left\{\begin{array}{c} \kappa_s \\ \kappa_c \end{array}\right\} =
 \left\{\begin{array}{c} \zeta_i\, \sn(\lambda_{\star},k_+) \\
 \cn (\lambda_{\star},k_+) \end{array}\right\}\, . $$

\noindent
(iv) Finally, the mapping which we look for is given by
\beq%vvvvvvvvvvvvvvvvvvvvvvvvvvvvvvvvvvvvvvvvvvvvvvvvvvvvvvvvvvvvvvvvvvvv
 u_f = \mu \left( 1 + e {\cal T}_{[{\rm sgn}(\kappa_s - {\cal S})]}
 \right) , \qquad \eta_f = \zeta_i\,\eta_i \, ,
\eeq%^^^^^^^^^^^^^^^^^^^^^^^^^^^^^^^^^^^^^^^^^^^^^^^^^^^^^^^^^^^^^^^^^^^^
where we have used
$$ {\cal T}_{\pm} = \gamma_{+}^{-1} \kappa_c^{-2} \left[ 2e\mu \pm
 \sqrt{4e^2 \mu^2 + \gamma_{+} \kappa_c^2 \left(\gamma_{+} \kappa_s^2
 - \gamma_{-} \right)} \right] , $$
$$ {\cal S} = -\sqrt{1-2 (6\mu -1)\gamma_{+}^{-1}} \, . $$
These orbits are unstable in the sense that they reach the singularity
at $r=0$ and thus in the Newtonian limit there are no analogous orbits
corresponding to this case.

Eqs.~(A14) and (A21) give the mapping for astrophysically
interesting cases of geodesics in the Schwarzschild geometry.

\parskip 10pt
\parindent 0pt
\bigskip\bigskip

{\large\bf FIGURE CAPTIONS}

{\bf Figure 1. }The sequence of successive intersections with the disc.
We plot radial coordinate $r$ of the intersection
on the ordinate and the number of intersections $N$
on the abscissa. For $N\gta 5\,{\rm x}\,10^4$ the upper
and lower boundaries of the
distribution of intersections in the Figure get closer to each other,
which means that the eccentricity of the orbit decreases; simultaneously
as the orbit is circularized it is also ground to the plane of the disc.
In this case, $a=0$ and $\alpha_\star=0.9999.$
The initial pericentre distance is $7$, eccentricity $0.7,$ inclination
$I=103\dg.$ The arrow indicates the moment when the orbit changes
its character from retrograde to prograde $(I=90\dg).$
\smallskip

{\bf Figure 2. }As in Fig.~1 but for the second model of the
star-disc interaction [Eq.~(40)].
Initial pericentre distance is $30$ and eccentricity
$e = 0.83$. Two initial inclinations are compared---(a) $I=35\dg$,
(b) $I=80\dg$. The proportionality constant in
(40) is taken $10^{-5}$. Originally less inclined orbit (a)
settles on the circular orbit in the disc with radius of $\approx 53$, while
the more inclined orbit (b) grinds to the circular orbit at radius
$\approx 17.7\,.$ We verified, as an example,
that keeping the initial pericentre at
$30$ the results are not very sensitive to the initial
eccentricity provided it is  $\gta 0.75$.
\smallskip

{\bf Figure 3. }The graph (a) shows a
similar orbit as are those in Fig.~2, but now the initial trajectory
is retrograde with an inclination of $130\dg$. As commented in the text,
it is captured by the central black hole. The graph (b) shows the
`Newtonian analogue' of (a). Only the `adiabatic evolution'
of the apocentre (upper curve) and the pericentre (lower curve) is
seen and there is no shift of pericentre.
\smallskip

{\bf Figure 4. }Orbits analogous to those in Fig.~2, but with a
nearly extreme Kerr metric ($a=0.9981$). The initial pericentre
distance is $15$ (the same as for the Fig.~2 orbits if expressed in
gravitational radii of the central black hole). The initial
inclinations are again chosen to be $I=35\dg$ (a) and $I=80\dg$ (b).

\newpage

\setcounter{equation}{0}
\renewcommand{\theequation}{\arabic{equation}}
\hyphenation{Schw-arz-sch-ild}
% -----------------------------------------------------------------------
\normalsize

{\Large\bf Relativistic precession of the orbit of a star
 near a supermassive black hole}
\bigskip\bigskip

\noindent
{\large\bf ABSTRACT}

We study the gravitomagnetic (Lense-Thirring) precession
of the trajectory (approximated by a geodesic) of a star orbiting
a supermassive rotating (Kerr) black hole. We do not
assume any particular value for the eccentricity or inclination of the
orbit or the angular momentum of the black hole.
We also discuss the periodicity related to the
relativistic shift of the pericenter.

The Seyfert galaxy NGC~6814 is an example of the
object for which effects of relativistic precession could be detectable
and we discuss the relevant precession frequency for this case.
The remarkably stable phase of several patterns and their position
in the light curve impose strong restrictions on the model of this object.
We conclude that, according to our present knowledge,
a star colliding with an accretion disk is somewhat
improbable though not completely excluded as a model of NGC~6814.
Our arguments are independent of studies of the period stability.
\medskip

\noindent
{\it Subject headings}: galaxies: active --- galaxies: individual
 (NGC 6814) --- general relativity --- black holes
\bigskip

%\newpage
\centerline{\large\bf 1.\quad INTRODUCTION AND MOTIVATION}
%%%%%%%%%%%%%%%%%%%%%%%%%%%%%%%%%%%%%%%%%%%%%%%%%%%%%%%%%%%%%%%%%%%%%%%%%%%%%%%
\medskip

It is widely accepted that supermassive black holes (SBH) are located in cores
of active galaxies and quasars (Begelman, Blandford \& Rees 1984; Shlosman,
Begelman~\& Frank 1990).
The mass of the SBH is usually estimated to be
in the range $M\approx 10^6$--$10^{11}\,M_\odot.$
Anomalous energy output of active galactic nuclei (AGN) may result from
an accretion process: the matter is attracted from the surroundings of
AGN or it
comes from tidally disrupted stars passing too close to the SBH. The accretion
disk is formed and the matter eventually falls onto the black hole.
Although this scenario can be called standard, the evidence for both
accretion disks and SBH in AGN is only indirect (Frank, King~\& Raine 1985;
Blandford, Netzer~\& Woltjer 1990; Falcke \etal 1993).
The best evidence comes from studies of
the central surface brightness of the nuclei, stellar velocity dispersion,
spatial distribution of stars and X-ray emission
(Young \etal 1978; Sargent \etal 1978; Binney~\& Tremaine 1987;
Dressler~\& Richstone 1988, 1990; Kormendy 1988a, b;
Halpern~\& Filippenko 1988; Dressler 1989).
General relativistic effects may have important consequences for the axial
symmetry and stability of accretion disks (Bardeen~\& Petterson 1975;
Abramowicz 1987). However, the presence of black holes
in AGN is largely masked by violent plasma processes in the surrounding
medium. Electromagnetic models of energy extraction assume that the SBH rotates
(Blandford~\& Znajek 1977; Macdonald~\& Thorne 1982; Phinney 1983;
Kaburaki~\& Okamoto 1991; Okamoto~\& Kaburaki 1991 and references cited
therein). In this scenario, the energy of an AGN comes, at least partially,
from the rotational energy of the SBH. Evolution of the angular
momentum of the black hole under such process was studied by Park~\&
Vishniac (1991). It appears that the angular momentum determines the
energy output of the AGN in a nontrivial manner,
with the maximum at some particular value (Bi\v{c}\'ak~\& Jani\v{s} 1985).
Unfortunately, it is unclear how the value of the angular momentum of the SBH
could be determined by an independent observation. The present paper deals
with this problem.

It has been proposed that a star could be captured in a bound orbit
around the SBH by the tidal distortion and associated dissipation
of energy (Fabian, Pringle~\& Rees 1975;
Frank~\& Rees 1976; Press~\& Teukolsky 1977; Lee~\& Ostriker 1986;
Rees 1988), tidal disruption of a binary
star (Hills 1988) or a cluster (Novikov, Pethick~\& Polnarev 1992), or by
cumulative effects of interactions with an accretion disk
(Syer, Clarke~\& Rees 1991).
Various aspects of star-disk collisions were studied by
Ostriker (1983), Zentsova (1983), Syer \etal (1991), Zurek,
Siemiginowska~\& Colgate (1992),
Sikora~\& Begelman (1992) and Vokrouhlick\'y~\& Karas (1993).
We assumed that the SBH forms such a binary system with a low
mass star in an orbit inclined with respect to the plane of an accretion
disk. The disk is presumably thin and its axis is aligned with the rotation
axis of the black hole (Bardeen~\& Petterson 1975; Kumar~\& Pringle 1985)
but the model can easily be generalized to a more complicated geometry.
Radiation from the disk is periodically
modulated each time the star crosses the disk. We did not attempt to specify
any particular mechanism of the modulation. We also ignored
secular changes of the orbital parameters of the star due to
collisions with the disk and corresponding
contributions of the collisions to the precession frequency.
We concentrated on a gravitomagnetically induced precession of the orbit
of the star. This precession is a general relativistic effect; it
occurs if the central black hole rotates and it becomes
important when the star revolves close to it.
In particular we attempted to pick up relevant frequencies in
the power spectrum of the simulated signal which are
independent of very complex and poorly understood details of the
star-disk interaction.
Given the model of interaction our approach can easily be adapted.
We took into account all the relativistic effects
affecting the motion and energy of photons arriving
from the source to a distant observer.

We assume that the star moves along a geodesic around
a Kerr black hole. The gravitomagnetic effect was originally
treated by Lense~\& Thirring (1918) in the weak-field limit and by Wilkins
(1972) in the case
of a spherical orbit, $r=\const$, around a black hole with the extreme value of
the angular momentum parameter, $a=1.$ According to our knowledge,
a generalization to an eccentric
and inclined orbit around a black hole with arbitrary value of parameter
$a\in(0,1)$ has not yet been discussed in the literature.
As a consequence of the gravitomagnetic effect,
orbital nodes are dragged in the sense of rotation of the black hole.
This dragging affects the position of the source with respect to a distant
observer. We try to {\it extract the information which might
help us to determine whether the central SBH rotates or not---provided other
details of star-disk collisions are known.}
The weak-field limit of the angular velocity of the gravitomagnetic dragging is
\beq %vvvvvvvvvvvvvvvvvvvvvvvvvvvvvvvvvvvvvvvvvvvvvvvvvvvvvvvvvvvvvvvvvvvvvvvvv
 \tilde{\Omega}_{\rm LT}\simeq
 4\times 10^5\,\frac{M_\odot}{M}\frac{a}{r^3}\quad{\rm s^{-1}},
\eeq %^^^^^^^^^^^^^^^^^^^^^^^^^^^^^^^^^^^^^^^^^^^^^^^^^^^^^^^^^^^^^^^^^^^^^
where $r$ and $a$ are the radius of the orbit and the angular momentum
parameter of SBH measured in dimensionless geometrized units,
$r=6.7\times 10^{-4}(\tilde{r}/1\;{\rm cm}).$
Several experiments to confirm the dragging effect in the limit of a
weak gravitational field have been proposed
but they have not been carried out as yet (Everitt 1974; Will 1981;
Braginskij, Polnarev~\& Thorne 1984;
Ciufolini 1986). The strong field regime of the spin-orbital interaction
is also intensively investigated for a relatively broad
class of gravity theories within the framework of the
parametrised post-Keplerian formalism. The main applications of this
approach are directed at the interpretation
of the binary pulsar data (e.g. Damour~\& Taylor 1992).
Though not yet confirmed by direct observations, the gravitomagnetic
effect is considered as a firm consequence of general relativity.
In the strong-field region close to the black hole the
gravitomagnetic precession depends on four parameters---the pericenter
distance of the orbit, eccentricity of the orbit, inclination with respect
to the equatorial plane of the black hole, and, of course, the angular
momentum parameter. The above mentioned processes of tidal interaction
can set the star on an initially very eccentric trajectory and we therefore
could not restrict ourselves to circular orbits. Star-disk collisions (Syer
\etal
1991; Vokrouhlick\'y~\& Karas 1993), tidal effects (Press, Wiita~\&
Smarr 1975;
Lecar, Wheeler~\& McKee 1976; Boyle~\& Walker 1986; Zahn 1977 and 1989;
Zahn~\& Bouchet 1989; Tassoul 1988)
and gravitational radiation (Peters~\& Mathews 1963; Zeldovich~\& Novikov
1971) tend to gradually circularize the elliptic
orbit, however. Naturally, these effects modify the
precession frequency, but we assumed that the
star is a dwarf or a low-mass compact object and ignored the corrections.
We also ignore all possible effects of the magnetic field on the
space-time geometry.
For the discussion of exact solutions of Einstein-Maxwell equations
describing a black hole in a magnetic field cf. Ernst (1976), Karas (1991),
Manko~\& Sibgatullin (1992), and references cited therein.

In other words, we ignored all effects which could induce the precession
of the star's orbit except the gravitomagnetic effect. We believe it is an
appropriate approach in solving the problem if these effects (like
star-disk interactions or gravitational radiation) are also weak.
We will discuss other effects elsewhere so that the whole subject is
treated in steps.

We determined the value of the precession frequency (or alternatively
the nodal shift per one revolution, $\delta\phi$) by direct integration
of the geodesic equation in terms of elliptic integrals.
Gravitational radiation losses can be neglected provided the change of
the energy is small; for a circular orbit
with energy $\calE$ one obtains the dimensionless
estimate (e.g. Rees, Ruffini~\& Wheeler 1974)
\beq %vvvvvvvvvvvvvvvvvvvvvvvvvvvvvvvvvvvvvvvvvvvvvvvvvvvvvvvvvvvvvvvvvvvv
 \mid\dot{\calE}\mid_{\rm grav}\approx
 6.4\left(\frac{M}{r}\right)^5\left(\frac{M_{\star}}{M}\right)^2\ll 1,
 \label{gr}
\eeq %^^^^^^^^^^^^^^^^^^^^^^^^^^^^^^^^^^^^^^^^^^^^^^^^^^^^^^^^^^^^^^^^^^^^^
where $M_{\star}$ is the mass of the star. Suppose we estimate $M,$ $r$ and
an upper limit on the change of orbital period $\dot{P}$ from observation.
Then equation (\ref{gr}) provides us with the
upper limit on the mass of the companion
star [cf. Sikora~\& Begelman (1992) and King~\& Done (1993)
for the application to
NGC~6814]. Analogously, if $R_{\star}$ is the radius of
the star one can obtain a constraint on the pericenter
distance $R_{\rm p}$ which is based on the
tidal limit (Carter~\& Luminet 1983; Luminet~\& Marck 1985; Rees 1988):
\beq %vvvvvvvvvvvvvvvvvvvvvvvvvvvvvvvvvvvvvvvvvvvvvvvvvvvvvvvvvvvvvvvvvvvv
 R_{\rm p}\gg R_{\star}\left(\frac{M_{\star}}{M}\right)^{1/3}.
\eeq %^^^^^^^^^^^^^^^^^^^^^^^^^^^^^^^^^^^^^^^^^^^^^^^^^^^^^^^^^^^^^^^^^^^^^

This paper is organized as follows. In Section~2 we derive the
azimuthal shift of orbital nodes. An unambiguous value can only be given in
the case of spherical trajectories. The shift oscillates
between the maximum and the minimum values, $\delta\phi_{\rm max}$ and
$\delta\phi_{\rm min},$ if the orbit is eccentric.
We define a suitable probability distribution for
$\delta\phi$ and compute, numerically, a mean value $\langle\delta\phi\rangle$
which determines the gravitomagnetic precession averaged over a large
number of revolutions of the star on an eccentric trajectory around the SBH.
Corresponding formulas are rather
complex and, therefore, we also present a table of numerical values in
Appendix. In Section~3 we present simple examples to illustrate
how the gravitomagnetic frequency could be extracted from observational
data. (In the present contribution, we use simulated data
from a simple model rather than real data.)
Naturally, the frequency cannot be too small so that the data
cover at least several periods if the effect is to be detectable.
The estimate of the time interval between successive collisions with the
disk [see eq.~(\ref{p}) below] leads us to assume that the star
crosses the accretion disk near the horizon (typically a few tens of the
gravitational radius of the SBH) in the region where the collisions can
modulate the disk radiation in the optical/X-ray bands.
Possible physical mechanisms for the modulation were discussed by
Mushotzky (1982), Guilbert, Fabian \& Ross (1982) and Zentsova (1985).
The flux of
radiation is modulated by the orbital period (short time-scale) and
by the perihelion shift and the Lense-Thirring period
(longer time-scale). In order to compute observable effects we consider both
the
time delay and the focusing of photons coming from the disk region
to a distant observer. The focusing effect strongly enhances
the influence of the
Lense-Thirring precession for observers with large inclination (edge on, with
respect to the accretion disk). Finally, we apply the
model to the case of the Seyfert galaxy NGC~6814. Periodic modulation
of the X-ray emission from this galaxy has been confirmed on the
time-scale $\approx 12,200\;$s (Mittaz~\& Branduardi-Raymont 1989;
Fiore, Massaro~\& Barone 1992;
Done \etal 1992) and possible mechanisms were proposed by several authors
(Abramowicz \etal 1989 and 1992;
Done \etal 1991; Honma \etal 1991;
Wallinder 1991 and 1992; Abramowicz 1992; Bao 1992a,~b;
Sikora~\& Begelman 1992; Rees 1993; Vio \etal 1993; King~\& Done 1993).

\bigskip\bigskip

\centerline{\large\bf 2.\quad PRECESSION FREQUENCY---DETAILS OF THE}

\centerline{\large\bf CALCULATION}
%%%%%%%%%%%%%%%%%%%%%%%%%%%%%%%%%%%%%%%%%%%%%%%%%%%%%%%%%%%%%%%%%%%%%%%%%%%%%
\bigskip

\centerline{2.1 \it Gravitomagnetic precession} %%%%%%%%%%%%%%%%%%%%%%%%%%%%%%
\smallskip

Geodesic motion in the Kerr space-time
can be integrated in terms of elliptic integrals (Carter 1968).
The appropriate form can be found in Vokrouhlick\'y~\& Karas (1993).
We use the standard notation for the Kerr metric (Bardeen 1973) and
we refer the reader not familiar with the Kerr metric to this reference.
We assume that the motion of the star is quasi-elliptic with
the pericenter outside the black hole horizon,
$r=R_+\equiv 1+\sqrt{1-a^2},$ and with positive energy,
$0<\calE<1.$ The locations of the pericenter $r=R_{\rm p}$ and the apocenter
$r=R_{\rm a}$ coincide with the two upper roots of the polynomial
\beq %vvvvvvvvvvvvvvvvvvvvvvvvvvvvvvvvvvvvvvvvvvvvvvvvvvvvvvvvvvvvvvvvvvvvvvv
 R(r)=(\calE^2-1)r^4+2r^3+\left[ (\calE^2-1)a^2-\Phi^2-{\calQ}\right]r^2
 +2\calK r-a^2{\calQ},
\eeq %^^^^^^^^^^^^^^^^^^^^^^^^^^^^^^^^^^^^^^^^^^^^^^^^^^^^^^^^^^^^^^^^^^^^^^^
where $\calE\equiv-p_t,$ $\Phi\equiv p_\phi$ and $\calK$ are
the usual constants of motion,
$\calQ\equiv\calK+(\Phi-a\calE)^2.$ In our case, all roots of $R(r)$ are real.
We denote them by $R_{\rm a}\geq R_{\rm p}\geq R_3\geq R_4;$ $R_3>R_+.$
Other possible combinations---e.g. two complex conjugated roots---are
excluded by our previous assumptions on energy and location of the
pericenter above the horizon (Stewart~\& Walker 1973).
We further assume that the star periodically crosses the equatorial plane,
$\theta=\pi/2;$ latitudal motion is restricted by
$\mid\cos\theta\mid\leq\mu_-,$
where $\mu_-$ is the lower of the two positive
roots $\mu_\pm$ of the polynomial
\beq %vvvvvvvvvvvvvvvvvvvvvvvvvvvvvvvvvvvvvvvvvvvvvvvvvvvvvvvvvvvvvvvvvvvvvv
 \Theta(\mu)=a^2\left(1-\calE^2\right)\mu^4-\left[{\calQ}+
 a^2\left(1-\calE^2\right)+\Phi^2\right]\mu^2+{\calQ};
 \label{th}
\eeq %^^^^^^^^^^^^^^^^^^^^^^^^^^^^^^^^^^^^^^^^^^^^^^^^^^^^^^^^^^^^^^^^^^^^^^
For $r\gg R_+$ one can interpret $\arccos\mu_-$ as the inclination of
stellar orbit.

A non-equatorial geodesic around a rotating black hole is not planar;
intersections with the $\theta=\pi/2$ plane are dragged
in the sense of rotation. At first, we apply the mapping
\beq %vvvvvvvvvvvvvvvvvvvvvvvvvvvvvvvvvvvvvvvvvvvvvvvvvvvvvvvvvvvvvvvvvvvvvvv
 \left[r,\phi,t,{\rm sign}(\dot{r})\right]_n\longrightarrow
 \left[r,\phi,t,{\rm sign}(\dot{r})\right]_{n+1}
 \label{map}
\eeq %^^^^^^^^^^^^^^^^^^^^^^^^^^^^^^^^^^^^^^^^^^^^^^^^^^^^^^^^^^^^^^^^^^^^^^^
which expresses coordinates of the $(n+1)$-st intersection in terms of
coordinates of the previous, $n$-th intersection with the equatorial plane.
The azimuthal shift of the orbital node $\delta\phi$ per one revolution is
then defined by
\beq %vvvvvvvvvvvvvvvvvvvvvvvvvvvvvvvvvvvvvvvvvvvvvvvvvvvvvvvvvvvvvvvvvvvvvvv
 \delta\phi=\phi_{n+2}-\phi_n-2\pi
 \label{uze}
\eeq %^^^^^^^^^^^^^^^^^^^^^^^^^^^^^^^^^^^^^^^^^^^^^^^^^^^^^^^^^^^^^^^^^^^^^^^
($\phi$-coordinate is not restricted to $\langle 0,2\pi\rangle$
in this convention). One can find
\beq %vvvvvvvvvvvvvvvvvvvvvvvvvvvvvvvvvvvvvvvvvvvvvvvvvvvvvvvvvvvvvvvvvvvvvvv
 r_{n+1}=
 \frac{(R_{\rm a}-R_3)R_{\rm p}+(R_{\rm a}-R_{\rm p})R_3\sigma}
 {R_{\rm a}-R_3-(R_{\rm a}-R_{\rm p})\sigma}\,,
 \label{r}
\eeq %^^^^^^^^^^^^^^^^^^^^^^^^^^^^^^^^^^^^^^^^^^^^^^^^^^^^^^^^^^^^^^^^^^^^^^^
\beq %vvvvvvvvvvvvvvvvvvvvvvvvvvvvvvvvvvvvvvvvvvvvvvvvvvvvvvvvvvvvvvvvvvvvvvv
 \phi_{n+1}-\phi_n=\left[2(a\calE-\Phi)A_++\Phi B_+\right]I_++
 \left[2(a\calE-\Phi)A_-+\Phi B_-\right]I_-+\Phi J_\mu\,,
 \label{phi}
\eeq %^^^^^^^^^^^^^^^^^^^^^^^^^^^^^^^^^^^^^^^^^^^^^^^^^^^^^^^^^^^^^^^^^^^^^^^
\begin{eqnarray} %vvvvvvvvvvvvvvvvvvvvvvvvvvvvvvvvvvvvvvvvvvvvvvvvvvvvvvvvvvv
 \lefteqn{\nonumber
 t_{n+1}-t_n=\calE\left(J_r+2K_r\right)+
 2\left[B_+R_+\calE+a(a\calE-\Phi)A_+\right]I_+ }\\
 & & +2\left[B_-R_-\calE+a(a\calE-\Phi)A_-\right]I_-+
 4\calE I_r+a^2\calE K_\mu\,.
 \label{t}
\end{eqnarray} %^^^^^^^^^^^^^^^^^^^^^^^^^^^^^^^^^^^^^^^^^^^^^^^^^^^^^^^^^^^^^^
We denoted
\beq %vvvvvvvvvvvvvvvvvvvvvvvvvvvvvvvvvvvvvvvvvvvvvvvvvvvvvvvvvvvvvvvvvvvvvvvv
 \sigma=\sn^2(u,k_1)\,,
\eeq %^^^^^^^^^^^^^^^^^^^^^^^^^^^^^^^^^^^^^^^^^^^^^^^^^^^^^^^^^^^^^^^^^^^^^^^^
\beq %vvvvvvvvvvvvvvvvvvvvvvvvvvvvvvvvvvvvvvvvvvvvvvvvvvvvvvvvvvvvvvvvvvvvvvvv
 u=\frac{1}{2}\sqrt{(R_{\rm a}-R_3)(R_{\rm p}-R_4)(1-\calE^2)}
 \left(I_\mu-\tilde{I}_r\right),
 \label{u}
\eeq %^^^^^^^^^^^^^^^^^^^^^^^^^^^^^^^^^^^^^^^^^^^^^^^^^^^^^^^^^^^^^^^^^^^^^^^^
$$k_1=\frac{(R_{\rm a}-R_{\rm p})(R_3-R_4)}{(R_{\rm a}-R_3)(R_{\rm p}-R_4)}\,,
 $$
$$A_\pm=\pm\frac{R_\pm}{R_+-R_-}\,,
 \qquad B_\pm=\pm\frac{2R_\pm-a^2}{R_+-R_-}\,,$$
$$R_\pm=1\pm\sqrt{1-a^2}\,,$$
\beq %vvvvvvvvvvvvvvvvvvvvvvvvvvvvvvvvvvvvvvvvvvvvvvvvvvvvvvvvvvvvvvvvvvvvvvv
I_\mu=\frac{2}{a\mu_+\sqrt{1-\calE^2}}K(\mu_-/\mu_+),
\eeq %^^^^^^^^^^^^^^^^^^^^^^^^^^^^^^^^^^^^^^^^^^^^^^^^^^^^^^^^^^^^^^^^^^^^^^^
\beq %vvvvvvvvvvvvvvvvvvvvvvvvvvvvvvvvvvvvvvvvvvvvvvvvvvvvvvvvvvvvvvvvvvvvvvv
I_\pm=\kappa_\pm\left[(R_{\rm p}-R_3)\Pi\left(\varphi,n_\pm,k_1\right)+
 (R_\pm-R_{\rm p})F(\varphi,k_1)\right]+\tilde{I}_\pm,
 \label{ipm}
\eeq %^^^^^^^^^^^^^^^^^^^^^^^^^^^^^^^^^^^^^^^^^^^^^^^^^^^^^^^^^^^^^^^^^^^^^^^
$$n_\pm=\frac{(R_{\rm p}-R_{\rm a})(R_\pm-R_3)}
 {(R_{\rm a}-R_3)(R_\pm-R_{\rm p})}\,;$$
$F(\varphi,k)$ is the incomplete elliptic integral of the first kind,
$K(k)\equiv F(\pi/2,k)$ is the corresponding complete integral and
$\sn(u,k)$ is the Jacobian elliptic function (e.g. Byrd~\& Friedman 1971;
Gradshteyn~\& Ryzhik 1980). Analogously, $\Pi(\varphi,n,k)$ is the
elliptic integral of the third kind. In equations (\ref{r})--(\ref{t}),
\begin{eqnarray} %vvvvvvvvvvvvvvvvvvvvvvvvvvvvvvvvvvvvvvvvvvvvvvvvvvvvvvvvvvv
\lefteqn{\nonumber
J_r+2K_r=\kappa R_{\rm p}\left[\left(\frac{R_{\rm p}\alpha_1^4}{\alpha^4}+
 2\frac{\alpha_1^2}{\alpha^2}\right)U+
 2\frac{\alpha^2-\alpha_1^2}{\alpha^2}\,V_1   \right.}\\
& & \left. +2R_{\rm p}\alpha_1^2\frac{\alpha^2-\alpha_1^2}{\alpha^4}\,V_1+
 R_{\rm p}\frac{(\alpha^2-\alpha_1^2)^2}{\alpha^4}\,V_2\right]+\tilde{J}_r
+2\tilde{K}_r,
\label{jk}
\end{eqnarray} %^^^^^^^^^^^^^^^^^^^^^^^^^^^^^^^^^^^^^^^^^^^^^^^^^^^^^^^^^^^^^
with $U,$ $V_1$ and $V_2$ defined by
$$U=F(\varphi,k_1)\,,$$
$$V_1=\Pi\left(\varphi,-\alpha^2,k_1\right),$$
\begin{eqnarray*} %vvvvvvvvvvvvvvvvvvvvvvvvvvvvvvvvvvvvvvvvvvvvvvvvvvvvvvvvvvv
\lefteqn{
V_2=\frac{1}{2(\alpha^2-1)(k_1^2-\alpha^2)}\Biggl[\alpha^2E(\varphi,k_1)+
 \left(2\alpha^2k_1^2+2\alpha^2-\alpha^4-
 3k_1^2\right)V_1    }\\
 & &  +\left(k_1^2-\alpha^2\right)U
 -\sn(U,k_1)\,\cn(U,k_1)\,\dn(U,k_1)\,\frac{\alpha^4}
 {1-\alpha^2\sn^2(U,k_1)}\Biggr],
\end{eqnarray*} %^^^^^^^^^^^^^^^^^^^^^^^^^^^^^^^^^^^^^^^^^^^^^^^^^^^^^^^^^^^^^
and
$$\alpha^2=\frac{R_{\rm a}-R_{\rm p}}{R_{\rm a}-R_3}\,,\qquad
 \alpha_1^2=\frac{R_3(R_{\rm a}-R_{\rm p})}{R_{\rm p}(R_{\rm a}-R_3)}\,,$$
$$\sin^2\varphi=\frac{(R_{\rm a}-R_3)(r_{n}-R_{\rm p})}
 {(R_{\rm a}-R_{\rm p})(r_{n}-R_3)}\,,$$
$$\kappa=\frac{2}{\sqrt{(R_1-R_3)(R_2-R_4)(1-\calE^2)}}\,,$$
$$\kappa_\pm=\frac{\kappa}{(R_\pm-R_3)(R_{\rm p}-R_\pm)}\,.$$
$E(\varphi,k)$ denotes the incomplete elliptic integral of the
second kind, $\cn(u,k)$ and $\dn(u,k)$ are the Jacobian elliptic
functions,
\beq %vvvvvvvvvvvvvvvvvvvvvvvvvvvvvvvvvvvvvvvvvvvvvvvvvvvvvvvvvvvvvvvvvvvvvvv
J_\mu=\frac{2}{a\mu_+\sqrt{1-\calE^2}}\Pi\left(-\mu_-^2,\mu_-/\mu_+\right),
\eeq %^^^^^^^^^^^^^^^^^^^^^^^^^^^^^^^^^^^^^^^^^^^^^^^^^^^^^^^^^^^^^^^^^^^^^^^
\beq %vvvvvvvvvvvvvvvvvvvvvvvvvvvvvvvvvvvvvvvvvvvvvvvvvvvvvvvvvvvvvvvvvvvvvvv
K_\mu=\frac{2R_{\rm a}}{a\sqrt{1-\calE^2}}
 \left[K(\mu_-/\mu_+)-E(\mu_-/\mu_+)\right]
\eeq %^^^^^^^^^^^^^^^^^^^^^^^^^^^^^^^^^^^^^^^^^^^^^^^^^^^^^^^^^^^^^^^^^^^^^^^
and $\Pi(n,k)\equiv\Pi(\pi/2,n,k).$
$\tilde{I}_r,$ $\tilde{I}_\pm,$ $\tilde{J}_r$ and $\tilde{K}_r,$
in equations (\ref{u}), (\ref{ipm}) and (\ref{jk}) are the
integration constants.
They depend on the number of turning points in $r$-coordinate between
successive
intersections with the equatorial plane $(m)$ and on the
sign of the radial velocity [sign$(\dot{r})]$ at the intersections. For
example,
\beq %vvvvvvvvvvvvvvvvvvvvvvvvvvvvvvvvvvvvvvvvvvvvvvvvvvvvvvvvvvvvvvvvvvvv
 \tilde{I}_r=
  \kappa\left[mK(k_1)-{\rm sign}(\dot{r}_{n+1})\,F(\varphi,k_1)\right].
\eeq %^^^^^^^^^^^^^^^^^^^^^^^^^^^^^^^^^^^^^^^^^^^^^^^^^^^^^^^^^^^^^^^^^^^^^
Further details of the derivation are given in Appendix of
Vokrouhlick\'y~\& Karas (1993)
where the orbits under consideration are called the ``Case~II orbits."

The above expressions can be considerably simplified in the
special case of spherical orbits
with $r$-coordinate constant. (It can be seen that $r=\const$ orbits
do satisfy the geodesic equation due to integrability and
separability of the geodesic motion in the Kerr spacetime;
Chandrasekhar 1983). The gravitomagnetic precession of
the spherical orbits
around an extreme $(a=1)$ Kerr black hole was
studied by Wilkins (1972). One can generalize his results to the case
of spherical orbits $r\equiv R_{\rm s}=\const$ with an arbitrary value
of the angular momentum parameter, $\mid a\mid < 1$. Taking into
account equation (\ref{phi}), we obtain
\begin{eqnarray} %vvvvvvvvvvvvvvvvvvvvvvvvvvvvvvvvvvvvvvvvvvvvvvvvvvvvvvvvv
 \left(\delta \phi\right)_{r=R_{\rm s}} & = & \frac{4}{a\mu_+\sqrt{1-\calE^2}}
     \left\{\Phi\Pi\left(-\mu_-^2,\mu_-/\mu_+\right)
                                                         \right. \nonumber \\
 & & \left.+a\left[\left(\calE (r^2+a^2) -a\Phi\right)\Delta^{-1}-
     \calE\right]K(\mu_-/\mu_+)\right\}-2\pi,
 \label{dphi}
\end{eqnarray} %^^^^^^^^^^^^^^^^^^^^^^^^^^^^^^^^^^^^^^^^^^^^^^^^^^^^^^^^^^^
where $\Delta=r^2-2r+a^2.$
The corresponding orbital period is
\beq %vvvvvvvvvvvvvvvvvvvvvvvvvvvvvvvvvvvvvvvvvvvvvvvvvvvvvvvvvvvvvvvvvvvvv
P_{r=R_{\rm s}}=
 2\left[\frac{r^2\left(r^2+a^2\right)\calE+2ar\left(a\calE-\Phi\right)}
 {\Delta}I_\mu+a^2\calE K_\mu\right]
 \label{p}
\eeq %^^^^^^^^^^^^^^^^^^^^^^^^^^^^^^^^^^^^^^^^^^^^^^^^^^^^^^^^^^^^^^^^^^^^^
and the precession frequency is
\beq %vvvvvvvvvvvvvvvvvvvvvvvvvvvvvvvvvvvvvvvvvvvvvvvvvvvvvvvvvvvvvvvvvvvvv
 \left(\Omega_{\rm LT}\right)_{r=R_{\rm s}}=
 \left(\frac{\delta\phi}{P}\right)_{r=R_{\rm s}}.
 \label{omlt}
\eeq %^^^^^^^^^^^^^^^^^^^^^^^^^^^^^^^^^^^^^^^^^^^^^^^^^^^^^^^^^^^^^^^^^^^^^
Setting $a=0$ in equations (\ref{dphi})--(\ref{p}) one arrives at
$\delta\phi=0$
and $P=2\pi R_{\rm s}^{3/2}$
which corresponds to Keplerian motion with vanishing nodal shift in the
Schwarzschild metric, as expected. On the other hand, if $a\neq 0$ the
dominant term in the asymptotic expansion of equation (\ref{dphi}) for
$r\longrightarrow\infty$ coincides with the
well-known result of Lense~\& Thirring (1918):
$\delta\phi=2\pi aR_{\rm s}^{-3/2}$. The dependence of period $P$ on the
inclination $\mu$ is only weak and equation (\ref{p}) can be
approximated by its value for equatorial orbits:
\beq %vvvvvvvvvvvvvvvvvvvvvvvvvvvvvvvvvvvvvvvvvvvvvvvvvvvvvvvvvvvvvvvvvvvv
 P\approx 2\pi \left(r^{3/2}\pm a\right),
 \label{pap}
\eeq %^^^^^^^^^^^^^^^^^^^^^^^^^^^^^^^^^^^^^^^^^^^^^^^^^^^^^^^^^^^^^^^^^^^^^
where $+/-$ signs correspond to direct/retrograde orbits with respect
to the rotating SBH. Analogously, the nodal shift of spherical
polar orbits $(\Phi=0)$ can be obtained from equation
(\ref{dphi}) with $\mu_-=1.$
Polar orbits were studied by Stoghianidis~\& Tsoubelis (1987).

\bigskip

\centerline{2.2. \it Motion of the pericenter} %%%%%%%%%%%%%%%%%%%%%%%%%%%%%%%
\smallskip

In the present paper we are mainly interested in the frequency of the
gravitomagnetic precession. In the data, however, there will also
be present a periodicity associated with another relativistic
effect---the pericenter shift. We thus need to estimate the corresponding
precession frequency $\Omega_{\rm P}.$ In the Schwarzschild case
\beq
 \Omega_{\rm P}=\delta\phi/P,
 \label{omp}
\eeq
where
\beq
 \delta\phi=\frac{4}{\omega}K(k)-2\pi,
\eeq
\begin{eqnarray}
 P  & = &  \sqrt{\frac{u_2+u_1}{u_3-u_1}\left(1-2u_1\right)
           \left(1-2u_2\right)}
           \biggl\{\frac{4}{u_1}
           \Pi\left(\alpha^2,k{^\prime}\right)+\frac{8}{1-2u_1}
                                                              \nonumber \\
  &\times& \Pi\left(2\frac{u_2-u_1}{1-2u_1},k{^\prime}\right)
           +\frac{1}{u_1^2\alpha^2\left(k{^\prime}^2-\alpha^2\right)}
           \biggl[\alpha^2E(k{^\prime})
           + \left(k{^\prime}^2-\alpha^2\right)K(k{^\prime})
                                                              \nonumber \\
    & + &  \left(2\alpha^2k{^\prime}^2+2\alpha^2-\alpha^4
           -3k{^\prime}^2\right)
           \Pi\left(\alpha^2,k{^\prime}\right)\biggr] \biggr\}.
\end{eqnarray}
Here,
$$ \omega = \sqrt{1 - 2u_2 - 4u_1}, \qquad k^2 = 2 (u_2 - u_1)\omega^{-2}, $$
$$ \alpha^2=1-\frac{u_2}{u_1} \quad , \quad k{^\prime}^2=
 \frac{u_2-u_1}{u_3-u_1},$$
where $u_1\equiv 1/R_{\rm a}$, $u_2\equiv 1/R_{\rm p}$ and $u_3=\frac{1}{2}-
u_1-u_2$ are roots of the polynomial in $u=1/r:$
$$ U(u) = 2u^3 - u^2 + 2\calL^{-2}u - (1-\calE^2)\calL^{-2} $$
[cf. ``Case~I orbits" in Appendix of Vokrouhlick\'y~\& Karas (1993)].
The angular momentum $\calL$ is
defined with respect to the axis perpendicular to
the orbital plane of the star, not with respect
to the axis of the disk like $\Phi$ in the Kerr case.
This definition is naturally more advantageous in the Schwarzschild case
because the orbit is planar.

An analogous effect of the pericenter shift can be expected
in the Kerr case, too. Now, however, there is
no unambiguous value of $\Omega_{\rm P}$ because
the orbit does not remain planar
and the pericenter shift is mixed up with the gravitomagnetic precession.
We calculate $\Omega_{\rm P}$ numerically by applying the mapping algorithm
(\ref{map}) for the $r$-coordinate over a large number of revolutions of the
star around the SBH.

\bigskip

\centerline{2.3. \it The mean value of the precession frequencies}
 %%%%%%%%%%%%%%
\smallskip

A specific value of the nodal shift $\delta\phi$
and corresponding precession frequency
can only be associated with spherical orbits.
In this case $\delta\phi$ is specified by equation (\ref{dphi}) with three
parameters---$a,$ $\mu_-$ and $R_{\rm s}.$
The case of a quasi-elliptic orbit with nonzero eccentricity requires
knowledge of four parameters. The appropriate parameters are
$a,$ $\mu_-,$ pericenter distance $R_{\rm p},$ and eccentricity
$e\equiv(R_{\rm a}-R_{\rm p})/(R_{\rm a}+R_{\rm p}).$
The nodal shift oscillates between the maximum and the minimum
values, $\delta\phi_{\rm max}$ and $\delta\phi_{\rm min}.$
Figures 1 and 2 show the graphs of $\delta\phi(r)$ for some typical values of
the parameters (here $r$ denotes the radius of the intersection with the
equatorial plane).
For practical purposes one needs the mean value of the shift which
characterizes an average taken over a large number of revolutions.
We define
\beq %vvvvvvvvvvvvvvvvvvvvvvvvvvvvvvvvvvvvvvvvvvvvvvvvvvvvvvvvvvvvvvvvvvvvvvvv
 \langle\delta\phi\rangle=\int_{R_{\rm p}}^{R_{\rm a}}
 \delta\phi(x)\,\calP(x)_{\mid\calE,\Phi,\calQ,a}\,dx,
 \label{fc}
\eeq %^^^^^^^^^^^^^^^^^^^^^^^^^^^^^^^^^^^^^^^^^^^^^^^^^^^^^^^^^^^^^^^^^^^^^^^^
where $\calP(r)$ is the probability distribution for the radial
coordinate of the intersections of the specified trajectory with the disk.
$\calP$ is normalized to unity, $\int\calP(x)\,dx=1.$ We applied
the following numerical procedure for evaluating equation
(\ref{fc}): (i) at first,
for a given set of orbital parameters $(\calE,\Phi,\calQ)$,
we constructed the probability distribution by following the course
of an ensemble of orbits with identical orbital parameters
and randomly chosen initial conditions;
(ii) then we applied equation (\ref{uze})
and performed the integration in equation (\ref{fc}). The resulting values of
the mean nodal shift are tabulated in the Appendix.
One can see that the dispersion is often
quite small and the mean value is a satisfactory approximation to the
current value of $\delta\phi,$ except for very close orbits.

Analogously to $\langle\delta\phi\rangle,$ we
introduce the mean orbital period $\langle P\rangle$ of an eccentric orbit
and we calculate $\langle\Omega_{\rm P}\rangle$ by applying
the mapping algorithm (\ref{map}) for the $r$-coordinate over a large number of
revolutions of the star around the SBH.

\bigskip\bigskip

\centerline{\large\bf 3.\quad NUMERICAL MODELS}
\bigskip

\centerline{3.1. \it Simple examples} %%%%%%%%%%%%%%%%%%%%%%%%%%%%%%%%%%%%%%%
\smallskip

In this Section we will show how the relativistic precession becomes evident
in a simulated signal from the source.
We take into account general relativistic effects
on photons coming from the source to the observer with no approximation.
The method for integrating light
trajectories in the Kerr metric has been discussed by many authors
(Cunningham~\& Bardeen 1973; de~Felice, Nobili~\& Calvani 1974;
Cunningham 1975 and 1976; Asaoka 1989).
We employ an efficient approach described by
Karas, Vokrouhlick\'y~\& Polnarev 1992.
We do not include the noise
component which is of course present in a
real signal. The reason is that the form of the noise
depends on its particular model and it does not affect the periodicities
we are interested in.
The way to account for the noise in the synthetic light curve
is straightforward, in principle, if one adopts
a specific model of its origin. Such a model can be based on
mechanisms proposed by Moskalik~\& Sikora 1986,
Chagelishvili, Lominadze~\& Rogava 1989, Abramowicz \etal 1991, or Baring 1992.

The position of a distant observer is characterized
by his inclination $\theta_{\rm o}$ with respect to the symmetry axis.
We assume that the star-disk collisions modulate the disk radiation
at the moment when the star crashes through the disk from the far side
to the near side with respect to the observer, i.e. once per revolution.
The light travel time to the observer depends on the location of the
intersection and
it is thus periodically affected by the precession. Frequencies corresponding
to the periodic modulation of the signal are evident in the Fourier
transform of arrival times.

The lensing effect
enhances the radiation when the source is behind the black hole and thus
contributes to the periodic modulation of the signal.
As an alternative to previously
described Fourier transform of arrival times, one can detect relevant
frequencies in the power spectrum of the photometric radiative flux.
However, in this alternative approach the input signal for the Fourier
transform
depends on the model of the star-disk interaction (see below).
We assume, for simplicity,
that the shape of the observed signal from successive collisions has always
an identical profile: a sharp onset and then an
exponential decrease of the local luminosity in the static frame or,
alternatively, in the disk co-rotating frame. In the latter case both
the lensing and the Doppler effect contribute to the signal strength.
Figure~3 is an example of typical light curves. We
assumed a spherical orbit, $r=6,$ around a nearly-extreme
Kerr black hole.
It is not clear whether such a close orbit is astrophysically realistic
because the process of capture is not well understood (Rees 1993). A  more
plausible configuration is treated in the next paragraph.

The basic frequency in the power spectrum of the observed signal corresponds
to the orbital motion.
One can detect two fundamental lines in the power spectrum. The first line
at higher frequency $[\Omega_{\rm o}=2\pi/P,$ see equation (\ref{p})]
corresponds to the orbital motion and the second one at
lower frequency $[\Omega_{\rm LT},$ see equation (\ref{omlt})]
corresponds to the gravitomagnetic precession.
Figure~4 shows the normalized power spectrum of the light curve from
the previous Figure, strictly speaking
the coefficient of spectral correlation as defined by Ferraz-Mello (1981).
The power spectrum contains also linear combinations of fundamental
frequencies.

The shape of the light curve
naturally depends on the particular model and position of the
observer, however, we expect that
identical periodicities caused by the precession of the orbit
will be present in the signal independently of the
details. The power spectrum constructed from arrival times has a simple
form. In Figure~5 we applied the Fourier transform on time intervals
elapsed between successive flares in the light curve. Now, the shape of
the spectrum is determined by orbital parameters of the star and angular
momentum of the SBH but it depends only very weakly on other details of
the model---decay time of the flare, viewing angle of the observer, etc.
This advantageous property is easy to understand because
the arrival times of successive flares are directly related to the
precession of the orbit and we thus avoid the model-dependent
features in the light curve.

As mentioned above, several effects tend to circularize the orbit of the
star and for this reason spherical orbits are of special interest. However,
the star can initially be captured in an eccentric orbit. Therefore,
we also considered the case of quasi-elliptic orbits $(e\neq 0).$
The pericenter shift reveals itself as another line in the
power spectrum. In Figure~6 one can detect lines corresponding to the
orbital motion $(\Omega_{\rm o}),$ the
gravitomagnetic precession $(\Omega_{\rm LT}),$
and the pericenter shift $(\Omega_{\rm P}).$ The gravitomagnetic precession
disappears in the Schwarzschild case and the power spectrum of an eccentric
orbit is then rather similar to the case of a spherical orbit near
a Kerr black hole. To distinguish between the both extreme cases one needs an
independent estimate of the radius and eccentricity of the orbit
(Figure 7).

\bigskip

\centerline{3.2. \it The case of NGC 6814}
%%%%%%%%%%%%%%%%%%%%%%%%%%%%%%%%%%%%%
\smallskip

At present there is no generally accepted explanation of the periodicity
observed in the X-ray flux from the Seyfert galaxy NGC~6814.
Among the most promising models are those which relate the periodicity to
the orbital motion of some object (unspecified for the time being)
around the SBH.
These models can in a natural way deal with the enormous stability of the
period over several years.
(For a discussion of different mechanisms which have been proposed in the
literature see, e.g., Abramowicz 1992.)
The model of a star colliding with an accretion disk has been considered as a
viable model (Syer \etal 1991; Sikora~\& Begelman 1992; Rees 1993).
The basic periodicity at approximately 12,200~s is interpreted as the orbital
period of the star. It is a straightforward conclusion that if the SBH
rotates then another periodicity associated with the gravitomagnetic
effect should be present in the signal. This conclusion is independent
of the unknown details of the star-disk interaction and it only assumes
a nonzero viewing angle.
Data which are currently available do not cover sufficiently long interval of
time to reveal additional periodicities besides the basic one.
For this reason we were unable to give a definitive conclusion about the
general relativistic precession
in NGC~6814. This could be changed if data from {\it ROSAT}
also show up the periodic component.
However, already now one can see that
the model of star-disk interactions faces serious
difficulties (see below).
Considering the time scale of the periodicity of NGC~6814
and the estimate of the black hole mass one can write
$$ \tilde{P} = 4.038\times 10^{-4} M_6 \langle P\rangle, $$
$$ \tilde{P}_{\rm LT} = \tilde{P} \frac{2\pi}{\langle\delta\phi\rangle}, $$
where $\tilde{P}$ and $\tilde{P}_{\rm LT}$ are measured in units of 12,200~s
and $M_6\equiv M/(10^6M_\odot).$ Assuming $e\approx 0,$ $\tilde{P}\approx 1$
and $M_6\approx 1,$ one can estimate the radius of the orbit to be
$$ R_{\rm s} = \left(394 \mp a\right)^{2/3} \approx 53.7\,.  $$
The periods associated with the gravitomagnetic precession and the
pericenter shift are thus substantially longer than the orbital one,
although one should remember that
the estimate of the mass of the SBH in NGC~6814 and consequently
the above value of $R_{\rm s}$ are rather uncertain (Bao 1992b);
$\langle\delta\phi\rangle$ can be estimated from the tables which are
given in Appendix.

We constructed the predicted light curve in the way described in Fig.~3
and compared main frequencies in the power spectrum with those in
observational data.
The eccentricity was assumed to be limited to a very small value
by the effects of orbital circularization; the low eccentricity is also
required by the phase stability of the light curve, as discussed below.
As a consequence, the dependence of
$\Omega_{\rm P}$ on $e$ can be ignored. Also the inclination
of the orbit is a free parameter but the dependence
of all relevant frequencies on $\mu_-$ is very weak. One can see that
the period related to the pericenter shift is
in the range $16\,\tilde{P}\lta\tilde{P}_{\rm P}\lta 20\,\tilde{P}$
(the lower limit corresponds to $a=0$, the upper
limit corresponds to $a\longrightarrow 1$). Time scales related to
both the orbital motion and the pericenter shift are
significantly shorter than the one related to the
gravitomagnetic precession, $\tilde{P}_{\rm LT}\gta 215\tilde{P}$ (indicated
value is the lower limit corresponding to $a\longrightarrow 1$).

We considered the scenario in which the X-ray flares are associated with bright
spots in the disk corona occurring in the place where the star crashes through
the disk. After their creation the spots subsequently rotate with the
disk matter, radiate with decreasing luminosity in their local frame,
and eventually disappear after several revolutions.
As the spots rotate, their radiation is periodically enhanced by the
lensing effect which causes secondary maxima in the light curve.
Secondary maxima can also be related to the event of the star crashing
through the disk in the direction away from the observer.
By comparing the {\it EXOSAT\/} (Mittaz~\& Branduardi-Raymont 1989;
Fiore \etal 1992) and {\it Ginga\/} (Done \etal 1992) folded
light curves, Abramowicz \etal
(1993) demonstrated that the phase of important peaks in the light curve
remains remarkably stable. This fact imposes strong restrictions on the model
and if it is confirmed one can conclude that
{\it the constant phase of individual patterns in the light curve
can only be understood if the black hole does not rotate and the orbit
of the star is nearly circular.} The change in the shape of the light
curve is then related to instabilities in the accretion process which
affect the structure of the disk. A significant change in the accretion
rate is evident from the change of the X-ray flux of NGC~6814
during the interval between the {\it EXOSAT\/} and {\it Ginga\/}
observations.

The peaks in the light curve are not regularly spaced.
Within the framework of the discussed model
it is very difficult to understand this
non-regular (but constant in time) distribution of phases of the peaks
in the light curve which has been reported at different ranges of energy.
We remind that one of the peaks in the light curve is,
according to the above model,
associated with the creation of a bright spot at the moment of
the star-disk collision. Subsequent peaks are gravitationally
lensed images of the orbiting spot.
These two mechanisms---star-disk interaction and lensing---are necessary
because we deal with non-regular spacing of the peaks.
The peaks originating by different mechanisms should very probably
be distinguishable by comparing light curves at different energies.
The distinction is not apparent in data,
nevertheless, the model can definitively be
ruled out only if the physics of X-ray flares in the disk
corona is well understood; this subject is beyond the scope
of the present paper. Further restrictions on the model come from the
limits on $\dot{P}$ derived from the studies of the effects of
gravitational radiation and star-disk collisions (King~\& Done 1993).
Alternative explanations of the flares
are based either on occultations of the central disk regions by
matter ejected in the star-disk collisions or interaction of ejected
matter with a jet. Within such models we need more data to restrict
the angular momentum of the SBH and the eccentricity of the orbit,
but these models face even more severe
problems with the relatively narrow width of the profile of the peaks in
the light curve and they require a precisely tuned inclination
of the stellar orbit.

\bigskip\bigskip

\centerline{\large\bf 4.\quad CONCLUSIONS} %%%%%%%%%%%%%%%%%%%%%%%%%%%%%%%%%%%%
\smallskip

We considered the general relativistic precession of a
star orbiting a supermassive black hole
and colliding with an accretion disk in the nucleus of an AGN.
We derived formulae for the azimuthal shift due to the gravitomagnetic
precession and perihelion shift during the free (geodesic) part of motion
between subsequent interactions with the accretion disk.
No restrictions on the orbital parameters and the black hole angular momentum
were imposed except that they describe a stable bound trajectory around
the Kerr black hole.
Within the framework of this model our results restrict possible values of
the angular momentum of the central black hole.
To illustrate the precession effect, we adopted a simplified model
of the star-disk interaction and we determined relevant frequencies in
the power spectrum of the observed signal.
Both types of the relativistic precession which are relevant for our
problem---the precession related to the
pericenter shift and the gravitomagnetic
(Lense-Thirring) precession---expose themselves clearly in the power
spectrum. We found the analysis of arrival times
conceptually more trivial and at the same time more advantageous
than the analysis of the complete photometric curve.

Our conjecture is that the
typical character of the power spectrum will remain conserved
at some level in
astrophysically more realistic models of the interaction.
Such an assumption is well-founded provided the
orbital parameters of the star are not changed significantly
during several periods associated with the precession motion
$(\Omega_{\rm o}\gg \Omega_{\rm LT},\;\Omega_{\rm P}).$
More realistic models must include the effects of
tidal interactions, gravitational radiation and star-disk collisions
on the orbital parameters (the work in preparation).

\bigskip\bigskip

We thank M.~Abramowicz, C.~Done and A.~Lanza
for very useful discussions concerning the problem
of NGC~6814 and P.~Haines for careful reading of the manuscript.
We also thank the unknown referee and
participants of the Relativity Seminar in Prague
for helpful comments and suggestions which helped us to improve our work.
\bigskip\bigskip

\noindent
\centerline{\large\bf APPENDIX} %%%%%%%%%%%%%%%%%%%%%%%%%%%%%%%%%%%%%%%%%%%%%%%
\medskip

We tabulate the azimuthal shift of orbital nodes due to the
gravitomagnetic precession. The three tables correspond to
$a=0.33$ (Table~1), $a=0.67$ (Table~2), and $a=1$ (Table~3).
In the tables, $R_{\rm p}$ denotes the
pericenter distance in units of the gravitational radius $R_+$,
$e$ is the eccentricity of the orbit, and $\mu_-$ is the parameter
characterizing inclination in the asymptotic region, $r\gg R_+$
[see equation (\ref{th})]. The columns denoted by ``$+$" and ``$-$"
correspond to direct $(\Phi>0)$ and retrograde
$(\Phi<0)$ orbits, respectively. (The difference naturally
disappears for polar orbits
which are characterized by $\mu_-=1$ and $\Phi=0.)$
The nodal shift is given in the form
$$\langle\delta\phi\rangle^{+(\delta\phi_{\rm max}-\langle\delta\phi\rangle)}
 _{-(\langle\delta\phi\rangle-\delta\phi_{\rm min})}
 \quad{\rm \left[10^{-4}rad\right]},$$
where $\langle\delta\phi\rangle$ is the mean value of the shift and
$\delta\phi_{\rm max},$ $\delta\phi_{\rm min}$ are the maximum and
the minimum values of the shift (for details see the text.)
In particular, for spherical orbits $(r=\const, e=0)$ we obtain
$\delta\phi_{\rm max}=\langle\delta\phi\rangle=\delta\phi_{\rm min};$
In this case, $\langle\delta\phi\rangle$ is given by equation (\ref{dphi}).
An ellipsis ``\ldots" in Table~3 excludes those combinations of parameters
which do not correspond to a time-like geodesic.

\bigskip\bigskip
\normalsize
\pagestyle{plain}
\parindent=-15pt
\centerline{\large\bf REFERENCES}%%%%%%%%%%%%%%%%%%%%%%%%%%%%%%%%%%%%%%%%
\medskip
\parskip=0pt

Abramowicz, M. A. 1987, in General Relativity and Gravitation,
 Proceedings of the 11th International Conference on General Relativity and
 Gravitation, ed. M.~A.~H. MacCallum (Cambridge: Cambridge University Press),
 p.~1

Abramowicz, M. A. 1992, in Testing the AGN Paradigm,
 Proceedings of the 2nd Maryland Astrophysical Conference,
 eds. S.~S. Holt, S.~G. Neff~\& C.~M. Urry (New York: Am. Inst. Phys.),
 p.~69

%Abramowicz M. A., Lanza A., Spiegel E. A. \& Szuszkiewicz E. 1992,
% {\it Nature}, {\bf 356}, 41

Abramowicz, M. A., Bao, G., Fiore, F., Lanza,~A., Massaro,~E., Perola, G.~C.,
 Spiegel E.~A., \& Szuszkiewicz~E. 1992, in Physics of Active Galactic
 Nuclei, eds.  W.~J. Duschl~\& S.~J. Wagner (Heidelberg: Springer)

Abramowicz, M. A., Bao, G., Karas, V.~\& Lanza,~A. 1993, A\&A, 272, 400

Abramowicz, M. A., Bao, G., Lanza, A., \& Zhang, X.-H. 1989, in
 Proceedings of the 23rd ESLAB Symposium on Two Topics in X-ray
 Astronomy, eds. J.~Hunt~\& B.~Battrick (ESA SP-296), 871

Abramowicz, M. A., Bao, G., Lanza, A., \& Zhang, X.-H. 1991, A\&A, 245, 454

Asaoka, I. 1989, PASJ, 41, 763

Bao, G. 1992a, A\&A, 257, 594

Bao, G. 1992b, Ph. D. Dissertation (Trieste: International School for
 Advanced Studies)

Bardeen, J. M. 1973, in Black Holes, eds. C. DeWitt~\& B.~S. DeWitt
 (New York: Gordon and Breach), 215

Bardeen, J. M.,~\& Petterson, J. A. 1975, ApJ, 195, L65

%Bardeen J. M., Press W. H. \& Teukolsky S. A. 1972, Astrophys. J.,
% {\bf 178}, 347

Baring, M. G. 1992, Nature, 360, 109

Begelman, M. C., Blandford, R. D., \& Rees, M. J. 1984, Rev. Mod. Phys.,
 56, 255

Binney, J., \& Tremaine, S. D. 1987, Galactic Dynamics (Princeton:
 Princeton University Press)

Bi\v{c}\'ak, J., \& Jani\v{s}, V. 1985, MNRAS, 212, 899

Blandford, R. D., \& Znajek, R. L. 1977, MNRAS, 179, 433

Blandford, R. D., Netzer, H., \& Woltjer, L. 1990, Active Galactic Nuclei,
 eds. T. J.-L. Curvoisier \& M. Mayor (Berlin: Springer-Verlag)

Boyle, C. B., \& Walker, I. W. 1986, MNRAS, 222, 559

Braginskij, V. B., Polnarev, A. G., \& Thorne, K. S. 1984, Phys. Rev. Lett.,
 53, 863

Byrd, P. F., \& Friedman, M. D. 1971, Handbook of Elliptic Integrals
 for Engineers and Scientists (Berlin: Springer-Verlag)

Carter, B. 1968, Phys. Rev., 174, 1559

%Carter B. 1992, ApJ, {\bf 391}, L67

Carter, B., \& Luminet, J.-P. 1983, A\&A, 121, 97

Chandrasekhar, S. 1983, The Mathematical Theory of Black Holes
 (Oxford: Clarendon Press)

Chagelishvili, G. D., Lominadze, J. G., \& Rogava, A. D. 1989, ApJ, 347, 1100

Ciufolini, I. 1986, Phys. Rev. Lett., 56, 278

Cunningham, C. T. 1975, ApJ, 202, 788

Cunningham, C. T. 1976, ApJ, 208, 534

Cunningham, C. T., \& Bardeen, J. M. 1973, ApJ, 183, 237

Damour, T., \& Taylor, J. H. 1992, Phys. Rev.~D 45, 1840

de Felice, F., Nobili L., \& Calvani M. 1974, A\&A 30, 111

Done, C., Koyama, K., Kunieda, H., Madejski,~G., Mushotzky,~R., \& Turner,
T.~J.
 1990, in Structure and Emission Properties of Accretion Disks,
 IAU Colloquium No. 129,
 eds. C.~Bertout \etal (Paris: Editions Fronti\`eres), 417

Done, C., Madejski, G. M., Mushotzky, R. F., \& Turner, T. J. 1992,
 ApJ, 400, 138

Dressler, A. 1989, in Active Galactic Nuclei, IAU Symposium No. 134,
 eds. D.~E. Osterbrock~\& J.~S. Miller (Dordrecht: Kluwer), 217

Dressler, A., \& Richstone, D. O. 1988, ApJ, 324, 701

Dressler, A., \& Richstone, D. O. 1990, ApJ, 348, 120

Ernst, F. J. 1976, J. Math. Phys., 17, 54

Everitt, C. W. F. 1974, in Experimental Gravitation, Proceedings
 of Course 56 of the International School of Physics ``Enrico Fermi,"
 ed. B.~Bertotti (New York: Academic Press), 331

Fabian, A. C., Pringle, J. E., \& Rees, M. J. 1975, MNRAS, 172, 15P

%Fabian A. C., Rees M. J., Stella L. \& White N. E. 1989,
% {\it Mon. Not. R. astr. Soc.}, 238, 729

Falcke H., Biermann P. L., Duschl W. J., \& Mezger P. G. 1993, A\&A, 270, 102

Ferraz-Mello, S. 1981, AJ, 86, 619

Fiore, F., Massaro, E., \& Barone, P. 1992, A\&A, 261, 405

Frank, J., \& Rees, M. J. 1976, MNRAS, 176, 633

Frank, J., King, A. R., \& Raine, D. J. 1985, Accretion Power in Astrophysics
 (Cambridge: Cambridge University Press.)

Gradshteyn, I. S., \& Ryzhik, I. W. 1980, Table of Integrals, Series, and
 Products (New York: Academic)

Guilbert, P. W., Fabian, A. C., \& Ross R. R. 1982, MNRAS, 199, 763

%Gr\" obner W. \& Hofreiter N. 1965, {\it Integraltafel, Erster Teil,
%Unbestimmte Integrale} (Vienna: Springer-Verlag).

Halpern, J. P., \& Filippenko, A. V. 1988, Nature, 331, 46

Hills, J. G. 1988, Nature, 331, 687

Honma, F., Matsumoto R., Kato S.,~\& Abramowicz, M.~A. 1991, PASJ, 43, 261

Kaburaki, O., \& Okamoto, I. 1991, Phys. Rev. D, 43, 340

Karas, V. 1991, MNRAS, 249, 122

Karas, V., Vokrouhlick\'y, D., \& Polnarev, A. 1992, MNRAS, 259, 569

%Kojima Y. 1991, MNRAS, 250, 629

King, A. R., \& Done, C. 1993, MNRAS, in press

Kormendy, J. 1988a, ApJ, 325, 128

Kormendy, J. 1988b, ApJ, 335, 40

Kumar, S., \& Pringle, J. E. 1983 MNRAS, 213, 435

%Laor A. 1991, {\it Mon. Not. R. astr. Soc.}, 376, 90

%Laor A. \& Netzer H. 1989, {\it Mon. Not. R. astr. Soc.}, 238, 897

Lecar, M., Wheeler, J. C., \& McKee, C. F. 1976, ApJ, 205,  556

Lee, H.-M., \& Ostriker, J. P. 1986, ApJ, 310, 176

Lense, J., \& Thirring, H. 1918, Physik Z., 19, 156

Luminet, J.-P., \& Marck, J.-A. 1985, MNRAS, 212, 1029

Macdonald, D. A., \& Thorne, K. S. 1982, MNRAS, 198, 345

Manko, V. S., \& Sibgatullin, N. R. 1992, Phys. Rev. D, 46, 4122

Mittaz, J. P. D., \& Branduardi-Raymont, G. 1989, MNRAS, 238, 1029

Moskalik, P., \& Sikora, M. 1986, Nature, 319, 649

Mushotzky, R. F. 1982, ApJ, 256, 92

Novikov, I. D., Pethick, C. J., \& Polnarev, A. G. 1992, MNRAS, 255, 276

%Novikov I. D. \& Thorne K. S. 1973, in Black Holes, eds. C.~DeWitt,
%B.~S. DeWitt (New York: Gordon and Breach), p.~343

Okamoto, I., \& Kaburaki, O. 1991, MNRAS, 250, 300

Ostriker, J. P. 1983, ApJ, 273, 99

%Page D. N. \& Thorne K. S. 1974, ApJ, 191, 499

Park, S. J., \& Vishniac, E. T. 1989, ApJ, 337, 78

Peters, P. C., \& Mathews, J. 1963, Phys. Rev., 131, 435

Phinney, E. S. 1983, Ph. D. Dissertation (Cambridge: Cambridge
 University)

Press, W. H., \& Teukolsky, S. A. 1977, ApJ, 213, 183

%Press W. H. \& Teukolsky S. A. 1990, Computers in Physics, Jan/Feb, 92

%Press W. H., Flannery B. P., Teukolsky S. A. \& Vetterling W. T. 1986,
% {\it Numerical Recipes: The Art of Scientific Computing}
% (New York: Cambridge University Press)

Press, W. H., Wiita, P. J., \& Smarr, L. L. 1975, ApJ,  202, L135

Rees, M. J. 1988, Nature, 333, 523

Rees, M. J. 1993, in The Renaissance of General Relativity and Cosmology:
 A survey meeting to celebrate the 65th birthday of Dennis Sciama,
 eds. G.~F.~R. Ellis, A.~Lanza~\& J.~C. Miller (Cambridge: Cambridge
 University Press)

Rees, M. J., Ruffini, R., \& Wheeler, J. A. 1974, Black Holes,
 Gravitational Waves and Cosmology (New York: Gordon \& Breach)

Sargent, W. L., Young, P. J., Boksenberg,~A., Shortridge,~K., Lynds, C.~R.,~\&
 Hartwick, F.~D.~A. 1978, ApJ, 221, 731

Shlosman, I., Begelman, M. C., \& Frank, J. 1990, Nature, 345, 679

Sikora, M., \& Begelman, M. C. 1992, Nature, 356, 224

Stoghianidis, E., \& Tsoubelis, D. 1987, Gen. Rel. Grav., 19, 1235

Stewart, J., \& Walker, M. 1973, in Springer Tracts in Modern Physics,
 Vol.~69, ed. G. H\"ohler (Berlin: Springer-Verlag)

Syer, D., Clarke, C. J., \& Rees, M. J. 1992, MNRAS, 250, 505

Tassoul, J.-L. 1988, ApJ, 324, L71

%Thorne K. S. 1974, ApJ, 191, 507.

Vio, R., Turolla, R., Cristiani, S., \& Barbieri, C. 1993, ApJ, 405, 163

Vokrouhlick\'y, D., \& Karas, V. 1993, MNRAS, in press (this preprint)

%Wallinder F. H., Kato S. \& Abramowicz M. A. 1992, {\it Astron. Astrophys},
% in press.

Wallinder, F. H. 1991, MNRAS, 253, 184

Wallinder, F. H. 1992, NORDITA preprint

Wilkins, D. C. 1972, Phys. Rev. D, 5, 814

Will, C. M. 1981, Theory and Experiment in Gravitational Physics,
 (New York: Cambridge University Press)

Young, P. J., Westphal, J. A., Kristian~J., \& Wilson, C.~P. 1978,
 ApJ, 221, 721

Zahn, J.-P. 1977, A\&A, 57, 383

Zahn, J.-P. 1989, A\&A, 220, 112

Zahn, J.-P., \& Bouchet, L. 1989, A\&A, 223, 112

Zeldovich, Ya. B., \& Novikov, I. D. 1971, Stars and Relativity (Chicago:
 Chicago University Press)

Zentsova, A. S. 1983, Astrophys. Sp. Sci, 95, 11

Zentsova, A. S., 1985, Astron. Zh., 62, 1227

Zurek, W. H., Siemiginowska, A., \& Colgate, S. A. 1992, in
 Testing the AGN Paradigm, Proceedings of the 2nd Maryland Astrophysical
 Conference, eds. S.~S. Holt, S.~G. Neff~\& C.~M. Urry
 (New York: Am. Inst. Phys.), p.~564

\bigskip\bigskip

\parindent=0pt
\parskip=15pt
\centerline{\large\bf FIGURE CAPTIONS}

FIG. 1.---The nodal shift $\delta\phi(r)$ (in radians)
per revolution for different eccentrities of the orbit as a function
of the radial distance of the intersection with the equatorial plane. Solid
 lines
correspond to direct orbits, dashed lines correspond to retrograde orbits.
Numbers given with each curve indicate the eccentricity.
The inclination parameter is
$\arccos\mu_-=45\dg,$ $a=0.9981.$ Two values of $\delta\phi$ for given $r$
correspond to $\dot{r}>0$ and $\dot{r}<0,$ respectively. The case of
spherical orbits is denoted by a star, ``$\,*\,$".

FIG. 2.---As in Fig.~1 but for different inclinations and constant
eccentricity. Now, $e=0.5$ and the numbers indicate the
value of $\arccos\mu_-.$ In particular, $\arccos\mu_-=90\dg$ corresponds
to equatorial orbits. Naturally, the curves of direct and retrograde
polar orbits $(\arccos\mu_-=0)$ coincide.

FIG. 3.---The numerical simulation of the observed light curve.
Radiative flux is given in arbitrary units on the ordinate.
The flux is periodically affected by the lensing
effect (main peaks) when the source of radiation is behind the black hole.
In this example we consider a source of radiation which is located
in the equatorial plane in the place
where the star intersected the disk.
The position of the intersection is affected by the gravitomagnetic
precession provided the central supermassive black hole rotates.
As a consequence, the places of intersection are dragged in the azimuthal
direction; the Figure covers 17 periods.
The luminosity of the source is isotropic in the locally static frame and
decreases exponentially with e-fold time $\tau.$ The position of
the observer at infinity is characterized by the viewing angle
$\theta_{\rm o};$ $\theta_{\rm o}=0$ is the rotation axis of the
black hole. The angular momentum parameter is chosen to be $a=0.9981\,.$
The star moves in an inclined spherical orbit: $r=6,$ $\mu_-=0.5\,.$
The three cases correspond to ({\it a\/}) $\theta=80\dg,$ $\tau=0.2P,$
({\it b\/}) $\theta=40\dg,$ $\tau=0.2P,$
({\it c\/}) $\theta=80\dg,$ $\tau=1.5P$ where $P_{\mid r=6}$ is the
orbital period given by equation (\ref{pap}). Further details of the model
are described in the text.

FIG. 4.---Normalized power spectrum of the light curve from Fig.~3.
The coefficient of spectral correlation is on ordinate and the spectral
window on abscissa.

FIG. 5.---Normalized power spectrum of the
arrival times for the signal from
Fig.~3. The axes are as in Fig.~4. One can verify that the typical shape of the
spectrum is quite insensitive to the details of configuration in a
broad range of parameters.

FIG. 6.---As in Fig.~5. The orbit of the star is now eccentric:
$e=0.1,$ $R_{\rm p}=6,$ $\mu_-=0.5\,.$ Parameters of the three cases
({\it a\/})--({\it c\/}) as before.

FIG. 7.---Calculated light curves for the simple model described in \S~III$\;a$
The case ({\it a\/}) is for the Kerr black hole with $a=0.9981$ and
a circular trajectory of the star at $R_{\rm s}=40$
(the value consistent with current estimates for NGC~6814).
The long-term modulation
of the light curve (three main peaks) is due to the gravitomagnetic
precession (angular frequency $\Omega_{\rm LT}).$
The case ({\it b\/}) is for a nonrotating black hole and an
eccentric trajectory $(e=0.25,$ $R_{\rm p}=40).$ Now the modulation
is due to the pericenter shift and it has a different period which
is given by $\langle\Omega_{\rm P}\rangle.$

\newpage
\pagestyle{empty}
\tabcolsep=0.6ex
\scriptsize
%\begin{center}
%%%%%%%%%%%%%%%%%%%%%%%%%  table 1 %%%%%%%%%%%%%%%%%%%%%%%%%%%%%%%%%%%%%%%%%%%%
\vspace*{-2.6cm}
\hspace*{-1cm}
\begin{tabular}{cccccccccccc}
\multicolumn{12}{c}{\normalsize TABLE 1} \bigskip \\
\hline\hline
\multicolumn{1}{c}{$R_{\rm p}$} & \multicolumn{1}{c}{$e\;\rightarrow$} &
\multicolumn{2}{c}{0.} &
\multicolumn{2}{c}{.2} &
\multicolumn{2}{c}{.4} &
\multicolumn{2}{c}{.6} &
\multicolumn{2}{c}{.8} \\ \cline{2-12}
\multicolumn{1}{c}{ $\downarrow$ } & \multicolumn{1}{c}{
$\mu_-\;\downarrow$ } &$+$&$-$&$+$&$-$&$+$&$-$&$+$&$-$&$+$&$-$\\
\hline\hline
 &  0. & \rule{0mm}{3.5ex}$1311\pm    0$ &
\rule{0mm}{3.5ex}$1441\pm    0$ & \rule{0mm}{3.5ex}$1005\pm
69$ & \rule{0mm}{3.5ex}$1110\pm  126$ & \rule{0mm}{3.5ex}$
810\pm   89$ & \rule{0mm}{3.5ex}$ 900\pm  156$ &
\rule{0mm}{3.5ex}$ 677\pm   92$ & \rule{0mm}{3.5ex}$ 754^{+
155}_{- 159}$ & \rule{0mm}{3.5ex}$ 580\pm   89$ &
\rule{0mm}{3.5ex}$ 648^{+ 145}_{- 151}$ \\ &  .25 & $1313\pm
0$ & $1439\pm    0$ & $1006\pm   70$ & $1109\pm  125$ & $ 812\pm
90$ & $ 898^{+ 154}_{- 156}$ & $ 678\pm   93$ & $ 753^{+ 154}_{-
157}$ & $ 581\pm   90$ & $ 647^{+ 144}_{- 150}$ \\ 5 &  .5 &
$1321\pm    0$ & $1434\pm    0$ & $1012\pm   72$ & $1104\pm
121$ & $ 816\pm   92$ & $ 894^{+ 150}_{- 152}$ & $ 682^{+
95}_{-  97}$ & $ 749^{+ 150}_{- 153}$ & $ 585^{+  92}_{-  94}$ &
$ 643^{+ 140}_{- 146}$ \\ &  .75 & $1336\pm    0$ & $1422\pm
0$ & $1024\pm   77$ & $1093\pm  114$ & $ 826\pm   98$ & $ 885^{+
142}_{- 144}$ & $ 690\pm  101$ & $ 741^{+ 142}_{- 146}$ & $
592^{+  97}_{-  99}$ & $ 636^{+ 134}_{- 139}$ \\ &  1.  &
\multicolumn{2}{c}{ \rule[-1.5ex]{0mm}{3ex}$1381\pm    0$} &
\multicolumn{2}{c}{ \rule[-1.5ex]{0mm}{3ex}$1059\pm   94$} &
\multicolumn{2}{c}{ \rule[-1.5ex]{0mm}{3ex}$ 856^{+ 118}_{-
120}$} & \multicolumn{2}{c}{ \rule[-1.5ex]{0mm}{3ex}$ 715^{+
120}_{- 123}$} & \multicolumn{2}{c}{ \rule[-1.5ex]{0mm}{3ex}$
614^{+ 114}_{- 118}$} \\ \hline &  0. & \rule{0mm}{3.5ex}$
466\pm    0$ & \rule{0mm}{3.5ex}$ 510\pm    0$ &
\rule{0mm}{3.5ex}$ 356\pm   10$ & \rule{0mm}{3.5ex}$ 389\pm
15$ & \rule{0mm}{3.5ex}$ 285\pm   13$ & \rule{0mm}{3.5ex}$
310\pm   20$ & \rule{0mm}{3.5ex}$ 236\pm   14$ &
\rule{0mm}{3.5ex}$ 256\pm   20$ & \rule{0mm}{3.5ex}$ 200\pm
14$ & \rule{0mm}{3.5ex}$ 217\pm   20$ \\ &  .25 & $ 467\pm    0$
& $ 509\pm    0$ & $ 357\pm   10$ & $ 388\pm   15$ & $ 286\pm
14$ & $ 310\pm   20$ & $ 236\pm   14$ & $ 256\pm   20$ & $
201\pm   14$ & $ 217\pm   19$ \\ 10 &  .5 & $ 469\pm    0$ & $
507\pm    0$ & $ 359\pm   10$ & $ 387\pm   15$ & $ 287\pm   14$
& $ 309\pm   19$ & $ 238\pm   15$ & $ 255\pm   20$ & $ 201\pm
14$ & $ 216\pm   19$ \\ &  .75 & $ 474\pm    0$ & $ 503\pm    0$
& $ 362\pm   11$ & $ 383\pm   14$ & $ 290\pm   14$ & $ 306\pm
19$ & $ 240\pm   15$ & $ 253\pm   19$ & $ 203\pm   15$ & $
214\pm   18$ \\ &  1.  & \multicolumn{2}{c}{
\rule[-1.5ex]{0mm}{3ex}$ 488\pm    0$} & \multicolumn{2}{c}{
\rule[-1.5ex]{0mm}{3ex}$ 373\pm   13$} & \multicolumn{2}{c}{
\rule[-1.5ex]{0mm}{3ex}$ 298\pm   16$} & \multicolumn{2}{c}{
\rule[-1.5ex]{0mm}{3ex}$ 246\pm   17$} & \multicolumn{2}{c}{
\rule[-1.5ex]{0mm}{3ex}$ 209\pm   17$} \\ \hline &  0. &
\rule{0mm}{3.5ex}$ 255\pm    0$ & \rule{0mm}{3.5ex}$ 276\pm
0$ & \rule{0mm}{3.5ex}$ 195\pm    3$ & \rule{0mm}{3.5ex}$ 210\pm
5$ & \rule{0mm}{3.5ex}$ 156\pm    5$ & \rule{0mm}{3.5ex}$ 167\pm
6$ & \rule{0mm}{3.5ex}$ 129\pm    5$ & \rule{0mm}{3.5ex}$ 138\pm
7$ & \rule{0mm}{3.5ex}$ 109\pm    6$ & \rule{0mm}{3.5ex}$ 116\pm
6$ \\ &  .25 & $ 256\pm    0$ & $ 276\pm    0$ & $ 195\pm    4$
& $ 210\pm    5$ & $ 156\pm    5$ & $ 167\pm    6$ & $ 129\pm
5$ & $ 137\pm    7$ & $ 109\pm    5$ & $ 116\pm    6$ \\ 15 &
.5 & $ 257\pm    0$ & $ 275\pm    0$ & $ 196\pm    4$ & $ 209\pm
5$ & $ 157\pm    5$ & $ 167\pm    6$ & $ 129\pm    5$ & $ 137\pm
7$ & $ 109\pm    5$ & $ 116\pm    6$ \\ &  .75 & $ 259\pm    0$
& $ 273\pm    0$ & $ 198\pm    4$ & $ 208\pm    5$ & $ 158\pm
5$ & $ 165\pm    6$ & $ 130\pm    5$ & $ 136\pm    6$ & $ 110\pm
5$ & $ 115\pm    6$ \\ &  1.  & \multicolumn{2}{c}{
\rule[-1.5ex]{0mm}{3ex}$ 266\pm    0$} & \multicolumn{2}{c}{
\rule[-1.5ex]{0mm}{3ex}$ 203\pm    4$} & \multicolumn{2}{c}{
\rule[-1.5ex]{0mm}{3ex}$ 162\pm    6$} & \multicolumn{2}{c}{
\rule[-1.5ex]{0mm}{3ex}$ 133\pm    6$} & \multicolumn{2}{c}{
\rule[-1.5ex]{0mm}{3ex}$ 112\pm    6$} \\ \hline &  0. &
\rule{0mm}{3.5ex}$ 166\pm    0$ & \rule{0mm}{3.5ex}$ 179\pm
0$ & \rule{0mm}{3.5ex}$ 127\pm    2$ & \rule{0mm}{3.5ex}$ 136\pm
2$ & \rule{0mm}{3.5ex}$ 101\pm    2$ & \rule{0mm}{3.5ex}$ 108\pm
3$ & \rule{0mm}{3.5ex}$  84\pm    2$ & \rule{0mm}{3.5ex}$  89\pm
3$ & \rule{0mm}{3.5ex}$  70\pm    2$ & \rule{0mm}{3.5ex}$  75\pm
3$ \\ &  .25 & $ 167\pm    0$ & $ 179\pm    0$ & $ 127\pm    2$
& $ 136\pm    2$ & $ 102\pm    2$ & $ 108\pm    3$ & $  84\pm
2$ & $  89\pm    3$ & $  71\pm    2$ & $  75\pm    3$ \\ 20 &
.5 & $ 167\pm    0$ & $ 178\pm    0$ & $ 128\pm    2$ & $ 135\pm
2$ & $ 102\pm    2$ & $ 108\pm    3$ & $  84\pm    2$ & $  88\pm
3$ & $  71\pm    2$ & $  74\pm    3$ \\ &  .75 & $ 169\pm    0$
& $ 177\pm    0$ & $ 129\pm    2$ & $ 135\pm    2$ & $ 103\pm
2$ & $ 107\pm    3$ & $  84\pm    3$ & $  88\pm    3$ & $  71\pm
2$ & $  74\pm    3$ \\ &  1.  & \multicolumn{2}{c}{
\rule[-1.5ex]{0mm}{3ex}$ 173\pm    0$} & \multicolumn{2}{c}{
\rule[-1.5ex]{0mm}{3ex}$ 132\pm    2$} & \multicolumn{2}{c}{
\rule[-1.5ex]{0mm}{3ex}$ 105\pm    3$} & \multicolumn{2}{c}{
\rule[-1.5ex]{0mm}{3ex}$  86\pm    3$} & \multicolumn{2}{c}{
\rule[-1.5ex]{0mm}{3ex}$  73\pm    3$} \\ \hline &  0. &
\rule{0mm}{3.5ex}$ 120\pm    0$ & \rule{0mm}{3.5ex}$ 128\pm
0$ & \rule{0mm}{3.5ex}$  91\pm    1$ & \rule{0mm}{3.5ex}$  97\pm
1$ & \rule{0mm}{3.5ex}$  73\pm    1$ & \rule{0mm}{3.5ex}$  77\pm
2$ & \rule{0mm}{3.5ex}$  60\pm    1$ & \rule{0mm}{3.5ex}$  63\pm
2$ & \rule{0mm}{3.5ex}$  50\pm    1$ & \rule{0mm}{3.5ex}$  53\pm
2$ \\ &  .25 & $ 120\pm    0$ & $ 128\pm    0$ & $  91\pm    1$
& $  97\pm    1$ & $  73\pm    1$ & $  77\pm    2$ & $  60\pm
1$ & $  63\pm    2$ & $  50\pm    1$ & $  53\pm    2$ \\ 25 &
.5 & $ 120\pm    0$ & $ 127\pm    0$ & $  92\pm    1$ & $  97\pm
1$ & $  73\pm    1$ & $  77\pm    2$ & $  60\pm    1$ & $  63\pm
2$ & $  51\pm    1$ & $  53\pm    2$ \\ &  .75 & $ 121\pm    0$
& $ 126\pm    0$ & $  92\pm    1$ & $  96\pm    1$ & $  73\pm
1$ & $  76\pm    2$ & $  60\pm    1$ & $  63\pm    2$ & $  51\pm
1$ & $  53\pm    2$ \\ &  1.  & \multicolumn{2}{c}{
\rule[-1.5ex]{0mm}{3ex}$ 124\pm    0$} & \multicolumn{2}{c}{
\rule[-1.5ex]{0mm}{3ex}$  94\pm    1$} & \multicolumn{2}{c}{
\rule[-1.5ex]{0mm}{3ex}$  75\pm    1$} & \multicolumn{2}{c}{
\rule[-1.5ex]{0mm}{3ex}$  62\pm    2$} & \multicolumn{2}{c}{
\rule[-1.5ex]{0mm}{3ex}$  52\pm    1$} \\ \hline &  0. &
\rule{0mm}{3.5ex}$  91\pm    0$ & \rule{0mm}{3.5ex}$  97\pm
0$ & \rule{0mm}{3.5ex}$  70\pm    1$ & \rule{0mm}{3.5ex}$  74\pm
1$ & \rule{0mm}{3.5ex}$  55\pm    1$ & \rule{0mm}{3.5ex}$  58\pm
1$ & \rule{0mm}{3.5ex}$  46\pm    1$ & \rule{0mm}{3.5ex}$  48\pm
1$ & \rule{0mm}{3.5ex}$  38\pm    1$ & \rule{0mm}{3.5ex}$  40\pm
1$ \\ &  .25 & $  91\pm    0$ & $  97\pm    0$ & $  70\pm    1$
& $  74\pm    1$ & $  55\pm    1$ & $  58\pm    1$ & $  46\pm
1$ & $  48\pm    1$ & $  38\pm    1$ & $  40\pm    1$ \\ 30 &
.5 & $  92\pm    0$ & $  96\pm    0$ & $  70\pm    1$ & $  73\pm
1$ & $  56\pm    1$ & $  58\pm    1$ & $  46\pm    1$ & $  48\pm
1$ & $  38\pm    1$ & $  40\pm    1$ \\ &  .75 & $  92\pm    0$
& $  96\pm    0$ & $  70\pm    1$ & $  73\pm    1$ & $  56\pm
1$ & $  58\pm    1$ & $  46\pm    1$ & $  48\pm    1$ & $  39\pm
1$ & $  40\pm    1$ \\ &  1.  & \multicolumn{2}{c}{
\rule[-1.5ex]{0mm}{3ex}$  94\pm    0$} & \multicolumn{2}{c}{
\rule[-1.5ex]{0mm}{3ex}$  72\pm    1$} & \multicolumn{2}{c}{
\rule[-1.5ex]{0mm}{3ex}$  57\pm    1$} & \multicolumn{2}{c}{
\rule[-1.5ex]{0mm}{3ex}$  47\pm    1$} & \multicolumn{2}{c}{
\rule[-1.5ex]{0mm}{3ex}$  39\pm    1$} \\ \hline &  0. &
\rule{0mm}{3.5ex}$  72\pm    0$ & \rule{0mm}{3.5ex}$  77\pm
0$ & \rule{0mm}{3.5ex}$  55\pm    0$ & \rule{0mm}{3.5ex}$  58\pm
1$ & \rule{0mm}{3.5ex}$  44\pm    1$ & \rule{0mm}{3.5ex}$  46\pm
1$ & \rule{0mm}{3.5ex}$  36\pm    1$ & \rule{0mm}{3.5ex}$  38\pm
1$ & \rule{0mm}{3.5ex}$  30\pm    1$ & \rule{0mm}{3.5ex}$  32\pm
1$ \\ &  .25 & $  73\pm    0$ & $  77\pm    0$ & $  55\pm    0$
& $  58\pm    1$ & $  44\pm    1$ & $  46\pm    1$ & $  36\pm
1$ & $  38\pm    1$ & $  30\pm    1$ & $  32\pm    1$ \\ 35 &
.5 & $  73\pm    0$ & $  76\pm    0$ & $  56\pm    0$ & $  58\pm
0$ & $  44\pm    1$ & $  46\pm    1$ & $  36\pm    1$ & $  38\pm
1$ & $  31\pm    1$ & $  32\pm    1$ \\ &  .75 & $  73\pm    0$
& $  76\pm    0$ & $  56\pm    0$ & $  58\pm    0$ & $  44\pm
1$ & $  46\pm    1$ & $  36\pm    1$ & $  38\pm    1$ & $  31\pm
1$ & $  32\pm    1$ \\ &  1.  & \multicolumn{2}{c}{
\rule[-1.5ex]{0mm}{3ex}$  75\pm    0$} & \multicolumn{2}{c}{
\rule[-1.5ex]{0mm}{3ex}$  57\pm    0$} & \multicolumn{2}{c}{
\rule[-1.5ex]{0mm}{3ex}$  45\pm    1$} & \multicolumn{2}{c}{
\rule[-1.5ex]{0mm}{3ex}$  37\pm    1$} & \multicolumn{2}{c}{
\rule[-1.5ex]{0mm}{3ex}$  31\pm    1$} \\ \hline &  0. &
\rule{0mm}{3.5ex}$  59\pm    0$ & \rule{0mm}{3.5ex}$  63\pm
0$ & \rule{0mm}{3.5ex}$  45\pm    0$ & \rule{0mm}{3.5ex}$  48\pm
0$ & \rule{0mm}{3.5ex}$  36\pm    0$ & \rule{0mm}{3.5ex}$  38\pm
0$ & \rule{0mm}{3.5ex}$  30\pm    0$ & \rule{0mm}{3.5ex}$  31\pm
0$ & \rule{0mm}{3.5ex}$  25\pm    0$ & \rule{0mm}{3.5ex}$  26\pm
0$ \\ &  .25 & $  59\pm    0$ & $  63\pm    0$ & $  45\pm    0$
& $  48\pm    0$ & $  36\pm    0$ & $  38\pm    0$ & $  30\pm
0$ & $  31\pm    0$ & $  25\pm    0$ & $  26\pm    0$ \\ 40 &
.5 & $  60\pm    0$ & $  62\pm    0$ & $  45\pm    0$ & $  47\pm
0$ & $  36\pm    0$ & $  38\pm    0$ & $  30\pm    0$ & $  31\pm
0$ & $  25\pm    0$ & $  26\pm    0$ \\ &  .75 & $  60\pm    0$
& $  62\pm    0$ & $  46\pm    0$ & $  47\pm    0$ & $  36\pm
0$ & $  38\pm    0$ & $  30\pm    0$ & $  31\pm    0$ & $  25\pm
0$ & $  26\pm    0$ \\ &  1.  & \multicolumn{2}{c}{
\rule[-1.5ex]{0mm}{3ex}$  61\pm    0$} & \multicolumn{2}{c}{
\rule[-1.5ex]{0mm}{3ex}$  46\pm    0$} & \multicolumn{2}{c}{
\rule[-1.5ex]{0mm}{3ex}$  37\pm    0$} & \multicolumn{2}{c}{
\rule[-1.5ex]{0mm}{3ex}$  30\pm    0$} & \multicolumn{2}{c}{
\rule[-1.5ex]{0mm}{3ex}$  25\pm    0$} \\ \hline &  0. &
\rule{0mm}{3.5ex}$  50\pm    0$ & \rule{0mm}{3.5ex}$  52\pm
0$ & \rule{0mm}{3.5ex}$  38\pm    0$ & \rule{0mm}{3.5ex}$  40\pm
0$ & \rule{0mm}{3.5ex}$  30\pm    0$ & \rule{0mm}{3.5ex}$  32\pm
0$ & \rule{0mm}{3.5ex}$  25\pm    0$ & \rule{0mm}{3.5ex}$  26\pm
0$ & \rule{0mm}{3.5ex}$  21\pm    0$ & \rule{0mm}{3.5ex}$  22\pm
0$ \\ &  .25 & $  50\pm    0$ & $  52\pm    0$ & $  38\pm    0$
& $  40\pm    0$ & $  30\pm    0$ & $  32\pm    0$ & $  25\pm
0$ & $  26\pm    0$ & $  21\pm    0$ & $  22\pm    0$ \\ 45 &
.5 & $  50\pm    0$ & $  52\pm    0$ & $  38\pm    0$ & $  40\pm
0$ & $  30\pm    0$ & $  32\pm    0$ & $  25\pm    0$ & $  26\pm
0$ & $  21\pm    0$ & $  22\pm    0$ \\ &  .75 & $  50\pm    0$
& $  52\pm    0$ & $  38\pm    0$ & $  40\pm    0$ & $  31\pm
0$ & $  31\pm    0$ & $  25\pm    0$ & $  26\pm    0$ & $  21\pm
0$ & $  22\pm    0$ \\ &  1.  & \multicolumn{2}{c}{
\rule[-1.5ex]{0mm}{3ex}$  51\pm    0$} & \multicolumn{2}{c}{
\rule[-1.5ex]{0mm}{3ex}$  39\pm    0$} & \multicolumn{2}{c}{
\rule[-1.5ex]{0mm}{3ex}$  31\pm    0$} & \multicolumn{2}{c}{
\rule[-1.5ex]{0mm}{3ex}$  25\pm    0$} & \multicolumn{2}{c}{
\rule[-1.5ex]{0mm}{3ex}$  21\pm    0$} \\ \hline &  0. &
\rule{0mm}{3.5ex}$  43\pm    0$ & \rule{0mm}{3.5ex}$  45\pm
0$ & \rule{0mm}{3.5ex}$  33\pm    0$ & \rule{0mm}{3.5ex}$  34\pm
0$ & \rule{0mm}{3.5ex}$  26\pm    0$ & \rule{0mm}{3.5ex}$  27\pm
0$ & \rule{0mm}{3.5ex}$  21\pm    0$ & \rule{0mm}{3.5ex}$  22\pm
0$ & \rule{0mm}{3.5ex}$  18\pm    0$ & \rule{0mm}{3.5ex}$  19\pm
0$ \\ &  .25 & $  43\pm    0$ & $  45\pm    0$ & $  33\pm    0$
& $  34\pm    0$ & $  26\pm    0$ & $  27\pm    0$ & $  21\pm
0$ & $  22\pm    0$ & $  18\pm    0$ & $  19\pm    0$ \\ 50 &
.5 & $  43\pm    0$ & $  45\pm    0$ & $  33\pm    0$ & $  34\pm
0$ & $  26\pm    0$ & $  27\pm    0$ & $  21\pm    0$ & $  22\pm
0$ & $  18\pm    0$ & $  18\pm    0$ \\ &  .75 & $  43\pm    0$
& $  44\pm    0$ & $  33\pm    0$ & $  34\pm    0$ & $  26\pm
0$ & $  27\pm    0$ & $  21\pm    0$ & $  22\pm    0$ & $  18\pm
0$ & $  18\pm    0$ \\ &  1.  & \multicolumn{2}{c}{
\rule[-1.5ex]{0mm}{3ex}$  44\pm    0$} & \multicolumn{2}{c}{
\rule[-1.5ex]{0mm}{3ex}$  33\pm    0$} & \multicolumn{2}{c}{
\rule[-1.5ex]{0mm}{3ex}$  26\pm    0$} & \multicolumn{2}{c}{
\rule[-1.5ex]{0mm}{3ex}$  22\pm    0$} & \multicolumn{2}{c}{
\rule[-1.5ex]{0mm}{3ex}$  18\pm    0$} \\ \hline
\end{tabular}
\newpage

%%%%%%%%%%%%%%%%%%%%%%%%%  table 2 %%%%%%%%%%%%%%%%%%%%%%%%%%%%%%%%%%%%%%%%%%%%
\vspace*{-2.6cm}
\hspace*{-1.5cm}
\begin{tabular}{cccccccccccc}
\multicolumn{12}{c}{\normalsize TABLE 2} \bigskip \\
\hline\hline
\multicolumn{1}{c}{$R_{\rm p}$} & \multicolumn{1}{c}{$e\;\rightarrow$} &
\multicolumn{2}{c}{0.} &
\multicolumn{2}{c}{.2} &
\multicolumn{2}{c}{.4} &
\multicolumn{2}{c}{.6} &
\multicolumn{2}{c}{.8} \\ \cline{2-12}
\multicolumn{1}{c}{ $\downarrow$ } & \multicolumn{1}{c}{
$\mu_-\;\downarrow$ } &$+$&$-$&$+$&$-$&$+$&$-$&$+$&$-$&$+$&$-$\\
\hline\hline
 &  0. & \rule{0mm}{3.5ex}$2883\pm    0$ &
\rule{0mm}{3.5ex}$3480\pm    0$ & \rule{0mm}{3.5ex}$2212\pm
128$ & \rule{0mm}{3.5ex}$2721\pm  462$ &
\rule{0mm}{3.5ex}$1787\pm  171$ & \rule{0mm}{3.5ex}$2229^{+
565}_{- 563}$ & \rule{0mm}{3.5ex}$1496\pm  181$ &
\rule{0mm}{3.5ex}$1883^{+ 555}_{- 558}$ &
\rule{0mm}{3.5ex}$1285^{+ 179}_{- 176}$ &
\rule{0mm}{3.5ex}$1629^{+ 510}_{- 524}$ \\ &  .25 & $2895\pm
0$ & $3474\pm    0$ & $2221\pm  131$ & $2713\pm  454$ & $1794\pm
174$ & $2221^{+ 556}_{- 553}$ & $1502\pm  185$ & $1876^{+
546}_{- 550}$ & $1291^{+ 182}_{- 180}$ & $1623^{+ 502}_{- 516}$
\\ 5 &  .5 & $2935\pm    0$ & $3453\pm    0$ & $2251\pm  141$ &
$2689\pm  428$ & $1818\pm  186$ & $2199\pm  525$ & $1522\pm
196$ & $1855^{+ 517}_{- 522}$ & $1308^{+ 193}_{- 191}$ &
$1604^{+ 476}_{- 490}$ \\ &  .75 & $3013\pm    0$ & $3407\pm
0$ & $2310\pm  163$ & $2642\pm  381$ & $1865\pm  212$ & $2154\pm
468$ & $1561\pm  222$ & $1814^{+ 464}_{- 470}$ & $1341^{+
215}_{- 217}$ & $1566^{+ 430}_{- 444}$ \\ &  1.  &
\multicolumn{2}{c}{ \rule[-1.5ex]{0mm}{3ex}$3232\pm    0$} &
\multicolumn{2}{c}{ \rule[-1.5ex]{0mm}{3ex}$2485^{+ 261}_{-
254}$} & \multicolumn{2}{c}{ \rule[-1.5ex]{0mm}{3ex}$2012^{+
319}_{- 322}$} & \multicolumn{2}{c} { \rule[-1.5ex]
{0mm}{3ex}$1687^{+ 323}_{- 329}$} & \multicolumn{2}{c}{
\rule[-1.5ex]{0mm}{3ex}$1452^{+ 306}_{- 316}$} \\ \hline &  0. &
\rule{0mm}{3.5ex}$1035\pm    0$ & \rule{0mm}{3.5ex}$1246\pm
0$ & \rule{0mm}{3.5ex}$ 794\pm   21$ & \rule{0mm}{3.5ex}$ 950\pm
51$ & \rule{0mm}{3.5ex}$ 638\pm   29$ & \rule{0mm}{3.5ex}$
760\pm   65$ & \rule{0mm}{3.5ex}$ 529\pm   31$ &
\rule{0mm}{3.5ex}$ 628^{+  65}_{-  68}$ & \rule{0mm}{3.5ex}$
450\pm   31$ & \rule{0mm}{3.5ex}$ 532\pm   66$ \\ &  .25 &
$1039\pm    0$ & $1243\pm    0$ & $ 797\pm   21$ & $ 948\pm
51$ & $ 640\pm   29$ & $ 758^{+  64}_{-  66}$ & $ 531\pm   31$ &
$ 626^{+  65}_{-  67}$ & $ 452\pm   31$ & $ 531^{+  61}_{-  65}$
\\ 10 &  .5 & $1050\pm    0$ & $1233\pm    0$ & $ 806\pm   22$ &
$ 941\pm   49$ & $ 647\pm   30$ & $ 752\pm   62$ & $ 536\pm
33$ & $ 621^{+  62}_{-  65}$ & $ 456\pm   32$ & $ 527^{+  59}_{-
62}$ \\ &  .75 & $1074\pm    0$ & $1213\pm    0$ & $ 822\pm
25$ & $ 925\pm   45$ & $ 659\pm   33$ & $ 740\pm   57$ & $
546\pm   35$ & $ 612^{+  58}_{-  60}$ & $ 464\pm   35$ & $
519^{+  55}_{-  58}$ \\ &  1.  & \multicolumn{2}{c}{
\rule[-1.5ex]{0mm}{3ex}$1146\pm    0$} & \multicolumn{2}{c}{
\rule[-1.5ex]{0mm}{3ex}$ 875\pm   34$} & \multicolumn{2}{c}{
\rule[-1.5ex]{0mm}{3ex}$ 700\pm   44$} & \multicolumn{2}{c}{
\rule[-1.5ex]{0mm}{3ex}$ 579\pm   46$} & \multicolumn{2}{c}{
\rule[-1.5ex]{0mm}{3ex}$ 491\pm   44$} \\ \hline &  0. &
\rule{0mm}{3.5ex}$ 571\pm    0$ & \rule{0mm}{3.5ex}$ 674\pm
0$ & \rule{0mm}{3.5ex}$ 438\pm    8$ & \rule{0mm}{3.5ex}$ 512\pm
16$ & \rule{0mm}{3.5ex}$ 351\pm   10$ & \rule{0mm}{3.5ex}$
408\pm   20$ & \rule{0mm}{3.5ex}$ 290\pm   11$ &
\rule{0mm}{3.5ex}$ 335\pm   21$ & \rule{0mm}{3.5ex}$ 246\pm
11$ & \rule{0mm}{3.5ex}$ 283\pm   20$ \\ &  .25 & $ 573\pm    0$
& $ 673\pm    0$ & $ 439\pm    8$ & $ 511\pm   15$ & $ 352\pm
10$ & $ 407\pm   20$ & $ 291\pm   11$ & $ 334\pm   20$ & $
246\pm   11$ & $ 282\pm   19$ \\ 15 &  .5 & $ 578\pm    0$ & $
668\pm    0$ & $ 443\pm    8$ & $ 508\pm   15$ & $ 355\pm   11$
& $ 404\pm   19$ & $ 293\pm   12$ & $ 332\pm   20$ & $ 248\pm
12$ & $ 280\pm   19$ \\ &  .75 & $ 589\pm    0$ & $ 658\pm    0$
& $ 451\pm    9$ & $ 500\pm   14$ & $ 361\pm   12$ & $ 398\pm
18$ & $ 298\pm   12$ & $ 328\pm   19$ & $ 252\pm   12$ & $
276\pm   18$ \\ &  1.  & \multicolumn{2}{c}{
\rule[-1.5ex]{0mm}{3ex}$ 624\pm    0$} & \multicolumn{2}{c}{
\rule[-1.5ex]{0mm}{3ex}$ 476\pm   11$} & \multicolumn{2}{c}{
\rule[-1.5ex]{0mm}{3ex}$ 380\pm   15$} & \multicolumn{2}{c}{
\rule[-1.5ex]{0mm}{3ex}$ 313\pm   15$} & \multicolumn{2}{c}{
\rule[-1.5ex]{0mm}{3ex}$ 264\pm   15$} \\ \hline &  0. &
\rule{0mm}{3.5ex}$ 375\pm    0$ & \rule{0mm}{3.5ex}$ 435\pm
0$ & \rule{0mm}{3.5ex}$ 287\pm    4$ & \rule{0mm}{3.5ex}$ 330\pm
7$ & \rule{0mm}{3.5ex}$ 230\pm    5$ & \rule{0mm}{3.5ex}$ 262\pm
9$ & \rule{0mm}{3.5ex}$ 190\pm    6$ & \rule{0mm}{3.5ex}$ 215\pm
9$ & \rule{0mm}{3.5ex}$ 160\pm    6$ & \rule{0mm}{3.5ex}$ 181\pm
9$ \\ &  .25 & $ 376\pm    0$ & $ 434\pm    0$ & $ 288\pm    4$
& $ 330\pm    7$ & $ 230\pm    5$ & $ 262\pm    9$ & $ 190\pm
6$ & $ 215\pm    9$ & $ 161\pm    6$ & $ 181\pm    9$ \\ 20 &
.5 & $ 379\pm    0$ & $ 431\pm    0$ & $ 290\pm    4$ & $ 328\pm
7$ & $ 232\pm    5$ & $ 260\pm    9$ & $ 191\pm    6$ & $ 214\pm
9$ & $ 162\pm    6$ & $ 180\pm    9$ \\ &  .75 & $ 385\pm    0$
& $ 425\pm    0$ & $ 295\pm    4$ & $ 323\pm    6$ & $ 235\pm
6$ & $ 257\pm    8$ & $ 194\pm    6$ & $ 211\pm    9$ & $ 164\pm
6$ & $ 178\pm    8$ \\ &  1.  & \multicolumn{2}{c}{
\rule[-1.5ex]{0mm}{3ex}$ 405\pm    0$} & \multicolumn{2}{c}{
\rule[-1.5ex]{0mm}{3ex}$ 309\pm    5$} & \multicolumn{2}{c}{
\rule[-1.5ex]{0mm}{3ex}$ 246\pm    7$} & \multicolumn{2}{c}{
\rule[-1.5ex]{0mm}{3ex}$ 202\pm    7$} & \multicolumn{2}{c}{
\rule[-1.5ex]{0mm}{3ex}$ 171\pm    7$} \\ \hline &  0. &
\rule{0mm}{3.5ex}$ 270\pm    0$ & \rule{0mm}{3.5ex}$ 310\pm
0$ & \rule{0mm}{3.5ex}$ 207\pm    2$ & \rule{0mm}{3.5ex}$ 235\pm
4$ & \rule{0mm}{3.5ex}$ 165\pm    3$ & \rule{0mm}{3.5ex}$ 186\pm
5$ & \rule{0mm}{3.5ex}$ 136\pm    3$ & \rule{0mm}{3.5ex}$ 153\pm
5$ & \rule{0mm}{3.5ex}$ 115\pm    3$ & \rule{0mm}{3.5ex}$ 128\pm
5$ \\ &  .25 & $ 271\pm    0$ & $ 309\pm    0$ & $ 207\pm    2$
& $ 235\pm    4$ & $ 166\pm    3$ & $ 186\pm    5$ & $ 137\pm
3$ & $ 153\pm    5$ & $ 115\pm    3$ & $ 128\pm    5$ \\ 25 &
.5 & $ 273\pm    0$ & $ 307\pm    0$ & $ 209\pm    2$ & $ 233\pm
4$ & $ 167\pm    3$ & $ 185\pm    5$ & $ 137\pm    3$ & $ 152\pm
5$ & $ 116\pm    3$ & $ 127\pm    5$ \\ &  .75 & $ 277\pm    0$
& $ 303\pm    0$ & $ 212\pm    2$ & $ 230\pm    3$ & $ 169\pm
3$ & $ 183\pm    4$ & $ 139\pm    3$ & $ 150\pm    5$ & $ 117\pm
3$ & $ 126\pm    5$ \\ &  1.  & \multicolumn{2}{c}{
\rule[-1.5ex]{0mm}{3ex}$ 290\pm    0$} & \multicolumn{2}{c}{
\rule[-1.5ex]{0mm}{3ex}$ 221\pm    3$} & \multicolumn{2}{c}{
\rule[-1.5ex]{0mm}{3ex}$ 176\pm    4$} & \multicolumn{2}{c}{
\rule[-1.5ex]{0mm}{3ex}$ 145\pm    4$} & \multicolumn{2}{c}{
\rule[-1.5ex]{0mm}{3ex}$ 122\pm    4$} \\ \hline &  0. &
\rule{0mm}{3.5ex}$ 207\pm    0$ & \rule{0mm}{3.5ex}$ 235\pm
0$ & \rule{0mm}{3.5ex}$ 158\pm    1$ & \rule{0mm}{3.5ex}$ 178\pm
2$ & \rule{0mm}{3.5ex}$ 126\pm    2$ & \rule{0mm}{3.5ex}$ 141\pm
3$ & \rule{0mm}{3.5ex}$ 104\pm    2$ & \rule{0mm}{3.5ex}$ 116\pm
3$ & \rule{0mm}{3.5ex}$  88\pm    2$ & \rule{0mm}{3.5ex}$  97\pm
3$ \\ &  .25 & $ 207\pm    0$ & $ 234\pm    0$ & $ 158\pm    1$
& $ 178\pm    2$ & $ 127\pm    2$ & $ 141\pm    3$ & $ 104\pm
2$ & $ 115\pm    3$ & $  88\pm    2$ & $  97\pm    3$ \\ 30 &
.5 & $ 208\pm    0$ & $ 233\pm    0$ & $ 159\pm    1$ & $ 177\pm
2$ & $ 127\pm    2$ & $ 140\pm    3$ & $ 105\pm    2$ & $ 115\pm
3$ & $  88\pm    2$ & $  96\pm    3$ \\ &  .75 & $ 211\pm    0$
& $ 230\pm    0$ & $ 162\pm    1$ & $ 175\pm    2$ & $ 129\pm
2$ & $ 139\pm    3$ & $ 106\pm    2$ & $ 114\pm    3$ & $  89\pm
2$ & $  95\pm    3$ \\ &  1.  & \multicolumn{2}{c}{
\rule[-1.5ex]{0mm}{3ex}$ 221\pm    0$} & \multicolumn{2}{c}{
\rule[-1.5ex]{0mm}{3ex}$ 168\pm    2$} & \multicolumn{2}{c}{
\rule[-1.5ex]{0mm}{3ex}$ 134\pm    3$} & \multicolumn{2}{c}{
\rule[-1.5ex]{0mm}{3ex}$ 110\pm    3$} & \multicolumn{2}{c}{
\rule[-1.5ex]{0mm}{3ex}$  92\pm    2$} \\ \hline &  0. &
\rule{0mm}{3.5ex}$ 165\pm    0$ & \rule{0mm}{3.5ex}$ 186\pm
0$ & \rule{0mm}{3.5ex}$ 126\pm    1$ & \rule{0mm}{3.5ex}$ 141\pm
1$ & \rule{0mm}{3.5ex}$ 101\pm    1$ & \rule{0mm}{3.5ex}$ 112\pm
2$ & \rule{0mm}{3.5ex}$  83\pm    1$ & \rule{0mm}{3.5ex}$  91\pm
2$ & \rule{0mm}{3.5ex}$  70\pm    1$ & \rule{0mm}{3.5ex}$  77\pm
2$ \\ &  .25 & $ 165\pm    0$ & $ 185\pm    0$ & $ 126\pm    1$
& $ 140\pm    1$ & $ 101\pm    1$ & $ 111\pm    2$ & $  83\pm
1$ & $  91\pm    2$ & $  70\pm    1$ & $  77\pm    2$ \\ 35 &
.5 & $ 166\pm    0$ & $ 184\pm    0$ & $ 127\pm    1$ & $ 140\pm
1$ & $ 101\pm    1$ & $ 111\pm    2$ & $  83\pm    1$ & $  91\pm
2$ & $  70\pm    1$ & $  76\pm    2$ \\ &  .75 & $ 168\pm    0$
& $ 182\pm    0$ & $ 128\pm    1$ & $ 138\pm    1$ & $ 102\pm
1$ & $ 110\pm    2$ & $  84\pm    1$ & $  90\pm    2$ & $  71\pm
1$ & $  75\pm    2$ \\ &  1.  & \multicolumn{2}{c}{
\rule[-1.5ex]{0mm}{3ex}$ 175\pm    0$} & \multicolumn{2}{c}{
\rule[-1.5ex]{0mm}{3ex}$ 133\pm    1$} & \multicolumn{2}{c}{
\rule[-1.5ex]{0mm}{3ex}$ 106\pm    2$} & \multicolumn{2}{c}{
\rule[-1.5ex]{0mm}{3ex}$  87\pm    2$} & \multicolumn{2}{c}{
\rule[-1.5ex]{0mm}{3ex}$  73\pm    2$} \\ \hline &  0. &
\rule{0mm}{3.5ex}$ 135\pm    0$ & \rule{0mm}{3.5ex}$ 151\pm
0$ & \rule{0mm}{3.5ex}$ 103\pm    1$ & \rule{0mm}{3.5ex}$ 115\pm
1$ & \rule{0mm}{3.5ex}$  83\pm    1$ & \rule{0mm}{3.5ex}$  91\pm
1$ & \rule{0mm}{3.5ex}$  68\pm    1$ & \rule{0mm}{3.5ex}$  74\pm
1$ & \rule{0mm}{3.5ex}$  57\pm    1$ & \rule{0mm}{3.5ex}$  62\pm
1$ \\ &  .25 & $ 136\pm    0$ & $ 151\pm    0$ & $ 104\pm    1$
& $ 115\pm    1$ & $  83\pm    1$ & $  91\pm    1$ & $  68\pm
1$ & $  74\pm    1$ & $  57\pm    1$ & $  62\pm    1$ \\ 40 &
.5 & $ 136\pm    0$ & $ 150\pm    0$ & $ 104\pm    1$ & $ 114\pm
1$ & $  83\pm    1$ & $  90\pm    1$ & $  68\pm    1$ & $  74\pm
1$ & $  58\pm    1$ & $  62\pm    1$ \\ &  .75 & $ 138\pm    0$
& $ 149\pm    0$ & $ 105\pm    1$ & $ 113\pm    1$ & $  84\pm
1$ & $  90\pm    1$ & $  69\pm    1$ & $  73\pm    1$ & $  58\pm
1$ & $  62\pm    1$ \\ &  1.  & \multicolumn{2}{c}{
\rule[-1.5ex]{0mm}{3ex}$ 143\pm    0$} & \multicolumn{2}{c}{
\rule[-1.5ex]{0mm}{3ex}$ 109\pm    1$} & \multicolumn{2}{c}{
\rule[-1.5ex]{0mm}{3ex}$  87\pm    1$} & \multicolumn{2}{c}{
\rule[-1.5ex]{0mm}{3ex}$  71\pm    1$} & \multicolumn{2}{c}{
\rule[-1.5ex]{0mm}{3ex}$  60\pm    1$} \\ \hline &  0. &
\rule{0mm}{3.5ex}$ 114\pm    0$ & \rule{0mm}{3.5ex}$ 127\pm
0$ & \rule{0mm}{3.5ex}$  87\pm    1$ & \rule{0mm}{3.5ex}$  96\pm
1$ & \rule{0mm}{3.5ex}$  69\pm    1$ & \rule{0mm}{3.5ex}$  76\pm
1$ & \rule{0mm}{3.5ex}$  57\pm    1$ & \rule{0mm}{3.5ex}$  62\pm
1$ & \rule{0mm}{3.5ex}$  48\pm    1$ & \rule{0mm}{3.5ex}$  52\pm
1$ \\ &  .25 & $ 114\pm    0$ & $ 126\pm    0$ & $  87\pm    1$
& $  96\pm    1$ & $  69\pm    1$ & $  76\pm    1$ & $  57\pm
1$ & $  62\pm    1$ & $  48\pm    1$ & $  52\pm    1$ \\ 45 &
.5 & $ 115\pm    0$ & $ 126\pm    0$ & $  88\pm    1$ & $  95\pm
1$ & $  70\pm    1$ & $  76\pm    1$ & $  57\pm    1$ & $  62\pm
1$ & $  48\pm    1$ & $  52\pm    1$ \\ &  .75 & $ 116\pm    0$
& $ 124\pm    0$ & $  88\pm    1$ & $  94\pm    1$ & $  70\pm
1$ & $  75\pm    1$ & $  58\pm    1$ & $  61\pm    1$ & $  49\pm
1$ & $  51\pm    1$ \\ &  1.  & \multicolumn{2}{c}{
\rule[-1.5ex]{0mm}{3ex}$ 120\pm    0$} & \multicolumn{2}{c}{
\rule[-1.5ex]{0mm}{3ex}$  91\pm    1$} & \multicolumn{2}{c}{
\rule[-1.5ex]{0mm}{3ex}$  73\pm    1$} & \multicolumn{2}{c}{
\rule[-1.5ex]{0mm}{3ex}$  60\pm    1$} & \multicolumn{2}{c}{
\rule[-1.5ex]{0mm}{3ex}$  50\pm    1$} \\ \hline &  0. &
\rule{0mm}{3.5ex}$  97\pm    0$ & \rule{0mm}{3.5ex}$ 108\pm
0$ & \rule{0mm}{3.5ex}$  74\pm    0$ & \rule{0mm}{3.5ex}$  82\pm
1$ & \rule{0mm}{3.5ex}$  59\pm    1$ & \rule{0mm}{3.5ex}$  65\pm
1$ & \rule{0mm}{3.5ex}$  49\pm    1$ & \rule{0mm}{3.5ex}$  53\pm
1$ & \rule{0mm}{3.5ex}$  41\pm    1$ & \rule{0mm}{3.5ex}$  44\pm
1$ \\ &  .25 & $  98\pm    0$ & $ 108\pm    0$ & $  75\pm    0$
& $  82\pm    1$ & $  59\pm    1$ & $  65\pm    1$ & $  49\pm
1$ & $  53\pm    1$ & $  41\pm    1$ & $  44\pm    1$ \\ 50 &
.5 & $  98\pm    0$ & $ 107\pm    0$ & $  75\pm    0$ & $  81\pm
1$ & $  60\pm    1$ & $  64\pm    1$ & $  49\pm    1$ & $  53\pm
1$ & $  41\pm    1$ & $  44\pm    1$ \\ &  .75 & $  99\pm    0$
& $ 106\pm    0$ & $  76\pm    0$ & $  81\pm    1$ & $  60\pm
1$ & $  64\pm    1$ & $  50\pm    1$ & $  52\pm    1$ & $  42\pm
1$ & $  44\pm    1$ \\ &  1.  & \multicolumn{2}{c}{
\rule[-1.5ex]{0mm}{3ex}$ 103\pm    0$} & \multicolumn{2}{c}{
\rule[-1.5ex]{0mm}{3ex}$  78\pm    0$} & \multicolumn{2}{c}{
\rule[-1.5ex]{0mm}{3ex}$  62\pm    1$} & \multicolumn{2}{c}{
\rule[-1.5ex]{0mm}{3ex}$  51\pm    1$} & \multicolumn{2}{c}{
\rule[-1.5ex]{0mm}{3ex}$  43\pm    1$} \\ \hline
\end{tabular}
\newpage

%%%%%%%%%%%%%%%%%%%%%%%%%  table 3
%%%%%%%%%%%%%%%%%%%%%%%%%%%%%%%%%%%%%%%%%%%%%%%
\vspace*{-2.8cm}
\hspace*{-1.5cm}
\begin{tabular}{cccccccccccc}
\multicolumn{12}{c}{\normalsize TABLE 3} \bigskip \\
\hline\hline
\multicolumn{1}{c}{$R_{\rm p}$} & \multicolumn{1}{c}{$e\;\rightarrow$} &
\multicolumn{2}{c}{0.} &
\multicolumn{2}{c}{.2} &
\multicolumn{2}{c}{.4} &
\multicolumn{2}{c}{.6} &
\multicolumn{2}{c}{.8} \\ \cline{2-12}
\multicolumn{1}{c}{ $\downarrow$ } & \multicolumn{1}{c}{
$\mu_-\;\downarrow$ } &$+$&$-$&$+$&$-$&$+$&$-$&$+$&$-$&$+$&$-$\\
\hline\hline
 &  0. & \rule{0mm}{3.5ex}$8538\pm    0$ & \ldots &
\rule{0mm}{3.5ex}$6548^{+ 432}_{- 427}$ & \ldots &
\rule{0mm}{3.5ex}$5322^{+ 598}_{- 584}$ & \ldots &
\rule{0mm}{3.5ex}$4498^{+ 657}_{- 632}$ & \ldots &
\rule{0mm}{3.5ex}$3909^{+ 669}_{- 635}$ & \ldots \\ &  .25 &
$8623\pm    0$ & \ldots & $6609^{+ 454}_{- 450}$ & \ldots & $5370^{+
626}_{- 613}$ & \ldots & $4537^{+ 685}_{- 662}$ & \ldots & $3943^{+
694}_{- 663}$ & \ldots \\ 5 &  .5 & $8882\pm    0$ & \ldots & $6799^{+
532}_{- 530}$ & \ldots & $5520^{+ 722}_{- 714}$ & \ldots & $4662^{+
781}_{- 765}$ & \ldots & $4050^{+ 783}_{- 759}$ & \ldots \\ &  .75 &
$9341\pm    0$ & \ldots & $7151\pm  723$ & \ldots & $5808\pm  955$ &
\ldots & $4907^{+1008}_{-1005}$ & \ldots & $4265^{+ 991}_{- 987}$ &
\ldots \\ &  1.  & \multicolumn{2}{c}{ \rule[-1.5ex]{0mm}{3ex}
\ldots } & \multicolumn{2}{c}{
\rule[-1.5ex]{0mm}{3ex}$8143^{+1701}_{-1792}$} &
\multicolumn{2}{c}{
\rule[-1.5ex]{0mm}{3ex}$6732^{+2184}_{-2193}$} &
\multicolumn{2}{c}{
\rule[-1.5ex]{0mm}{3ex}$5752^{+2208}_{-2196}$} &
\multicolumn{2}{c}{ \rule[-1.5ex]{0mm}{3ex}$5035\pm 2075$} \\
\hline &  0. & \rule{0mm}{3.5ex}$3066\pm    0$ &
\rule{0mm}{3.5ex}$4132\pm    0$ & \rule{0mm}{3.5ex}$2366\pm
81$ & \rule{0mm}{3.5ex}$3216\pm  480$ &
\rule{0mm}{3.5ex}$1914^{+ 113}_{- 110}$ &
\rule{0mm}{3.5ex}$2622\pm  586$ & \rule{0mm}{3.5ex}$1601^{+
125}_{- 120}$ & \rule{0mm}{3.5ex}$2204^{+ 574}_{- 582}$ &
\rule{0mm}{3.5ex}$1373^{+ 127}_{- 121}$ &
\rule{0mm}{3.5ex}$1897^{+ 526}_{- 546}$ \\ &  .25 & $3089\pm
0$ & $4121\pm    0$ & $2382\pm   84$ & $3204^{+ 469}_{- 467}$ &
$1926^{+ 117}_{- 114}$ & $2611\pm  573$ & $1611^{+ 129}_{- 123}$
& $2194^{+ 561}_{- 570}$ & $1381^{+ 131}_{- 124}$ & $1888^{+
515}_{- 535}$ \\ 10 &  .5 & $3161\pm    0$ & $4084\pm    0$ &
$2434\pm   93$ & $3168\pm  434$ & $1966^{+ 128}_{- 126}$ &
$2576^{+ 531}_{- 533}$ & $1642^{+ 140}_{- 136}$ & $2163^{+
523}_{- 533}$ & $1407^{+ 141}_{- 136}$ & $1859^{+ 481}_{- 501}$
\\ &  .75 & $3300\pm    0$ & $4005\pm    0$ & $2535\pm  115$ &
$3094\pm  370$ & $2044\pm  155$ & $2508^{+ 457}_{- 460}$ &
$1706^{+ 166}_{- 164}$ & $2101^{+ 453}_{- 464}$ & $1460^{+
165}_{- 162}$ & $1804^{+ 420}_{- 439}$ \\ &  1.  &
\multicolumn{2}{c}{ \rule[-1.5ex]{0mm}{3ex}$3695\pm    0$} &
\multicolumn{2}{c}{ \rule[-1.5ex]{0mm}{3ex}$2834\pm  213$} &
\multicolumn{2}{c}{ \rule[-1.5ex]{0mm}{3ex}$2284^{+ 273}_{-
276}$} & \multicolumn{2}{c}{ \rule[-1.5ex]{0mm}{3ex}$1905^{+
280}_{- 285}$} & \multicolumn{2}{c}{
\rule[-1.5ex]{0mm}{3ex}$1630^{+ 268}_{- 276}$} \\ \hline &  0. &
\rule{0mm}{3.5ex}$1708\pm    0$ & \rule{0mm}{3.5ex}$2276\pm
0$ & \rule{0mm}{3.5ex}$1317\pm   31$ & \rule{0mm}{3.5ex}$1741\pm
130$ & \rule{0mm}{3.5ex}$1062\pm   43$ &
\rule{0mm}{3.5ex}$1396^{+ 161}_{- 165}$ & \rule{0mm}{3.5ex}$
884^{+  48}_{-  46}$ & \rule{0mm}{3.5ex}$1157^{+ 160}_{- 167}$ &
\rule{0mm}{3.5ex}$ 754^{+  49}_{-  46}$ & \rule{0mm}{3.5ex}$
983^{+ 149}_{- 159}$ \\ &  .25 & $1718\pm    0$ & $2269\pm    0$
& $1325\pm   32$ & $1736\pm  128$ & $1068\pm   44$ & $1391^{+
158}_{- 162}$ & $ 889^{+  49}_{-  47}$ & $1153^{+ 157}_{- 164}$
& $ 758^{+  50}_{-  47}$ & $ 980^{+ 147}_{- 156}$ \\ 15 &  .5 &
$1752\pm    0$ & $2245\pm    0$ & $1349\pm   35$ & $1716\pm
120$ & $1086\pm   48$ & $1375^{+ 149}_{- 153}$ & $ 903\pm   52$
& $1139^{+ 149}_{- 155}$ & $ 770^{+  53}_{-  51}$ & $ 968^{+
139}_{- 148}$ \\ &  .75 & $1818\pm    0$ & $2195\pm    0$ &
$1396\pm   41$ & $1677\pm  105$ & $1122\pm   56$ & $1343^{+
132}_{- 135}$ & $ 932\pm   60$ & $1112^{+ 133}_{- 138}$ & $
793\pm   60$ & $ 945^{+ 125}_{- 132}$ \\ &  1.  &
\multicolumn{2}{c}{ \rule[-1.5ex]{0mm}{3ex}$2018\pm    0$} &
\multicolumn{2}{c}{ \rule[-1.5ex]{0mm}{3ex}$1542\pm   67$} &
\multicolumn{2}{c}{ \rule[-1.5ex]{0mm}{3ex}$1235\pm   87$} &
\multicolumn{2}{c}{ \rule[-1.5ex]{0mm}{3ex}$1023\pm   91$} &
\multicolumn{2}{c}{ \rule[-1.5ex]{0mm}{3ex}$ 869^{+  88}_{-
90}$} \\ \hline &  0. & \rule{0mm}{3.5ex}$1129\pm    0$ &
\rule{0mm}{3.5ex}$1476\pm    0$ & \rule{0mm}{3.5ex}$ 870\pm
16$ & \rule{0mm}{3.5ex}$1123\pm   54$ & \rule{0mm}{3.5ex}$
700\pm   22$ & \rule{0mm}{3.5ex}$ 895^{+  67}_{-  69}$ &
\rule{0mm}{3.5ex}$ 581\pm   24$ & \rule{0mm}{3.5ex}$ 738^{+
68}_{-  71}$ & \rule{0mm}{3.5ex}$ 494^{+  25}_{-  23}$ &
\rule{0mm}{3.5ex}$ 623^{+  64}_{-  68}$ \\ &  .25 & $1135\pm
0$ & $1471\pm    0$ & $ 874\pm   16$ & $1119\pm   53$ & $ 703\pm
22$ & $ 892^{+  66}_{-  68}$ & $ 584\pm   25$ & $ 735^{+  67}_{-
70}$ & $ 496\pm   25$ & $ 621^{+  63}_{-  67}$ \\ 20 &  .5 &
$1154\pm    0$ & $1455\pm    0$ & $ 888\pm   17$ & $1107\pm
50$ & $ 714\pm   24$ & $ 883^{+  63}_{-  65}$ & $ 592\pm   26$ &
$ 727^{+  64}_{-  67}$ & $ 503\pm   26$ & $ 615^{+  60}_{-  64}$
\\ &  .75 & $1193\pm    0$ & $1423\pm    0$ & $ 915\pm   20$ &
$1083\pm   45$ & $ 734\pm   27$ & $ 864\pm   57$ & $ 608\pm
29$ & $ 712^{+  58}_{-  60}$ & $ 516^{+  33}_{-  29}$ & $ 602^{+
55}_{-  58}$ \\ &  1.  & \multicolumn{2}{c}{
\rule[-1.5ex]{0mm}{3ex}$1312\pm    0$} & \multicolumn{2}{c}{
\rule[-1.5ex]{0mm}{3ex}$1001\pm   30$} & \multicolumn{2}{c}{
\rule[-1.5ex]{0mm}{3ex}$ 800\pm   40$} & \multicolumn{2}{c}{
\rule[-1.5ex]{0mm}{3ex}$ 660\pm   42$} & \multicolumn{2}{c}{
\rule[-1.5ex]{0mm}{3ex}$ 559\pm   41$} \\ \hline &  0. &
\rule{0mm}{3.5ex}$ 818\pm    0$ & \rule{0mm}{3.5ex}$1051\pm
0$ & \rule{0mm}{3.5ex}$ 630\pm    9$ & \rule{0mm}{3.5ex}$ 798\pm
28$ & \rule{0mm}{3.5ex}$ 506\pm   13$ & \rule{0mm}{3.5ex}$
634\pm   35$ & \rule{0mm}{3.5ex}$ 420\pm   14$ &
\rule{0mm}{3.5ex}$ 521\pm   36$ & \rule{0mm}{3.5ex}$ 356\pm
14$ & \rule{0mm}{3.5ex}$ 439^{+  34}_{-  36}$ \\ &  .25 & $
822\pm    0$ & $1048\pm    0$ & $ 633\pm    9$ & $ 796\pm   27$
& $ 508\pm   13$ & $ 632\pm   35$ & $ 421\pm   14$ & $ 520^{+
35}_{-  37}$ & $ 358\pm   15$ & $ 438^{+  33}_{-  35}$ \\ 25 &
.5 & $ 835\pm    0$ & $1037\pm    0$ & $ 642\pm   10$ & $ 787\pm
26$ & $ 515\pm   14$ & $ 626\pm   33$ & $ 427\pm   15$ & $
515\pm   34$ & $ 362\pm   15$ & $ 434^{+  32}_{-  34}$ \\ &  .75
& $ 860\pm    0$ & $1014\pm    0$ & $ 660\pm   11$ & $ 771\pm
23$ & $ 529\pm   16$ & $ 613\pm   30$ & $ 437\pm   17$ & $
504\pm   31$ & $ 370\pm   17$ & $ 425^{+  29}_{-  31}$ \\ &  1.
& \multicolumn{2}{c}{ \rule[-1.5ex]{0mm}{3ex}$ 939\pm    0$} &
\multicolumn{2}{c}{ \rule[-1.5ex]{0mm}{3ex}$ 716\pm   17$} &
\multicolumn{2}{c}{ \rule[-1.5ex]{0mm}{3ex}$ 572\pm   22$} &
\multicolumn{2}{c}{ \rule[-1.5ex]{0mm}{3ex}$ 471\pm   23$} &
\multicolumn{2}{c}{ \rule[-1.5ex]{0mm}{3ex}$ 398\pm   22$} \\
\hline &  0. & \rule{0mm}{3.5ex}$ 629\pm    0$ &
\rule{0mm}{3.5ex}$ 796\pm    0$ & \rule{0mm}{3.5ex}$ 484\pm
6$ & \rule{0mm}{3.5ex}$ 603\pm   16$ & \rule{0mm}{3.5ex}$ 388\pm
8$ & \rule{0mm}{3.5ex}$ 479\pm   21$ & \rule{0mm}{3.5ex}$ 322\pm
9$ & \rule{0mm}{3.5ex}$ 393\pm   21$ & \rule{0mm}{3.5ex}$ 273\pm
9$ & \rule{0mm}{3.5ex}$ 330\pm   20$ \\ &  .25 & $ 632\pm    0$
& $ 793\pm    0$ & $ 486\pm    6$ & $ 602\pm   16$ & $ 390\pm
9$ & $ 477\pm   20$ & $ 323\pm    9$ & $ 392\pm   21$ & $ 273\pm
9$ & $ 329\pm   20$ \\ 30 &  .5 & $ 641\pm    0$ & $ 785\pm
0$ & $ 492\pm    6$ & $ 596\pm   15$ & $ 395\pm    9$ & $ 473\pm
20$ & $ 326\pm   10$ & $ 388\pm   20$ & $ 276\pm   10$ & $
327\pm   19$ \\ &  .75 & $ 659\pm    0$ & $ 769\pm    0$ & $
505\pm    7$ & $ 584\pm   14$ & $ 404\pm   10$ & $ 464\pm   18$
& $ 334\pm   11$ & $ 381\pm   19$ & $ 282\pm   11$ & $ 321\pm
18$ \\ &  1.  & \multicolumn{2}{c}{ \rule[-1.5ex]{0mm}{3ex}$
715\pm    0$} & \multicolumn{2}{c}{ \rule[-1.5ex]{0mm}{3ex}$
545\pm   10$} & \multicolumn{2}{c}{ \rule[-1.5ex]{0mm}{3ex}$
434\pm   14$} & \multicolumn{2}{c}{ \rule[-1.5ex]{0mm}{3ex}$
357\pm   14$} & \multicolumn{2}{c}{ \rule[-1.5ex]{0mm}{3ex}$
302\pm   15$} \\ \hline &  0. & \rule{0mm}{3.5ex}$ 503\pm    0$
& \rule{0mm}{3.5ex}$ 628\pm    0$ & \rule{0mm}{3.5ex}$ 387\pm
4$ & \rule{0mm}{3.5ex}$ 476\pm   10$ & \rule{0mm}{3.5ex}$ 310\pm
6$ & \rule{0mm}{3.5ex}$ 378\pm   13$ & \rule{0mm}{3.5ex}$ 257\pm
6$ & \rule{0mm}{3.5ex}$ 309\pm   14$ & \rule{0mm}{3.5ex}$ 217\pm
6$ & \rule{0mm}{3.5ex}$ 260\pm   13$ \\ &  .25 & $ 505\pm    0$
& $ 627\pm    0$ & $ 389\pm    4$ & $ 475\pm   10$ & $ 312\pm
6$ & $ 377\pm   13$ & $ 258\pm    6$ & $ 309\pm   14$ & $ 218\pm
7$ & $ 259\pm   13$ \\ 35 &  .5 & $ 512\pm    0$ & $ 620\pm
0$ & $ 393\pm    4$ & $ 470\pm   10$ & $ 315\pm    6$ & $ 373\pm
13$ & $ 260\pm    7$ & $ 306\pm   13$ & $ 220\pm    7$ & $
257\pm   13$ \\ &  .75 & $ 525\pm    0$ & $ 608\pm    0$ & $
403\pm    5$ & $ 462\pm    9$ & $ 322\pm    7$ & $ 366\pm   12$
& $ 266\pm    7$ & $ 301\pm   12$ & $ 225\pm    7$ & $ 253\pm
12$ \\ &  1.  & \multicolumn{2}{c}{ \rule[-1.5ex]{0mm}{3ex}$
567\pm    0$} & \multicolumn{2}{c}{ \rule[-1.5ex]{0mm}{3ex}$
432\pm    7$} & \multicolumn{2}{c}{ \rule[-1.5ex]{0mm}{3ex}$
344\pm    9$} & \multicolumn{2}{c}{ \rule[-1.5ex]{0mm}{3ex}$
283\pm   10$} & \multicolumn{2}{c}{ \rule[-1.5ex]{0mm}{3ex}$
239\pm   10$} \\ \hline &  0. & \rule{0mm}{3.5ex}$ 415\pm    0$
& \rule{0mm}{3.5ex}$ 512\pm    0$ & \rule{0mm}{3.5ex}$ 319\pm
3$ & \rule{0mm}{3.5ex}$ 388\pm    7$ & \rule{0mm}{3.5ex}$ 256\pm
4$ & \rule{0mm}{3.5ex}$ 307\pm    9$ & \rule{0mm}{3.5ex}$ 211\pm
5$ & \rule{0mm}{3.5ex}$ 252\pm    9$ & \rule{0mm}{3.5ex}$ 179\pm
5$ & \rule{0mm}{3.5ex}$ 211\pm    9$ \\ &  .25 & $ 416\pm    0$
& $ 511\pm    0$ & $ 320\pm    3$ & $ 387\pm    7$ & $ 256\pm
4$ & $ 307\pm    9$ & $ 212\pm    5$ & $ 251\pm    9$ & $ 179\pm
5$ & $ 211\pm    9$ \\ 40 &  .5 & $ 422\pm    0$ & $ 506\pm
0$ & $ 324\pm    3$ & $ 383\pm    7$ & $ 259\pm    4$ & $ 304\pm
9$ & $ 214\pm    5$ & $ 249\pm    9$ & $ 181\pm    5$ & $ 209\pm
9$ \\ &  .75 & $ 432\pm    0$ & $ 496\pm    0$ & $ 331\pm    4$
& $ 377\pm    6$ & $ 264\pm    5$ & $ 299\pm    8$ & $ 218\pm
5$ & $ 245\pm    8$ & $ 184\pm    5$ & $ 206\pm    8$ \\ &  1.
& \multicolumn{2}{c}{ \rule[-1.5ex]{0mm}{3ex}$ 464\pm    0$} &
\multicolumn{2}{c}{ \rule[-1.5ex]{0mm}{3ex}$ 354\pm    5$} &
\multicolumn{2}{c}{ \rule[-1.5ex]{0mm}{3ex}$ 282\pm    6$} &
\multicolumn{2}{c}{ \rule[-1.5ex]{0mm}{3ex}$ 232\pm    7$} &
\multicolumn{2}{c}{ \rule[-1.5ex]{0mm}{3ex}$ 195\pm    7$} \\
\hline &  0. & \rule{0mm}{3.5ex}$ 350\pm    0$ &
\rule{0mm}{3.5ex}$ 428\pm    0$ & \rule{0mm}{3.5ex}$ 269\pm
2$ & \rule{0mm}{3.5ex}$ 324\pm    5$ & \rule{0mm}{3.5ex}$ 215\pm
3$ & \rule{0mm}{3.5ex}$ 256\pm    7$ & \rule{0mm}{3.5ex}$ 178\pm
4$ & \rule{0mm}{3.5ex}$ 210\pm    7$ & \rule{0mm}{3.5ex}$ 150\pm
4$ & \rule{0mm}{3.5ex}$ 176\pm    7$ \\ &  .25 & $ 351\pm    0$
& $ 426\pm    0$ & $ 270\pm    2$ & $ 323\pm    5$ & $ 216\pm
3$ & $ 256\pm    7$ & $ 178\pm    4$ & $ 209\pm    7$ & $ 151\pm
4$ & $ 176\pm    7$ \\ 45 &  .5 & $ 355\pm    0$ & $ 422\pm
0$ & $ 272\pm    2$ & $ 320\pm    5$ & $ 218\pm    3$ & $ 254\pm
6$ & $ 180\pm    4$ & $ 208\pm    7$ & $ 152\pm    4$ & $ 174\pm
6$ \\ &  .75 & $ 363\pm    0$ & $ 415\pm    0$ & $ 278\pm    3$
& $ 315\pm    5$ & $ 222\pm    4$ & $ 250\pm    6$ & $ 183\pm
4$ & $ 204\pm    6$ & $ 155\pm    4$ & $ 172\pm    6$ \\ &  1.
& \multicolumn{2}{c}{ \rule[-1.5ex]{0mm}{3ex}$ 389\pm    0$} &
\multicolumn{2}{c}{ \rule[-1.5ex]{0mm}{3ex}$ 297\pm    4$} &
\multicolumn{2}{c}{ \rule[-1.5ex]{0mm}{3ex}$ 236\pm    5$} &
\multicolumn{2}{c}{ \rule[-1.5ex]{0mm}{3ex}$ 194\pm    5$} &
\multicolumn{2}{c}{ \rule[-1.5ex]{0mm}{3ex}$ 163\pm    5$} \\
\hline &  0. & \rule{0mm}{3.5ex}$ 300\pm    0$ &
\rule{0mm}{3.5ex}$ 364\pm    0$ & \rule{0mm}{3.5ex}$ 230\pm
2$ & \rule{0mm}{3.5ex}$ 275\pm    4$ & \rule{0mm}{3.5ex}$ 185\pm
2$ & \rule{0mm}{3.5ex}$ 218\pm    5$ & \rule{0mm}{3.5ex}$ 152\pm
3$ & \rule{0mm}{3.5ex}$ 178\pm    5$ & \rule{0mm}{3.5ex}$ 129\pm
3$ & \rule{0mm}{3.5ex}$ 150\pm    5$ \\ &  .25 & $ 301\pm    0$
& $ 363\pm    0$ & $ 231\pm    2$ & $ 275\pm    4$ & $ 185\pm
3$ & $ 218\pm    5$ & $ 153\pm    3$ & $ 178\pm    5$ & $ 129\pm
3$ & $ 149\pm    5$ \\ 50 &  .5 & $ 305\pm    0$ & $ 360\pm
0$ & $ 234\pm    2$ & $ 272\pm    4$ & $ 187\pm    3$ & $ 216\pm
5$ & $ 154\pm    3$ & $ 177\pm    5$ & $ 130\pm    3$ & $ 148\pm
5$ \\ &  .75 & $ 311\pm    0$ & $ 353\pm    0$ & $ 238\pm    2$
& $ 268\pm    3$ & $ 190\pm    3$ & $ 212\pm    4$ & $ 157\pm
3$ & $ 174\pm    5$ & $ 132\pm    3$ & $ 146\pm    5$ \\ &  1.
& \multicolumn{2}{c}{ \rule[-1.5ex]{0mm}{3ex}$ 332\pm    0$} &
\multicolumn{2}{c}{ \rule[-1.5ex]{0mm}{3ex}$ 253\pm    3$} &
\multicolumn{2}{c}{ \rule[-1.5ex]{0mm}{3ex}$ 201\pm    4$} &
\multicolumn{2}{c}{ \rule[-1.5ex]{0mm}{3ex}$ 165\pm    4$} &
\multicolumn{2}{c}{ \rule[-1.5ex]{0mm}{3ex}$ 139\pm    4$} \\
\hline
\end{tabular}

\end{document}